\documentclass[twocolumn,aps,pra,groupedaddress,longbibliography]{revtex4-1}
\usepackage[T1]{fontenc}
\usepackage[latin9]{inputenc}
\usepackage{prettyref}
\usepackage{bm}
\usepackage{amsmath}
\usepackage{amssymb}
\usepackage{graphicx}

\makeatletter
\usepackage{commath}
\usepackage{bm}
\usepackage{siunitx}

\renewcommand{\vec}{\bm}
\newcommand{\Ei}{\operatorname{Ei}}

\makeatother

\begin{document}
\global\long\def\vmiinfy{V_{-\infty}}%
\global\long\def\vmiinfyi{V_{-\infty,i}}%
\global\long\def\vinf{V_{\infty}}%
\global\long\def\vinfi{V_{\infty,i}}%
\global\long\def\vth{V_{\text{th}}}%
\global\long\def\vre{V_{\text{re}}}%
\global\long\def\vcut{V_{\text{cutoff}}}%
\global\long\def\ie{I^{\text{ext}}}%
\global\long\def\LIF{\text{LIF}}%
\global\long\def\XIF{\text{XIF}}%
\global\long\def\e{\mathrm{e}}%
\global\long\def\ei{\Ei}%
\global\long\def\mf{\text{mf}}%
\global\long\def\vout{V_{\text{out}}}%
\global\long\def\tout{t_{\text{out}}}%
\global\long\def\Err{\text{Err}}%

\title{Dynamics and computation in mixed networks containing neurons that
accelerate towards spiking}
\author{Paul Manz}
\author{Sven Goedeke}
\author{Raoul-Martin Memmesheimer}
\affiliation{Neural Network Dynamics and Computation, Institute for Genetics, University
of Bonn, 53115 Bonn, Germany.}
\begin{abstract}
Networks in the brain consist of different types of neurons. Here
we investigate the influence of neuron diversity on the dynamics,
phase space structure and computational capabilities of spiking neural
networks. We find that already a single neuron of a different type
can qualitatively change the network dynamics and that mixed networks
may combine the computational capabilities of ones with a single neuron
type. We study inhibitory networks of concave leaky (LIF) and convex
``anti-leaky'' (XIF) integrate-and-fire neurons that generalize
irregularly spiking non-chaotic LIF neuron networks. Endowed with
simple conductance-based synapses for XIF neurons, our networks can
generate a balanced state of irregular asynchronous spiking as well.
We determine the voltage probability distributions and self-consistent
firing rates assuming Poisson input with finite size spike impacts.
Further, we compute the full spectrum of Lyapunov exponents (LEs)
and the covariant Lyapunov vectors (CLVs) specifying the corresponding
perturbation directions. We find that there is approximately one positive
LE for each XIF neuron. This indicates in particular that a single
XIF neuron renders the network dynamics chaotic. A simple mean-field
approach, which can be justified by properties of the CLVs, explains
the finding. As an application, we propose a spike-based computing
scheme where our networks serve as computational reservoirs and their
different stability properties yield different computational capabilities.
\end{abstract}
\maketitle

\section{Introduction}

Biological neural networks consist of a large variety of interconnected
neurons, which communicate via short stereotypical electrical pulses
called action potentials or spikes. After a neuron has generated a
spike, this travels along the axon and is transmitted to other neurons
at synaptic contacts. The electrical membrane potential of the receiving
neuron is then changed by an excitatory or inhibitory current pulse.
Sufficiently many excitatory inputs in turn lead to spike generation
in a receiving neuron. Many biological neural networks generate irregular
and asynchronous spiking. This is likely caused by a dynamically balanced
network state, in which the average inhibitory and excitatory input
current to each neuron sum to a value that is insufficient for frequent
spike generation \citep{GM64,shadlen1994noise,Vreeswijk1996,deneve2016efficient}.
Spikes are caused by fluctuations in the inputs and the resulting
spiking dynamics appear random and irregular.

Irregular dynamics are often chaotic, implying that the dynamics are
sensitive to perturbations: initially small ones can strongly grow
with time, which results in ultimately large quantitative differences
between perturbed and unperturbed trajectories. A powerful tool to
quantify this sensitivity and therewith the local phase space structure
are the Lyapunov exponents (LEs) and associated with them the covariant
Lyapunov vectors (CLVs) \citep{pikovsky+politi,Kuptsov2012}. The
sign of the largest LE indicates whether the system is chaotic and
its magnitude equals the long-term average growth or decay rate of
generic infinitesimal perturbations. The spectrum of LEs describes
the long-term average evolution of volumes spanned by tangent vectors
and the change of infinitesimal perturbations in non-generic directions,
which are specified by the CLVs. To each LE, there is a CLV. The size
of a perturbation in the CLV's direction changes with an average rate
of plus or minus the corresponding LE for long-term forward or backward
time evolution, respectively. The CLVs thereby indicate the directions
of the unstable and stable manifolds along a trajectory. Furthermore,
the spectrum of LEs can be used to derive dynamical quantities such
as the Kaplan-Yorke fractal dimension of a chaotic attractor \citep{KY2}.

In our study, we consider purely inhibitory networks of current-based,
oscillating integrate-and-fire type neurons with post-synaptic currents
of infinitesimally short duration and instantaneous reset. It has
been shown numerically \citep{ZLPT06,ZBH09} and analytically \citep{JMT08,JMT09}
that if such networks contain only leaky integrate-and-fire (LIF)
neurons, the networks' irregular balanced state dynamics are stable
against infinitesimal and small finite size perturbations and are
thus not chaotic but a realization of stable chaos \citep{Politi_1993,Politi2010}.
The dynamics ultimately converge to a periodic orbit; the durations
of the preceding irregular transients, however, grow exponentially
with system size. The stability of the network dynamics is robust
against introducing excitatory connections and considering synaptic
currents of finite temporal extent \citep{ZBH09,JMT09} and there
is a smooth transition to chaos upon increasing the number of excitatory
connections and the duration of synaptic currents. The computational
abilities of the stable precise spiking dynamics have not yet been
explored, even though the specific structure of the phase space, which
is composed of ``flux tubes'', may be beneficial and exploitable
\citep{MW12}.

LIF neurons incorporate a leak current as found in biological neurons
\citep{DA01}. This increases linearly with increasing membrane potential
and leads to dissipation (contraction of phase space volume) in the
subthreshold dynamics. When driven by a constant depolarizing input
current, the membrane potential therefore has negative second derivative;
the neuron has a purely concave so-called rise function. In the considered
class of networks, this implies the stability of the microscopic dynamics
if only LIF neurons are present \citep{JMT08,JMT09}. In biological
neurons as well as in neuron models that explicitly model spike generation,
such as the quadratic and the exponential integrate-and-fire neuron
\citep{gerstner2014neuronal}, the membrane potential accelerates
towards a spike for larger membrane potentials. The rise function
thus has a convex part. Ref.~\citep{MW10} showed that networks of
quadratic integrate-and-fire neurons that are otherwise similar to
those considered in refs.~\citep{ZBH09,JMT08,JMT09} exhibit chaos.
Furthermore, ref.~\citep{MW10} computed the spectrum of LEs and
quantities that are derivable from them, as well as the statistics
of the first CLV, which points into the directions to which a generic
perturbation vector aligns in the long term.

Motivated by the above results and by the fact that there are many
different types of cortical inhibitory interneurons \citep{tremblay2016gabaergic},
in the present study we investigate the impact of inserting a different
type of neuron, with non-concave rise function, into inhibitory networks
of LIF neurons. To be specific, we insert ``anti-leaky'' integrate-and-fire
(XIF) neurons with purely convex rise function. We choose the letter
``X'' in the abbreviation to highlight this convexity and the expansion
of phase space volume by the flow of the subthreshold dynamics. XIF
neurons may be interpreted as a model for a class of biological neurons
whose membrane potential lingers in a region where it accelerates
towards spiking. Simultaneously, these neurons maintain similar analytical
tractability as their leaky counterparts because of their mostly linear
subthreshold dynamics. We describe our neuron and network models in
detail in the next section. Thereafter, self-consistent firing rates
and membrane potential probability distributions for both types of
neurons are analytically derived, assuming Poisson input with finite
size spike impacts. We then consider the dynamical stability properties
and local phase space structures of the network dynamics, computing
the entire spectra of LEs both numerically and analytically in a mean-field
approximation. We also compute their CLVs to investigate how the stable
and unstable directions are related to the different neuron types
within the network.

Finally, we consider computations in pure and mixed networks of the
considered types and show how the richer phase space structure in
mixed ones can be exploited. For this, we propose a reservoir computer
based entirely on precisely timed spikes. Reservoir computing has
been introduced several times at different levels of elaborateness
and in different flavors, in machine learning and in neuroscience
\citep{BM95,dominey1995complex,JH04,MNM02}. A reservoir computer
consists of a high dimensional, nonlinear dynamical system, the reservoir
or liquid, and a comparably simple readout. The reservoir ``echoes''
the input in a complicated, nonlinear way; it acts like a random filter
bank with finite memory as each of its units generates a nonlinearly
filtered version of the current input and its recent past while forgetting
more remote inputs \citep{BM95,westover2002linearly,MNM02,JH04}.
The simple, often linear readout can then be trained to extract the
desired results, while the reservoir is static. In our scheme, the
output neuron is spiking and thus nonlinear, the desired outputs are
trains of precisely timed spikes. The learning thus requires different
approaches than learning of conventional continuous targets; gradient-descent
based methods \citep{lecun2015deep} fail due to the discontinuity
at the threshold as well as methods that require errors to be small
but finite \citep{SA09}. A number of algorithms have been suggested
to learn precisely timed spikes \citep{PK10,Florian2012,MSM+12,XZZ13,MRS13,albers16learning,zenke2018superspike,huh18gradient},
mostly using heuristic approaches. For our readout neuron, we can
use the Finite Precision Learning scheme \citep{MRS13}. It has been
shown to generically converge if the input-output relation is realizable
at all, which explains its numerically found superior learning abilities
\citep{albers16learning}.

\section{Mixed networks of neurons with concave and convex rise function\label{sec:Mixed-networks}}

We consider a recurrent network with $N$ neurons. The $k$th spike
of neuron $j$, which is sent at time $t_{jk}$, generates a postsynaptic
current pulse $h_{i}(V_{i}^{-})C_{ij}\delta\del{t-t_{jk}}$ in neuron
$i$. Here $C_{ij}\leq0$ is the weight of the inhibitory connection
and $h_{i}(V_{i}^{-})$ is a possible voltage-dependent modulation,
which depends on the membrane potential of neuron $i$ just before
input arrival, given by the left-hand side limit $V_{i}^{-}=V_{i}(t^{-})=\lim_{\varepsilon\searrow0}V_{i}(t-\varepsilon)$.
We assume that all excitatory inputs to neuron $i$ can be gathered
into a constant excitatory external input current $I_{i}^{\mathrm{ext}}>0$
and that the remaining explicitly modeled recurrent inhibition is
fast \citep{JMT08,MW12,olmi2017exact}. We further assume that there
is a leak term with prefactor $\gamma_{i}\neq0$. Taken together,
we model the subthreshold membrane potential dynamics of neuron $i$
by
\begin{equation}
\dot{V}_{i}=-\gamma_{i}V_{i}+I_{i}^{\mathrm{ext}}+h_{i}(V_{i}^{-})\sum_{j=1}^{N}C_{ij}\sum_{k}\delta\del{t-t_{jk}}.\label{eq:IF}
\end{equation}
When $V_{i}$ reaches the spike threshold at time $t$, $V_{i}{}^{-}=\vth>0$,
it is reset, $V(t)=\vre=0$, and a spike is emitted. This, in turn,
generates in a postsynaptic neuron $l$ a current pulse as introduced
above, which causes $V_{l}$ to decrease in jump-like manner from
$V_{l}^{-}$ to $V_{l}^{-}+h_{l}(V_{l}^{-})C_{li}$. The rise function,
i.e.,~the membrane potential dynamics with $V_{i}(0)=0$ in absence
of recurrent inhibitory input \citep{MS90,MT06B}, reads
\begin{equation}
V_{i}(t)=\frac{I_{i}^{\mathrm{ext}}}{\gamma_{i}}\left[1-\exp(-\gamma_{i}t)\right].\label{eq:RiseFunc}
\end{equation}
It is concave for $\gamma_{i}>0$ and convex for $\gamma_{i}<0$.

\begin{figure}
\includegraphics[width=1\columnwidth]{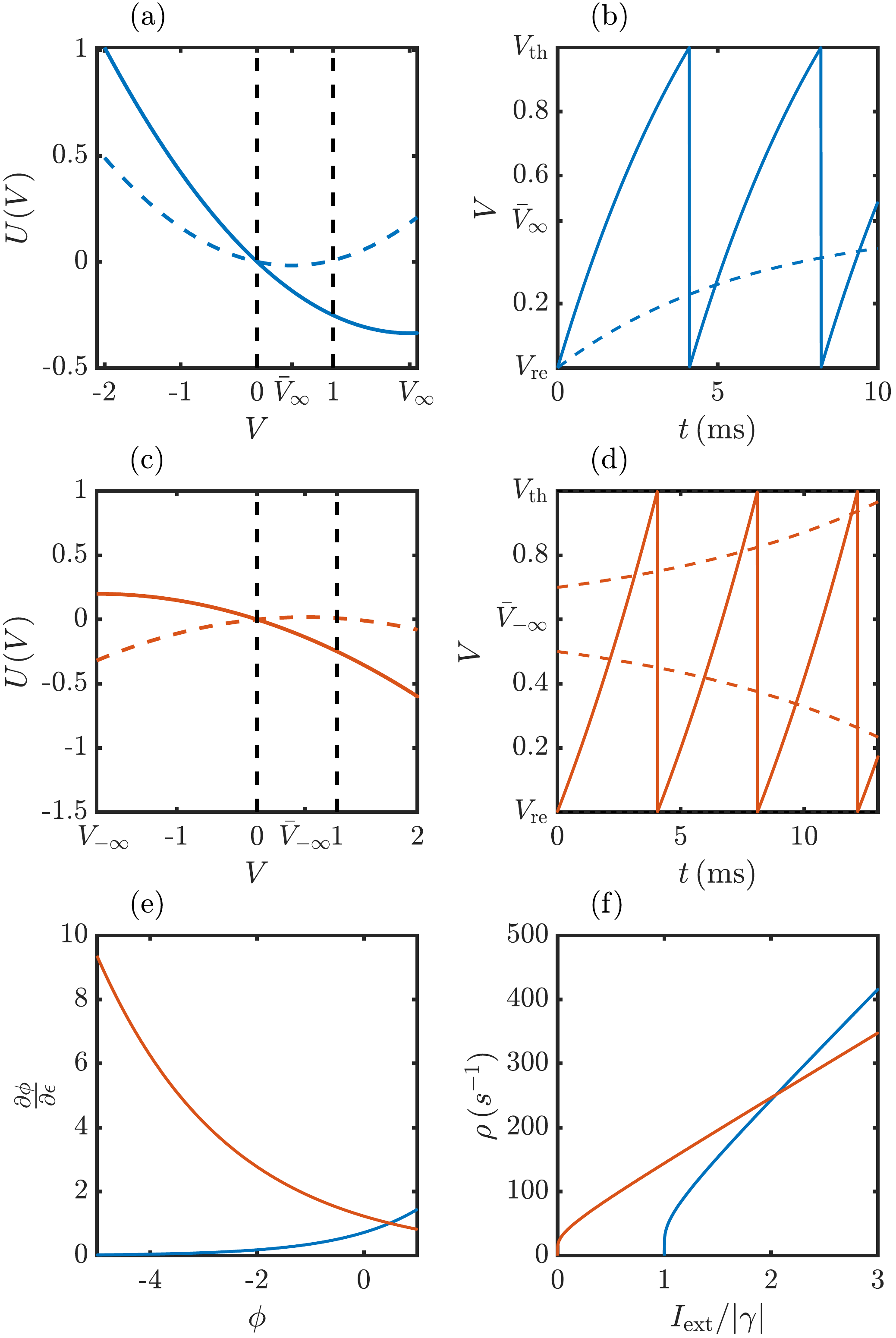}

\caption{LIF and XIF neuron dynamics with constant input. Blue and red indicate
LIF and XIF neurons, respectively. Solid curves in (a-d) indicate
excitatory constant input only, dashed lines inclusion of an average
inhibitory input current $I^{\text{inh}}$ (the cutoff for XIF inputs
is neglected). (a,c) Potential function $U(V)$ of the membrane potential
(voltage) $V$; $V$ follows the negative gradient of $U$, $\dot{V}(t)=-U'(V(t))$,
if there is no threshold. (a) $U$ for an LIF neuron is an upward
parabola; $V$ tends to the stable fixed point at $U$'s minimum (at
$V_{\infty}$ or $\bar{V}_{\infty}$), if there is no threshold. (c)
$U$ for an XIF neuron is a downward parabola; $V$ tends to $-\infty$
or $+\infty$ when starting left or right of $U$'s maximum (at $V_{-\infty}$
or $\bar{V}_{-\infty}$), if there is no threshold. A monotonically
decreasing potential function $U$ between reset potential and threshold
(left and right vertical dashed lines) indicates mean driven periodic
spiking (solid curves in (a,c)). In the balanced state the spiking
is fluctuation driven with (a) $U$'s minimum below threshold for
LIF neurons and (c) $U$'s maximum above reset for XIF neurons (dashed
curves). (b,d) Example trajectories for LIF and XIF dynamics including
threshold and reset. Without inhibition, $V$ is periodically driven
over the threshold and reset. Once averaged inhibition is included,
the LIF voltage (b) converges to the subthreshold fixed point at $\bar{V}_{\infty}$,
while the XIF voltage (d) is repelled from $\bar{V}_{-\infty}$. (e)
Infinitesimal phase response curves \citep{goel2002synchrony,smeal2010phase,viriyopase18analyzing}.
Inputs to LIF (XIF) neurons have a smaller (larger) spike delaying
effect, the lower $V$ is. (f) Rates of free LIF and XIF neurons at
different strengths of the normalized external drive. Neuron parameters
and (if applicable) values for excitatory drive and average inhibitory
input are as in our network simulations.\label{fig:.PotentialPlotsFigure}}
\end{figure}
There are two types of neurons in our networks: LIF neurons with dissipation
and concave rise function, which obey Eq.~\eqref{eq:IF} with $\gamma_{i}>0$,
and anti-leaky XIF neurons with convex rise function, which obey Eq.~\eqref{eq:IF}
with $\gamma_{i}<0$, see Fig.~\ref{fig:.PotentialPlotsFigure}.
The membrane potential dynamics of an LIF neuron has a globally attracting
fixed point at $\vinfi=I_{i}^{\mathrm{ext}}/\gamma_{i}$, if there
is no threshold for spike generation and no inhibitory input. We assume
$\vinfi>\vth$, so neurons without inhibitory input periodically spike
and reset. For our study it is sufficient to endow the LIF neurons
with a simple, current-based synapse model, setting $h_{i}(V_{i}^{-})=1$.
A coarse approximation of the membrane potential dynamics without
threshold and neglecting input fluctuations yields $\dot{\bar{V}}_{i}=-\gamma_{i}\bar{V}_{i}+I_{i}^{\mathrm{ext}}+I_{i}^{\text{inh}}$,
where $I_{i}^{\text{inh}}$ is the average inhibitory input current.
In the balanced state, its attractor at $\bar{V}_{\infty,i}=\left(I_{i}^{\mathrm{ext}}+I_{i}^{\text{inh}}\right)/\gamma_{i}$
is below or close to the spike threshold, such that spikes are always
or typically generated by input fluctuations, more specifically by
periods of less than average inhibition.

In the absence of inhibitory input XIF neurons have an unstable, repelling
fixed point at $\vmiinfyi=I_{i}^{\mathrm{ext}}/\gamma_{i}<0$. If
the membrane potential starts above this separatrix, it increases
exponentially towards the threshold. When it reaches there, the neuron
spikes, its membrane potential resets to zero, increases towards the
threshold again and so forth: XIF neurons oscillate and spike periodically
for any $I_{i}^{\mathrm{ext}}>0$ , if there is no inhibitory input.
If the membrane potential starts below the separatrix, it decreases
exponentially to $-\infty$. Also in the presence of recurrent inhibitory
inputs an XIF neuron is unrecoverably switched off once its membrane
potential falls below $I_{i}^{\mathrm{ext}}/\gamma_{i}$, since the
inputs only decrease the membrane potential further. Averaging over
the inhibitory inputs as before yields an effective separatrix at
$\bar{V}_{-\infty,i}=\left(I_{i}^{\mathrm{ext}}+I_{i}^{\text{inh}}\right)/\gamma_{i}$
. Membrane potentials falling below it have a tendency to further
decrease, causing the neuron to effectively switch off. This can be
also seen from the phase response curve of XIF neurons, which gets
steeper for negative phases, in contrast to that of LIF neurons which
becomes flatter, see Fig.~\ref{fig:.PotentialPlotsFigure}c. In other
words, in XIF neurons an incoming inhibitory input at a low potential
still above the separatrix (and thus at a low phase) has a larger
effect in the sense that it delays the next spiking more than the
same input arriving at a higher potential. As a consequence, we observe
in networks containing XIF neurons with purely current-based input
{[}$h(V_{i})=1${]} that many of these neurons are first effectively
and then unrecoverably switched off, if the network dynamics are irregular
and the inhibitory inputs are therefore strongly fluctuating. In order
to prevent this biologically implausible phenomenon, we introduce
a voltage dependence
\begin{equation}
h(V_{i}^{-})=\Theta\del{V_{i}^{-}-V_{\text{cutoff}}}\label{eq:Voltage dependent coupling}
\end{equation}
of the input coupling strength, where $\Theta$ is the Heaviside theta
function. Inhibitory inputs arriving at a membrane potential lower
than $V_{\text{cutoff}}$ then do not induce a further decrease. This
provides a simple conductance-based model for the synapses, where
the driving force of the current vanishes below $V_{i}=V_{\text{cutoff}}$
and is constant above. We assume $\vmiinfyi<\vcut+C_{ij}$ for all
$j$ to exclude unrecoverable switching off and $\vcut\leq\vre$.
We exemplarily checked that the overall network dynamics and their
stability properties remain qualitatively unchanged, if we also endow
the LIF neurons with these synapses.

For simplicity, we choose the parameters of all LIF and of all XIF
neurons identical, i.e., $\gamma_{i}=\gamma_{\LIF},$ $I_{i}=I_{\LIF}$,
etc., if neuron $i$ is an LIF neuron, and $\gamma_{i}=\gamma_{\XIF},$
$I_{i}=I_{\XIF}$, etc., if neuron $i$ is an XIF neuron. The spike
threshold and reset potentials are $\vth=1$ and $\vre=0$, independent
of the neuron type. We set $\vcut=\vre$ to avoid any effective switching
off of XIF neurons. Coupling strengths are homogeneous, $C_{ij}=C$
if the coupling is present. To keep the number of relevant parameters
small, we further choose $I_{\LIF}^{\text{ext}}/\gamma_{\LIF}=V_{\infty,\LIF}=-V_{-\infty,\XIF}=-I_{\XIF}^{\text{ext}}/\gamma_{\XIF}$.
The additional choice $\gamma_{\XIF}=-\gamma_{\LIF}$ leads already
in absence of recurrent inhibition to a higher spike rate $\rho_{\text{\text{free},XIF}}$
in XIF neurons, since
\begin{equation}
\rho_{\text{\text{free},XIF}}=\frac{-\gamma_{\XIF}}{\ln\left(\frac{-V_{-\infty,\XIF}+V_{\text{th}}}{-V_{-\infty,\XIF}}\right)},\label{eq:ffreeXIF}
\end{equation}
whereas in LIF neurons,
\begin{equation}
\rho_{\text{\text{free},\text{LIF}}}=\frac{\gamma_{\LIF}}{\ln\left(\frac{V_{\infty,\LIF}}{V_{\infty,\LIF}-V_{\text{th}}}\right)}.\label{eq:ffreeLIF}
\end{equation}
As a consequence, we observe that in a mixed network the XIF suppress
the LIF neurons, which become quiescent. Using the analytical results
of the next section, we therefore rescale $\gamma_{\LIF}$ such that
the spike rates in both populations are identical. Further, we fix
the neurons' indegree to the same number $K$, implying that $\sum_{j}C_{ij}$
is identical for each neuron $i$. This reduces quenched noise \citep{Vreeswijk1998}
and avoids strong differences in average spike rates and switched
off neurons.

\begin{figure}
\includegraphics[width=1\columnwidth]{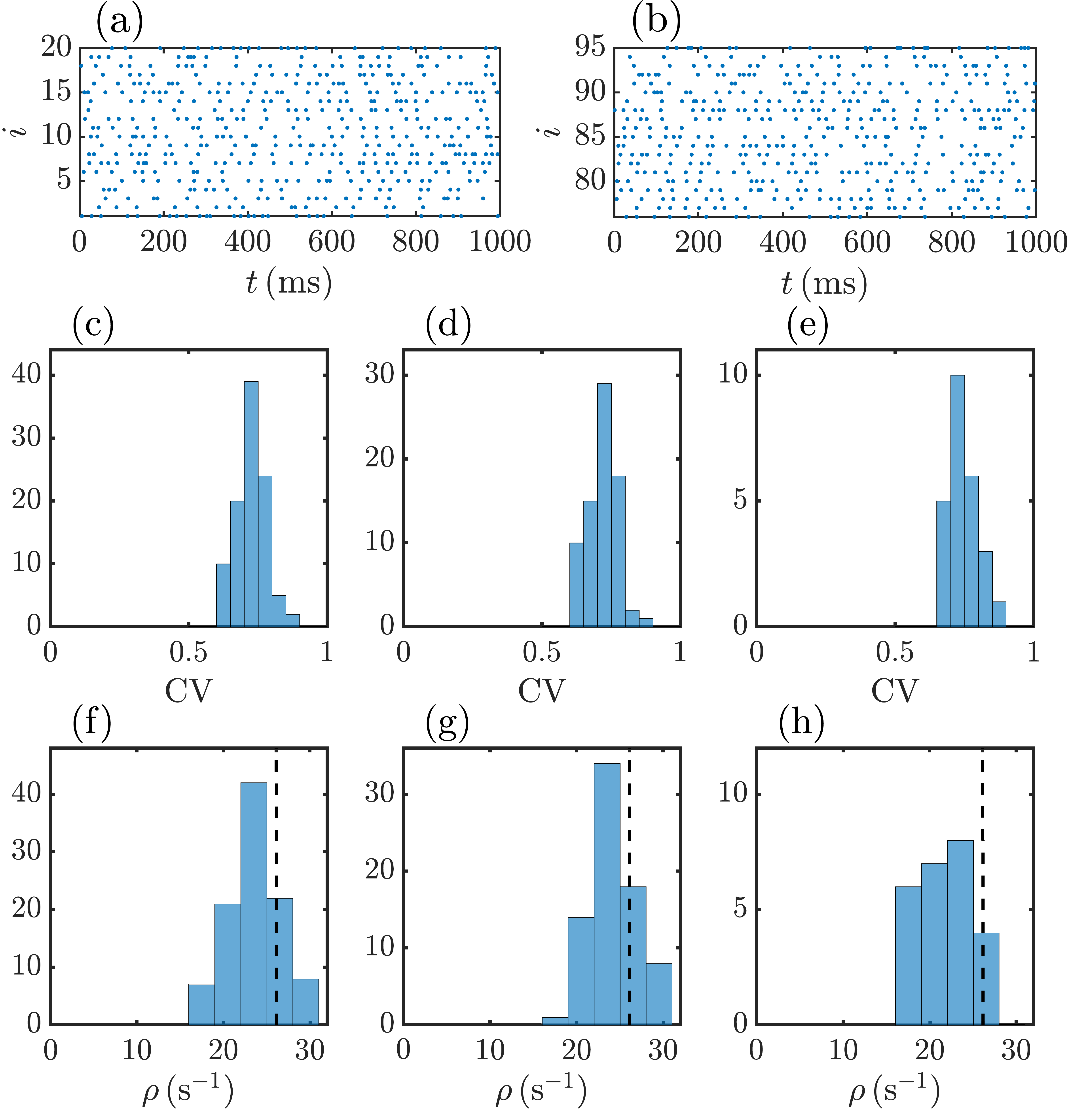} \caption{\label{fig:rasterplot}Mixed networks of LIF neurons with concave
rise function and of XIF neurons with convex rise function can exhibit
a balanced state with asynchronous irregular activity in both types
of neurons. (a,b) Spiking activity for a subset of the LIF (a) and
the XIF (b) neurons in a network with $75$ LIF and $25$ XIF neurons
($N=100$). (c-e) Distribution of coefficients of variation of inter-spike
intervals for all neurons (c) and for LIF (d) and XIF (e) neurons
separately. (f-h) Distribution of the average spike rates of all neurons
(f) and of LIF (g) and XIF (h) neurons separately. The analytically
derived rate $\rho\approx\SI{26.1}{\per\second}$ (Eqs.~\eqref{eq:GLIF self-consistent},\eqref{eq:GXIF self-consistent})
is indicated by a black dashed vertical line. We use $\gamma_{\protect\XIF}=-\SI{0.1}{\per\milli\second}$,
$\gamma_{\protect\LIF}=\SI{0.169}{\per\milli\second}$, $V_{\infty,\protect\LIF}=-V_{-\infty,\protect\XIF}=2$
and a randomly connected network with fixed indegrees $K=50$ and
non-zero synaptic strength $C_{ij}=C=-0.2$.}
\end{figure}

With the described network model setup, we observe balanced states
of asynchronous irregular spiking activity for any ratio of neurons
with concave and convex rise function, see Fig.~\ref{fig:rasterplot}
for an illustration.

\section{Network firing rate and membrane potential distributions\label{sec:Network-firing-rate}}

Mean-field theories have been developed in statistical physics \citep{kadanoff2009more}
and are frequently used in computational neuroscience, see, for example,
refs.~\citep{Vreeswijk1998,Brunel00,BTM13,schuecker18optimal}. The
basic idea is to average the interactions in a high-dimensional system
to obtain for each element an effective action, which is not influenced
by this element anymore. One can thereby reduce a high-dimensional
problem to low-dimensional ones. In this section we analytically determine
the steady-state firing rate and the voltage probability densities
for LIF and XIF neurons in mixed networks using a mean-field approximation.
We use the results to obtain neuron parameters that lead to the same
average firing rates for both neuron types and thus to homogeneous
firing rates in the entire network. In addition we employ the firing
rates to analytically approximate the Lyapunov spectrum of the network
dynamics using a mean-field approach in Sec.~\ref{sec:Meanfield}.

We approximate the superposed input spike trains to a neuron by a
Poisson spike train with a given rate, i.e.~we assume that all input
spikes are sent independently of each other. A common approach is
to additionally consider the limit of a large number of small inputs.
The neuron dynamics can then be approximated by a diffusion process,
which allows to compute firing rates and membrane potential distributions
\citep{T88b,B06a}. This diffusion approximation assumes that the
inputs have (infinitesimally) small amplitude and arrive at (infinitely)
high rate. Here we use a shot noise approach, which accounts for the
finite input rate and size of individual inputs \citep{T88b,B06a},
in the recent formulation of refs.~\citep{richardson,olmi2017exact}.
This allows to more accurately obtain the firing rates and membrane
potential distributions. In particular, the fact that in our networks
the voltage probability density does not go to zero at threshold is
reflected. We shortly review the approach for LIF neurons \citep{richardson,olmi2017exact,angulo-garcia}
and then extend it to XIF neurons with the voltage-dependent coupling
Eq.~\eqref{eq:Voltage dependent coupling}.

The shot-noise approach (like the diffusion approximation) is based
on the continuity equation for the voltage probability density $p(V,t)$.
For our neuron models it reads
\begin{equation}
\frac{\partial p}{\partial t}+\frac{\partial j}{\partial V}=\sigma_{\text{inh}}+\sigma_{\text{reset}},\label{eq:Continuity equation}
\end{equation}
where $j(V,t)=\dot{V}(V)p(V,t)$ is the drift probability current
with velocity $\dot{V}(V)=-\gamma V+I^{\text{ext}}$. $\sigma_{\text{inh}}(V,t)$
and $\sigma_{\text{reset}}(V,t)$ are source terms incorporating the
effects of inputs and resets of the neuron's membrane potential $V$.

For the LIF neuron without the voltage-dependent input, inhibitory
input spikes arriving when the considered neuron is at a voltage $V$
give rise to a sink at $V$, whereas spikes arriving when the neuron
is at a voltage $V-C>V$ give rise to a source at $V$. We therefore
have a first source term
\begin{equation}
\sigma_{\text{inh}}(V,t)=r(t)\left[p(V-C,t)-p(V,t)\right]\label{eq:sigma inh}
\end{equation}
with the rate $r(t)$ of input spikes. We note that refs.~\citep{richardson,olmi2017exact,angulo-garcia}
include this term in the probability current. The second source term
is due to the spike and reset mechanism of the neuron model. Its threshold
and reset act as Dirac delta sink and source at the corresponding
discrete voltages,
\begin{equation}
\sigma_{\text{reset}}(V,t)=\rho(t)\left[\delta(V-\vre)-\delta(V-\vth)\right].\label{eq:sigma reset}
\end{equation}
This term is proportional to the instantaneous firing rate $\rho(t)$
of the stochastic neuron dynamics or, in other words, to the probability
current through the threshold {[}$\rho(t)=j(V_{\text{th}},t)\geq0${]}.

We investigate stationary network dynamics, which are described by
constant $r$ and $\rho$ and time-independent $p(V)$. For these
Eq.~\eqref{eq:Continuity equation} reduces to the linear delay differential
equation (or differential-difference equation)
\begin{multline}
\frac{d}{dV}p(V)\left(-\gamma V+I^{\text{ext}}\right)=r\left[p(V-C)-p(V)\right]\\
+\rho\left[\delta(V-\vre)-\delta(V-\vth)\right].\label{eq:Continuity equation stationary}
\end{multline}
Dividing Eq.~\eqref{eq:Continuity equation stationary} by $\rho>0$
yields an equation for the rescaled density $q(V)=p(V)/\rho$, which
is independent of the unknown steady-state firing rate $\rho$. This
equation can be integrated for example with the method of steps \citep{hale1993introduction}.
The integration starts with the ``initial conditions'' $q(V)=0$
for $V>\vth$ and thus $q(\vth)=1/\left(-\gamma\vth+I^{\text{ext}}\right)$
slightly below $\vth$. The normalization of $p(V)$ allows us to
compute $\rho$ via
\begin{equation}
\frac{1}{\rho}=\int_{-\infty}^{\infty}q(V)dV.\label{eq:rho q}
\end{equation}

To obtain an analytic expression for $\rho$, one applies a bilateral
Laplace transform $\tilde{f}(s)=\int_{-\infty}^{\infty}f(V)e^{sV}dV$.
We can focus on $s\geq0$; $\tilde{q}(0)$ yields $\rho^{-1}$. The
Laplace transform of the rescaled Eq.~\eqref{eq:Continuity equation stationary}
results in a linear first-order ordinary differential equation for
$\tilde{q}(s)$,
\begin{equation}
\frac{d}{ds}\tilde{q}(s)=\left[\frac{I^{\text{ext}}}{\gamma}+\frac{r\left(\e^{Cs}-1\right)}{\gamma s}\right]\tilde{q}(s)+\frac{\e^{\vre s}-\e^{\vth s}}{\gamma s}.\label{eq:Laplace continuity equation}
\end{equation}
It can be solved by variation of constants. The solution of the homogeneous
equation is
\begin{equation}
Z_{0}(s)=A\e^{\Psi(s)}\label{eq:Solution homogeneous qtilde(s)}
\end{equation}
with an arbitrary constant $A$ and
\begin{align}
\Psi(s) & =\frac{I^{\text{ext}}}{\gamma}s+\frac{r}{\gamma}\int_{0}^{s}\frac{\e^{Cu}-1}{u}du\nonumber \\
 & =\frac{I^{\text{ext}}}{\gamma}s+\frac{r}{\gamma}\left[\ei(Cs)-\log(-Cs)-\Gamma\right].\label{eq:psi}
\end{align}
Here, $\ei(x)$ is the exponential integral $\ei(x)=-\int_{-x}^{\infty}\frac{\e^{-t}}{t}dx$
and $\Gamma$ is the Euler-Mascheroni constant. The solution of the
full equation then reads
\begin{equation}
\tilde{q}(s)=\e^{\Psi(s)}\left[\tilde{q}(0)-\int_{0}^{s}\e^{-\Psi(u)}\frac{\e^{\vth u}-\e^{\vre u}}{\gamma u}du\right].\label{eq:Full qtilde equation solution}
\end{equation}
Since the support of $q(V)$ is bounded from above by $\vth$, $\tilde{q}(s)=\int_{-\infty}^{\infty}q(V)\e^{Vs}dV\leq\e^{\vth s}/\rho_{0}$.
To balance the faster exponential growth $\sim\exp(\ie s/\gamma)$
of its prefactor $\exp\left[\Psi(s)\right]$, the bracket on the right
hand side of Eq.~\eqref{eq:Full qtilde equation solution} needs
to vanish for large $s$. We thus have
\begin{equation}
\tilde{q}(0)=\int_{0}^{\infty}\e^{-\Psi(u)}\frac{\e^{\vth u}-\e^{\vre u}}{\gamma u}du=\frac{1}{\rho}.\label{eq:rho analytic final result}
\end{equation}

For an XIF neuron without voltage-dependent synapses there is no stationary
membrane potential probability density $p(V).$ This is because for
any time $t>0$ there is a finite probability that the membrane potential
of a neuron jumps below $I^{\text{ext}}/\gamma$ and thereafter tends
to minus infinity. In contrast, for an XIF neuron with the voltage
dependence Eq.~\eqref{eq:Voltage dependent coupling}, $p(V)$ exists
and we may use the same approach as for the LIF neuron to determine
it together with the firing rate. Since membrane potentials do not
drop below $\vcut+C$, we focus on the interval $[\vcut+C,\vth]$,
where $p(V)$ can be nonzero. The couplings' voltage dependence enters
the source term $\sigma_{\text{inh}}$ in Eq.~\eqref{eq:Continuity equation}:
If $V$ is below $\vcut$, incoming spikes have no effect and the
sink term due to them vanishes. Eq.~\eqref{eq:sigma inh} therefore
changes to
\begin{equation}
\sigma_{\text{inh}}(V,t)=r(t)\left[p(V-C,t)-h(V)p(V,t)\right],\label{eq:sigma inh voltage dependent}
\end{equation}
where we used that $V-C\geq\vcut$ in the relevant voltage range such
that a modification of the source term is unnecessary. The stationary
continuity equation becomes
\begin{multline}
\frac{d}{dV}p(V)\left(-\gamma V+I^{\text{ext}}\right)=r\left[p(V-C)-h(V)p(V)\right]\\
+\rho\left[\delta(V-\vre)-\delta(V-\vth)\right],\label{eq:Continuity equation stationary voltage dependent coupling}
\end{multline}
which can be rescaled and integrated using the method of steps to
obtain $q(V)$, $\rho$ and $p(V)$ as before. The nonlinear prefactor
$h(V),$ however, impedes the derivation of $\rho$ via Laplace transform.

We apply the above results to find mixed networks in which $\LIF$
and $\XIF$ neurons have similar firing rates. Eq.~\eqref{eq:rho analytic final result}
provides a map $G_{\LIF}$ from the input to the output rate, $G_{\LIF}(r)=\rho$.
Eq.~\eqref{eq:Continuity equation stationary voltage dependent coupling}
implicitly defines such a map $G_{\XIF}$ for XIF neurons. The firing
rate $\rho$ of the LIF and XIF neurons in the desired mixed network
needs to solve both self-consistency equations 
\begin{align}
G_{\LIF}(K\rho) & =\rho,\label{eq:GLIF self-consistent}\\
G_{\XIF}(K\rho) & =\rho,\label{eq:GXIF self-consistent}
\end{align}
with the neurons' indegree $K$. We employ Eq.~\eqref{eq:GXIF self-consistent}
to compute $\rho$ for XIF neurons. Thereafter, we adapt the parameters
of Eq.~\eqref{eq:GLIF self-consistent} such that the same $\rho$
becomes a solution. Specifically, we solve for $\gamma_{\LIF}$, keeping
the other parameters fixed.

\begin{figure}
\includegraphics[width=1\columnwidth]{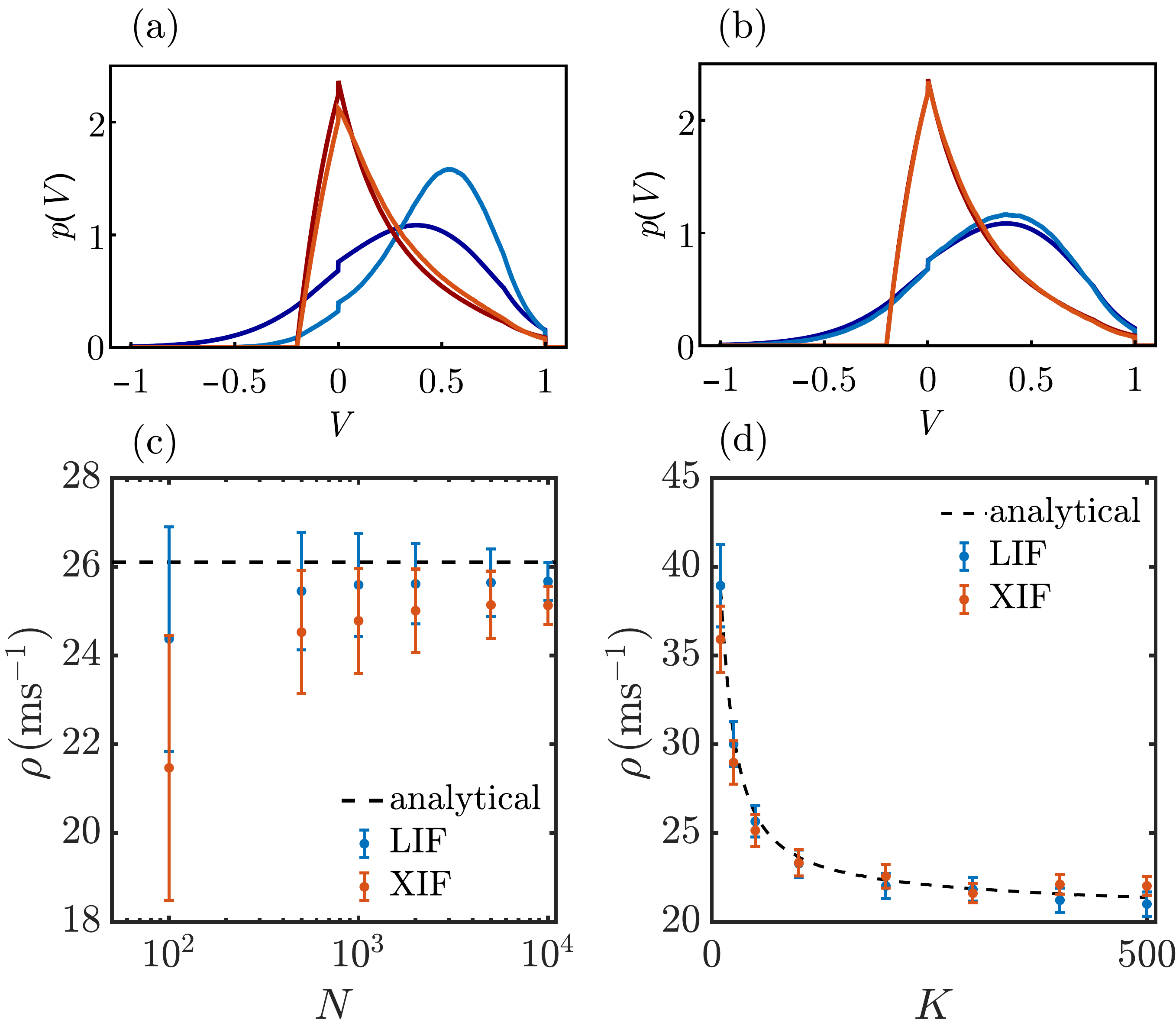} \caption{\label{fig:probdensity}Analytically and numerically estimated voltage
probability densities and spike rates of LIF and XIF neurons. (a,b)
Voltage probability densities for networks of (a) $N=100$ and (b)
$N=10000$ neurons. The dark blue and dark red curves show the analytical
results Eqs.~\eqref{eq:Continuity equation stationary} and \eqref{eq:Continuity equation stationary voltage dependent coupling}
of $p(V)$ for LIF and XIF neurons, where $\rho$ is obtained self-consistently
from Eqs.~\eqref{eq:GLIF self-consistent} and \eqref{eq:GXIF self-consistent}.
The light blue and light red curves show representative numerically
sampled voltage densities of an LIF and an XIF neuron, where the input
spike trains are superpositions of simultaneous output spike trains
of $K$ neurons in the recurrent network. (c,d) Spike rates of neurons
in networks of (c) different size $N$ and indegree $K=50$ and (d)
size $N=10000$ and different indegree $K$. In (d) the presynaptic
weights are scaled with $1/K$ such that their sum is independent
of $K$. Numerically measured average spike rates of LIF and XIF neurons
in the different networks are shown by blue and red dots. Error bars
display the standard deviations of the rate distributions. Analytical
results obtained from Eqs.~\eqref{eq:GLIF self-consistent} and \eqref{eq:GXIF self-consistent}
are displayed by dashed black lines. Remaining parameters are as in
Fig.~\ref{fig:rasterplot}.}
\end{figure}

Fig.~\ref{fig:probdensity} compares the voltage densities $p(V)$
and rates $\rho$ obtained from the shot noise approach with those
of an LIF and an XIF neuron that receive input spike trains as they
are generated in the recurrent network of Fig.~\ref{fig:rasterplot}.
There is a pronounced discrepancy between the densities and rates
for an LIF neuron for $K=50$ and small $N$, because both the individual
(see Fig.~\ref{fig:rasterplot}a-e) and the superposed input spike
trains in these dense networks are more regular than Poisson spike
trains. Removing spatial correlations for example by increasing $N$
reduces the discrepancy, see Fig.~\ref{fig:probdensity}b-d and App.~\ref{subsec:Voltage-probability-distribution}
for further analysis. Such input spike trains reduce the variance
of the voltage and generate a $p(V)$ that is more concentrated around
the value $\left(I_{i}^{\mathrm{ext}}+I_{i}^{\text{inh}}\right)/\gamma_{i}$,
where $I_{i}^{\text{inh}}$ is the average inhibitory input current
as discussed in Sec.~\ref{sec:Mixed-networks}. For the XIF neuron,
the input spike train statistics has less impact on $p(V)$. Presumably,
this is because voltage excursions due to input fluctuations are anyways
suppressed by the voltage dependence of the input strength (for potentials
near $\vcut$) and by the drive towards threshold (for larger potentials).
We note that the assumption of Poisson input spike trains is the only
approximation in the chosen approach, such that sampled membrane potential
distributions of neurons with Poisson input match the analytical ones
up to the sampling noise as shown in App.~\ref{subsec:Voltage-probability-distribution}.

\section{Growth of dynamical perturbations}

\subsection{Mean-field approach}

\label{sec:Meanfield}
\begin{figure}
\includegraphics[width=0.45\textwidth]{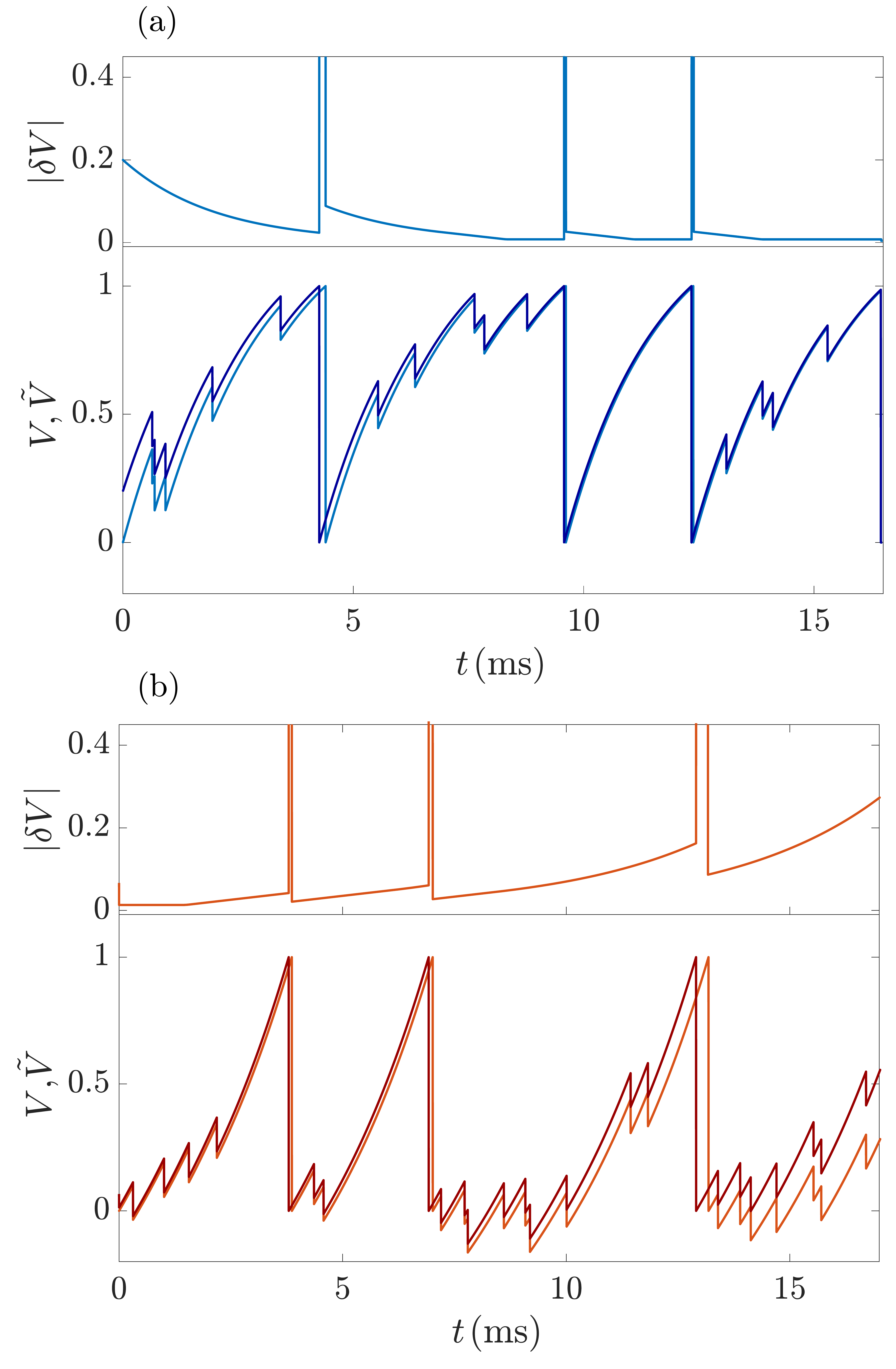} \caption{\label{fig:intuition}Evolution of perturbations during subthreshold
evolution and spiking in (a) an LIF and (b) an XIF neuron. During
subthreshold evolution the distance (perturbation) $\left|\delta V(t)\right|$
between two neighboring trajectories shrinks for LIF and grows for
XIF neurons, while spike generation partially resets it. Due to the
receiving of inhibitory spikes, the intervals between spike generations
are generally longer than for freely oscillating neurons. The impact
of the subthreshold dynamics therefore dominates and overall the perturbation
in the LIF neuron decays while that in the XIF neuron grows. Here
we use $V_{\infty,\protect\LIF}=1.33$, $V_{-\infty,\protect\XIF}=-1$,
$\gamma_{\protect\LIF}=\SI{0.5}{\per\milli\second}$, and $\gamma_{\protect\XIF}=-\SI{0.3}{\per\milli\second}$
for better illustration of the mechanism.}
\end{figure}

After obtaining the spike rates and membrane
potential distributions using a statistical mean-field theory, we
investigate the mixed network dynamics from a dynamical systems perspective.
We first analytically determine the Lyapunov spectrum using again
a mean-field approach. It focuses on the evolution of perturbations
to a single neuron and treats the input from other neurons as external.
Specifically, we disregard perturbations of the rest of the network
including those generated by the considered neuron's changed spiking.
Inputs thus arrive at the same times in the perturbed and in the unperturbed
system and do not change the neuron's perturbation. Fig.~\ref{fig:intuition}
illustrates this and compares the resulting evolution of a perturbation
of an XIF neuron and of an LIF neuron: The perturbation of the XIF
neuron gradually increases as long as it is not spiking, while that
of the LIF neuron decreases. Conversely, in the XIF neuron spiking
and resetting reduces perturbations, while it increases them in the
LIF neuron; compare the values of the (finite size) distance $\left|\delta V(t)\right|=\lvert\tilde{V}(t)-V(t)\rvert$
between two neighboring trajectories $\tilde{V}(t)$ and $V(t)$ in
Fig.~\ref{fig:intuition}a,b before and after a spike event has taken
place in both the perturbed and the unperturbed dynamics. To assess
the influence of these two processes, we first note that in a freely
oscillating neuron they need to cancel each other such that perturbations
persist on average and the LE is zero. We then note that the inhibitory
inputs do not affect perturbations but prolong the subthreshold evolution
between spikes. Its impact therefore dominates, and perturbations
in XIF neurons grow over time, while they shrink in LIF neurons. This
does not depend on the specifics of the LIF and XIF dynamics but is
a consequence of the curvature of the rise function and the inhibitory
inputs.

In App.~\ref{subsec:Meanfield-1}, we make the gained intuitive understanding
precise by quantifying the growth of perturbations and the resulting
LE. For this, we describe the dynamics by a sequence of discrete maps
from the state at a time (infinitesimally) shortly after generation
of a spike to the state at a time shortly after generation of the
next spike. The discrete time dynamics of small perturbations are
then given by the ``single spike Jacobians'' \citep{MW10,MW12}
$J(k)$. For the effective single neuron dynamics here, they reduce
to scalar factors
\begin{equation}
J(k)=\frac{\partial V(t_{k+1}^{+})}{\partial V(t_{k}^{+})}=\exp\left(\frac{\gamma}{\rho_{\text{free}}}-\gamma(t_{k+1}-t_{k})\right)\label{eq:ssjac-MF}
\end{equation}
with the free firing rate $\rho_{\text{free}}$ (Eq.~\eqref{eq:ffreeXIF}
or \eqref{eq:ffreeLIF}) of the neuron.

The growth rate of perturbations and thus the mean-field LE are given
by the long-term average of Eq.~\eqref{eq:ssjac-MF},
\begin{align}
\lambda_{\mf} & =\lim_{L\to\infty}\frac{1}{t_{L}}\sum_{k=0}^{L-1}\ln\left|J_{\mf}(k)\right|\nonumber \\
 & =-\gamma\left(1-\frac{\rho}{\rho_{\text{free}}}\right).\label{eq:meanfieldexponent}
\end{align}
This expression confirms the intuitive understanding that without
perturbed inputs the growth rate depends (i) on the growth rate during
subthreshold evolution and (ii) on the prevalence of subthreshold
evolution ($\rho<\rho_{\text{free}}$) or spike sending ($\rho>\rho_{\text{free}}$)
relative to the free neuron case. In particular, without input we
have $\lambda_{\mf}=0$ and if the neuron is silenced $\lambda_{\mf}=-\gamma$.
In our inhibitory networks we have $\rho<\rho_{\text{free}}$ such
that $\lambda_{\mf}>0$ for XIF and $\lambda_{\mf}<0$ for LIF neurons.
In networks in the balanced state, the actual spike rate is much smaller
than the spike rate of a neuron if only excitation is present. Since
in our networks the latter equals the spike rate of the freely oscillating
neuron, we have $\rho/\rho_{\text{free}}\ll1$. Thus the mean-field
approach indicates that the growth rate of perturbations is mainly
given by the subthreshold growth. The mean-field approach further
indicates that a single XIF neuron renders the entire network dynamics
unstable and that the number of unstable directions equals the number
of XIF neurons in the network, while the number of stable directions
equals the number of LIF neurons. This, however, does not give rise
to a zero LE, which occurs in the full autonomous network due to time-translation
symmetry. The mean-field spectrum and the rule for the number of stable
and unstable directions can thus only be an approximation to the exact
results.

\begin{figure*}
\includegraphics[width=1\textwidth]{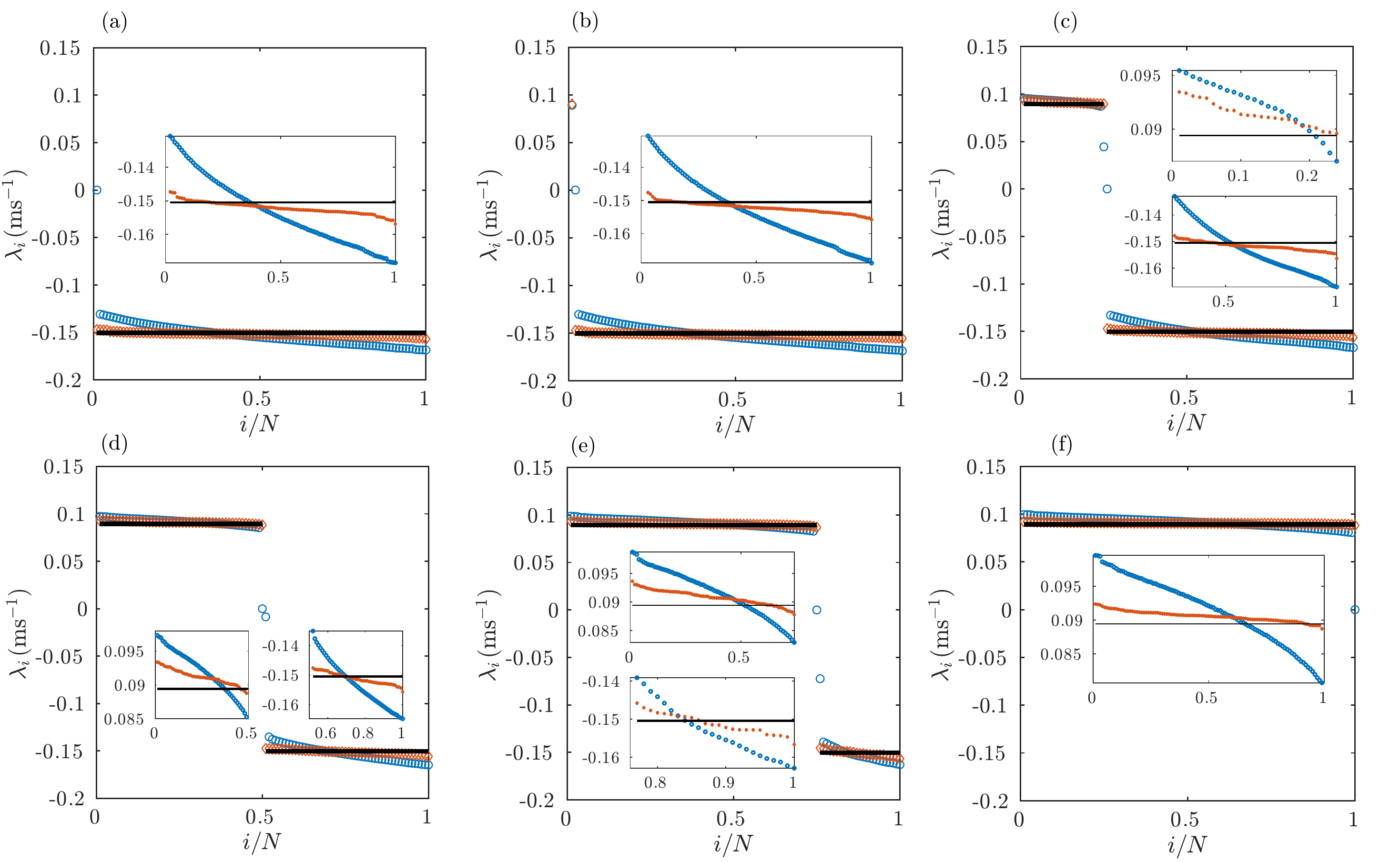} \caption{\label{fig:spectra}Lyapunov spectra of mixed networks. The numbers
of LIF and XIF neurons in the networks are (a) $N_{\protect\LIF}=100$
and $N_{\protect\XIF}=0$, (b) $N_{\protect\LIF}=99$ and $N_{\protect\XIF}=1$,
(c) $N_{\protect\LIF}=75$ and $N_{\protect\XIF}=25$, (d) $N_{\protect\LIF}=50$
and $N_{\protect\XIF}=50$, (e) $N_{\protect\LIF}=75$ and $N_{\protect\XIF}=25$,
(f) $N_{\protect\LIF}=0$ and $N_{\protect\XIF}=100$. Blue circles
display the numerically computed Lyapunov spectra using Eq.~\eqref{eq:Single spike Jacobian}.
Red diamonds display the mean-field result Eq.~\eqref{eq:meanfieldexponent}
with numerically measured neuron rates and black lines display the
mean-field result Eq.~\eqref{eq:meanfieldexponent} using the analytically
obtained rates Eqs.~\eqref{eq:GLIF self-consistent} and \eqref{eq:GXIF self-consistent}.
Insets show closeups of the positive and/or negative parts of the
spectra.}
\end{figure*}

Eq.~\eqref{eq:meanfieldexponent} together with the analytical results
Eqs.~\eqref{eq:GLIF self-consistent},\eqref{eq:GXIF self-consistent}
for $\rho$ give a fully analytical estimate of the Lyapunov spectrum.
Since all LIF or XIF neurons have the same analytical rate estimates
and leak strengths, the spectrum consists of $N_{\LIF}$ identical
negative and $N_{\XIF}$ identical positive exponents, see Fig.~\ref{fig:spectra}.
Due to quenched noise from random coupling, the rates in the actual
network are distributed. We can account for this by inserting the
numerically measured rates into Eq.~\eqref{eq:meanfieldexponent},
see Fig.~\ref{fig:spectra}.

\subsection{Network single spike Jacobian\label{subsec:Single-Spike-Jacobian}}

To derive exact Lyapunov spectra we need to take into account the
spreading of perturbations in the network. For this, we compute the
full single spike Jacobian $J(k)$, which is a map from tangent vectors
at the point $\vec{V}(t_{k}^{+})$ in phase space to tangent vectors
at $\vec{V}(t_{k+1}^{+})$, where $\vec{V}(t)=\left(V_{1}(t),\ldots,V_{N}(t)\right)^{T}$
is the state of the system at time $t$. The resulting components
of $J(k)$ read
\begin{align}
\begin{alignedat}{1}J_{ij}(k) & =\end{alignedat}
 & \frac{\partial V_{i}(t_{k+1}^{+})}{\partial V_{j}(t_{k}^{+})}\nonumber \\
= & \ \delta_{ij}\mathrm{e}^{-\gamma_{i}(t_{k+1}-t_{k})}\nonumber \\
 & +\delta_{jl}\frac{\gamma_{i}}{\gamma_{l}}\frac{\delta_{il}\vth-C_{il}h_{i}(V_{i}(t_{k+1}^{-}))}{\frac{I_{l}^{\mathrm{ext}}}{\gamma_{l}}-V_{l}(t_{k}^{+})},\label{eq:Single spike Jacobian}
\end{align}
for an LIF or an XIF neuron $i$, where $l$ is the index of the neuron
sending the $(k+1)$th spike, see App.~\prettyref{subsec:Single-Spike-Jacobian-1}
for details. We note that the mean-field theory accounts for the diagonal
terms of this Jacobian.

\subsection{Volume contraction}

Owing to the simple form of the single spike Jacobians we can find
an analytical expression for the full network dynamics' expansion
rate of infinitesimal phase space volumes or, equivalently, for the
sum of the LEs. The result in terms of the neuronal spike rates in
the network is exact. It allows to analytically compute the Lyapunov
spectra for two neuron systems and offers a test for the accuracy
of their numerical estimates in larger networks.

The volume expansion and the sum of LEs are given by the time averaged
logarithms of the determinants of the Jacobians \citep{pikovsky+politi}.
We thus have
\begin{equation}
\sum_{i=1}^{N}\lambda_{i}=\lim_{L\to\infty}\frac{1}{t_{L}}\sum_{k=0}^{L-1}\ln\left|\det J(k)\right|\label{eq:sum of Lyap}
\end{equation}
in terms of single spike Jacobians \citep{MW12}. In App.~\ref{subsec:Volume-contraction}
we exploit the specific form of $J(k)$ to compute $\det J(k)$ with
the matrix determinant lemma. The subsequent time averaging yields
\begin{align}
\sum_{i=1}^{N}\lambda_{i} & =-\sum_{j=1}^{N}\gamma_{j}\left(1-\frac{\rho_{j}}{\rho_{\text{free},j}}\right).\label{eq:sum of Lyap final}
\end{align}
Notably, this shows that our mean-field theory yields an exact expression
for the volume contraction rate and the sum of LEs: the estimate $\sum_{l=1}^{N}\lambda_{\mf,l}$
with Eq.~\eqref{eq:meanfieldexponent} agrees with the exact expression
Eq.~\eqref{eq:sum of Lyap final}.

\subsection{Numerical computation of the Lyapunov spectrum\label{subsec:Lyapunov spectrum numerics}}

The single spike Jacobians \eqref{eq:Single spike Jacobian} allow
us to iteratively compute the largest LE and the full Lyapunov spectrum
\citep{MW10,pikovsky+politi,engelken2017chaotic}, see also App.~\ref{sec:CLValgorithm}.
In short, for the largest LE, one iterates an initial random perturbation
vector by the single spike Jacobians, stores its growth every few
steps and thereafter renormalizes it to its initial magnitude. The
long-term average of the growth rate equals $\lambda_{1}$. For the
full spectrum, one iterates a system of $N$ orthogonal perturbation
vectors with the single spike Jacobians. Every few steps, one records
the growth of the different vectors. Thereafter one reorthogonalizes,
always in the same order, and finally renormalizes the vectors. The
long-term average growth rate of the first vector then equals $\lambda_{1}$,
that of the second equals $\lambda_{2}$ etc. Ref.~\citep{engelken2017chaotic}
suggested an efficient method to compute the Lyapunov spectrum and
applied it to large networks; we use some of the ideas in our implementation.

For networks consisting only of LIF neurons we find in agreement with
previous work \citep{ZLPT06,JMT08,JMT09,ZBH09} and our mean-field
theory that the largest nontrivial LE is negative, see Fig.~\ref{fig:spectra}a.
However, we also find that already the presence of a single XIF neuron
renders the largest LE positive, see Fig.~\ref{fig:spectra}b, indicating
chaos in agreement with the mean-field theory. The computations also
confirm that the destabilization of a network by a single XIF neuron
is a special case of a general rule, namely that each XIF neuron introduces
about one positive LE. This holds independently of $N$ and $K$,
see Fig.~\ref{fig:spectra} and App.~\ref{sec:LyapSpect-Dependence-N-K}.
The trivial (zero) exponent is an exception to the rule. Our numerical
results indicate that it replaces a negative exponent if there are
more LIF than XIF neurons in the network and a positive exponent otherwise.
There is also good quantitative agreement with the mean-field spectrum,
in particular the exponents are close to $\gamma_{\text{LIF }}$ and
$\gamma_{\text{XIF}}.$ However, also when inserting the measured
spike rates into Eq.~\eqref{eq:meanfieldexponent} some discrepancy
remains, showing that the spread of perturbations in the network and
their transfer between neurons has a pronounced effect on their growth.

\section{Stable and unstable directions\label{sec:CLV}}

\subsection{Lyapunov vectors and perturbation growth}

\begin{figure*}
\includegraphics[width=1\textwidth]{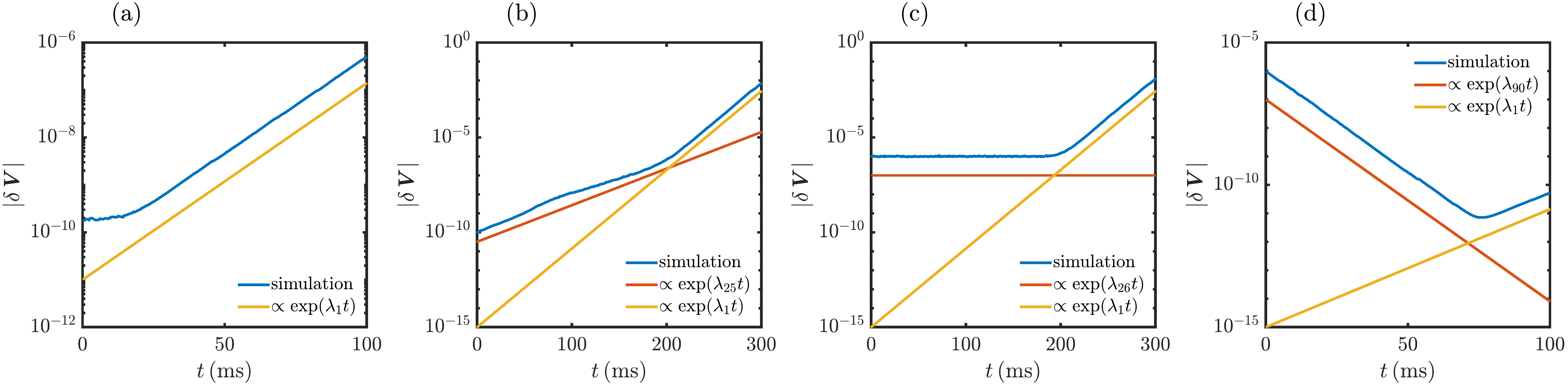} \caption{\label{fig:pertgrowth} Tracking the growth or shrinkage of finite
perturbations. We consider the same network as in Figs.~\ref{fig:rasterplot}
and \ref{fig:spectra}c and explicitly perturb it in different directions.
The semilogarithmic plots display the time evolution of distances
between original and perturbed trajectories (blue) and compare them
to exponential functions with growth rates equal to the relevant LEs
(yellow, red). Transient large perturbations due to different event
times in the two systems (cf.~Fig.~\ref{fig:intuition}) are excluded.
We perturb in (a) in a generic direction, in (b) in the direction
of the unstable CLV $\vec{v}_{25}$ corresponding to LE $\lambda_{25}$
(closest to the trivial one), in (c) in the direction of the trivial
CLV $\vec{v}_{26}$ (in the direction of the trajectory) and in (d)
in the direction of the stable CLV $\vec{v}_{90}$. In (a) after a
short equilibration time we have growth with the largest LE. In (b),(c)
and (d) the perturbation grows initially with the LE of the CLV. The
small numerical error of the CLVs grows exponentially with the largest
LE and eventually dominates the evolution.}
\end{figure*}

To further elucidate the local phase space structure, we numerically
investigate the characteristics of the perturbations that grow according
to the individual LEs, i.e.~how they are distributed across neurons
and how they change during evolution. This will, in particular, allow
us to understand why the mean-field theory works well. The directions
of the perturbations are given by the CLVs or, in other words, by
the stable and unstable manifolds along the trajectory \citep{Kuptsov2012,pikovsky+politi}.

The $i$th CLV $\vec{v}_{i}(\vec{V}_{0})$ at a point $\vec{V}_{0}$
in phase space is a normalized tangent vector that grows with long-term
average rates $\lambda_{i}$ and $-\lambda_{i}$ when evolved forward
and backward in time. We call it a stable CLV if $\lambda_{i}<0$
and an unstable one if $\lambda_{i}>0$. We assume for simplicity
that all LEs are different; the vector is then unique up to its orientation.
Consider a trajectory $\vec{V}(t)$ that reaches shortly after the
spike time $t_{0}$ the state $\vec{V}(t_{0}^{+})=\vec{V}_{0}$. Using
the single spike Jacobians $J(k)$, $\vec{v}_{i}(\vec{V}_{0})$ may
be defined as the tangent vector satisfying
\begin{align}
\Bigl|\prod_{k=0}^{L}J(k)\,\vec{v}_{i}(\vec{V}_{0})\Bigr| & \sim\mathrm{e}^{\lambda_{i}t_{L}},\label{eq:clvdef1}\\
\Bigl|\prod_{k=-M}^{-1}J^{-1}(k)\,\vec{v}_{i}(\vec{V}_{0})\Bigr| & \sim\mathrm{e}^{\lambda_{i}t_{-M}},\label{eq:clvdef2}
\end{align}
where $L$ and $M$ are chosen sufficiently large. The definition
can be straightforwardly extended to states between spiking events.
Both of its parts are important: The first part alone does not uniquely
define the direction of $\vec{v}_{i}(\vec{V}_{0})$, since adding
any vector with growth rate less than $\lambda_{i}$ yields the same
asymptotics. The second part excludes such an addition, since its
shrinkage rate is slower than $-\lambda_{i}$ and thus yields a different
dominant asymptotics of backward evolution. As anticipated by the
notation, the vector depends only on the state but not on the time
when $\vec{V}(t)$ reaches it. Furthermore, the definition ensures
covariance, that is the evolution of infinitesimal perturbations (the
tangent flow) maps CLVs to CLVs. At subsequent spike times we thus
have
\begin{equation}
J(k)\vec{v}_{i}(\vec{V}(t_{k}^{+}))\propto\vec{v}_{i}(\vec{V}(t_{k+1}^{+})).\label{eq:covariant}
\end{equation}
The extension to states between spike times is again straightforward:
the covariance implies that we can obtain CLVs $\vec{v}_{i}(\vec{V}(t))$
at a state between spike times $t_{k}$ and $t_{k+1}$ by propagating
$\vec{v}_{i}(\vec{V}(t_{k}^{+}))$ forward with the Jacobian $\hat{J}_{ij}(t-t_{k})=\delta_{ij}\exp(-\gamma_{i}\del{t-t_{k}})$
of subthreshold evolution.

We compute the CLVs in a dynamical manner by forward and restricted
backward propagating sets of vectors, following refs.~\citep{ginelli,pikovsky+politi}.
Appendix \ref{sec:CLValgorithm} provides a short description of the
method. We note that the dynamics of our system are not invertible:
given a state there is no unique way of propagating it back in time.
This is because an ambiguity can arise at states where one neuron
is at the reset potential; we generally cannot tell whether it was
reset or crossed the reset potential from below (unless some postsynaptic
neuron is too near to threshold to be able to have just received a
spike). It is, however, still possible to compute the Lyapunov vectors
by backward propagating along the trajectory that was previously taken
for the forward propagation \citep{ginelli}.

\subsection{Stable and unstable directions in mixed networks}

\begin{figure*}
\includegraphics[width=1\textwidth]{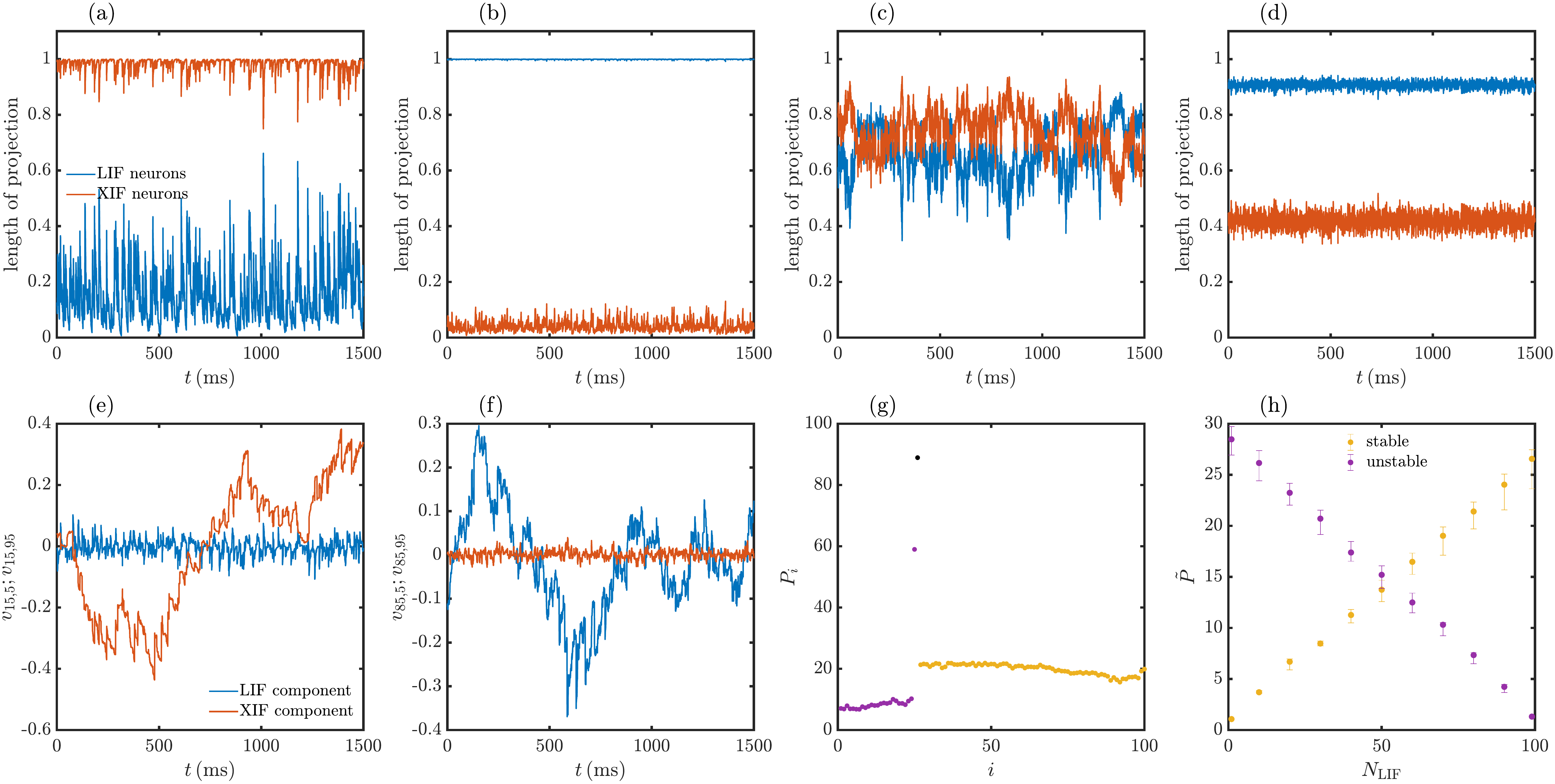} \caption{\label{fig:projection} Stable and unstable CLVs are mostly confined
to the subspaces of (perturbations to) LIF or XIF neurons, respectively.
(a-d) show the time evolution of the lengths of the projections of
different CLVs onto the subspaces of LIF (blue) and XIF (red) neurons:
(a) unstable CLV $\vec{v}_{6}$, (b) stable CLV $\vec{v}_{88}$, (c)
unstable CLV $\vec{v}_{25}$ (corresponding to the LE closest to the
trivial one) and (d) trivial CLV $\vec{v}_{26}$. (e,f) show the time
evolutions of the projections of different CLVs onto the subspace
of a single LIF neuron (neuron index $n=5$, blue) and a single XIF
neuron ($n=95$, red): (e) unstable vector $(\bm{v}_{15}$), (f) stable
vector ($\bm{v}_{85}$). Values are plotted at event times $t=t_{k}$.
(g) shows the participation ratios $P_{i}$ of the unstable (purple),
trivial (black) and stable (yellow) CLVs averaged over 10000 events.
The network is the same as in Figs.~\ref{fig:rasterplot}, \ref{fig:spectra}c
and \ref{fig:pertgrowth}. (h) displays the median $\tilde{P}$ of
the participation ratios of the stable (yellow) and unstable (purple)
CLVs, in networks as in (a-g) with $N=100$ neurons but with different
fractions of LIF and XIF neurons. Bars indicate the first and third
quartiles of the distribution.}
\end{figure*}

The CLVs yield the directions in which small but finite perturbations
evolve according to the different LEs as shown in Fig.~\ref{fig:pertgrowth}.
We find that they generally contain perturbations to a variety of
neurons and that they strongly change their direction during evolution.
More specifically, we observe that the stable and unstable CLVs stay
approximately confined to the subspaces of (strictly speaking: perturbations
to) LIF and XIF neurons, respectively. Fig.~\ref{fig:projection}a-d
illustrates this by displaying the lengths $\sqrt{\sum_{j=1}^{N_{\LIF}}v_{i,j}^{2}(\vec{V}(t))}$
and $\sqrt{\sum_{j=N_{\LIF}+1}^{N}v_{i,j}^{2}(\vec{V}(t))}$ of the
projections of different CLVs $\vec{v}_{i}$ onto the subspaces of
LIF and XIF neurons. Here and in the following we assume that the
LIF and XIF neurons have the indices $1,...,N_{\LIF}$ and $N_{\LIF+1},...,N$,
respectively. Fig.~\ref{fig:projection}e,f further illustrates the
confinement and shows the large temporal variability of single CLV
components $v_{i,j}$ that are not close to zero. The confinement
does not hold exactly since perturbations of LIF neurons usually also
give rise to perturbations of XIF neurons and vice versa. In networks
with inhomogeneous spike rates, we observe that single neurons that
are strongly suppressed by inhibition have CLVs more aligned to them,
because their perturbation spreads less in the network due to their
lack of spiking.

We further quantify the localization of the CLVs using an inverse
participation ratio (number) \citep{kramer1993localization,ginelli},
which we define for the $i$th CLV as
\begin{equation}
P_{i}^{-1}=\left\langle \sum_{j=1}^{N}v_{i,j}^{4}(\vec{V}(t_{k}))\right\rangle _{k}.
\end{equation}
Here, $\left\langle .\right\rangle _{k}$ is an average over sufficiently
many events and we use that the CLVs are normalized, $\sum_{j=1}^{N}v_{i,j}^{2}(\vec{V}(t_{k}))=1$.
The participation ratio $P_{i}$ measures how many components contribute
to a vector. If, for example, the vector $\vec{v}_{i}(\vec{V}(t_{k}))$
always has only one nonzero component, $P_{i}=1$. If there are always
$m$ nonzero components of equal size, $P_{i}=m$. We observe that
the participation ratio of unstable CLVs increases approximately linearly
with the number of XIF neurons starting with $P_{i}\approx1$ at $N_{\XIF}=1$,
consistent with a delocalization of these CLVs between the present
XIF neurons, see Fig.~\ref{fig:projection}g,h. $P_{i}$ for stable
CLVs increases likewise with the number of LIF neurons. The trivial
CLV has a participation ratio close to $N$, because the components
of the tangential vector $dV_{j}(t^{+})/dt$ and thus the components
of the CLV have roughly similar size.

Our mean-field approach uses the assumption that each LE is independently
generated by the growth or shrinkage of a single neuron perturbation,
with negligible influence of the perturbation's spread and backreaction
in the network. Its suitability can now be understood as follows:
The approximately $N_{\LIF}$ stable CLVs are confined to the $N_{\LIF}$-dimensional
subspace of perturbations to LIF neurons. The stable CLVs thus form
a basis of the subspace of perturbations to LIF neurons. Likewise
the unstable CLVs form a basis of the subspace of perturbations to
XIF neurons. At each time point, a perturbation to a single LIF neuron
can therefore be expressed as a linear combination of stable CLVs,
while a perturbation to an XIF neuron can be expressed as a linear
combination of unstable CLVs. The stable CLVs have similar decay rates
(negative LEs) and the unstable CLVs have similar growth rates (positive
LEs), see Fig.~\ref{fig:spectra}. Any linear combination of only
stable or only unstable CLVs inherits this decay or growth rate. This
holds in particular for the perturbation of a single neuron. At each
time point the perturbation to a single LIF or XIF neuron thus grows
according to the negative or according to the positive LEs, respectively.
The mean-field approach therefore yields good results.

\section{Computations with precisely timed spikes\label{sec:Computations-with-precisely}}

\subsection{Network architecture and task design}

\begin{figure*}
\includegraphics[width=1\textwidth]{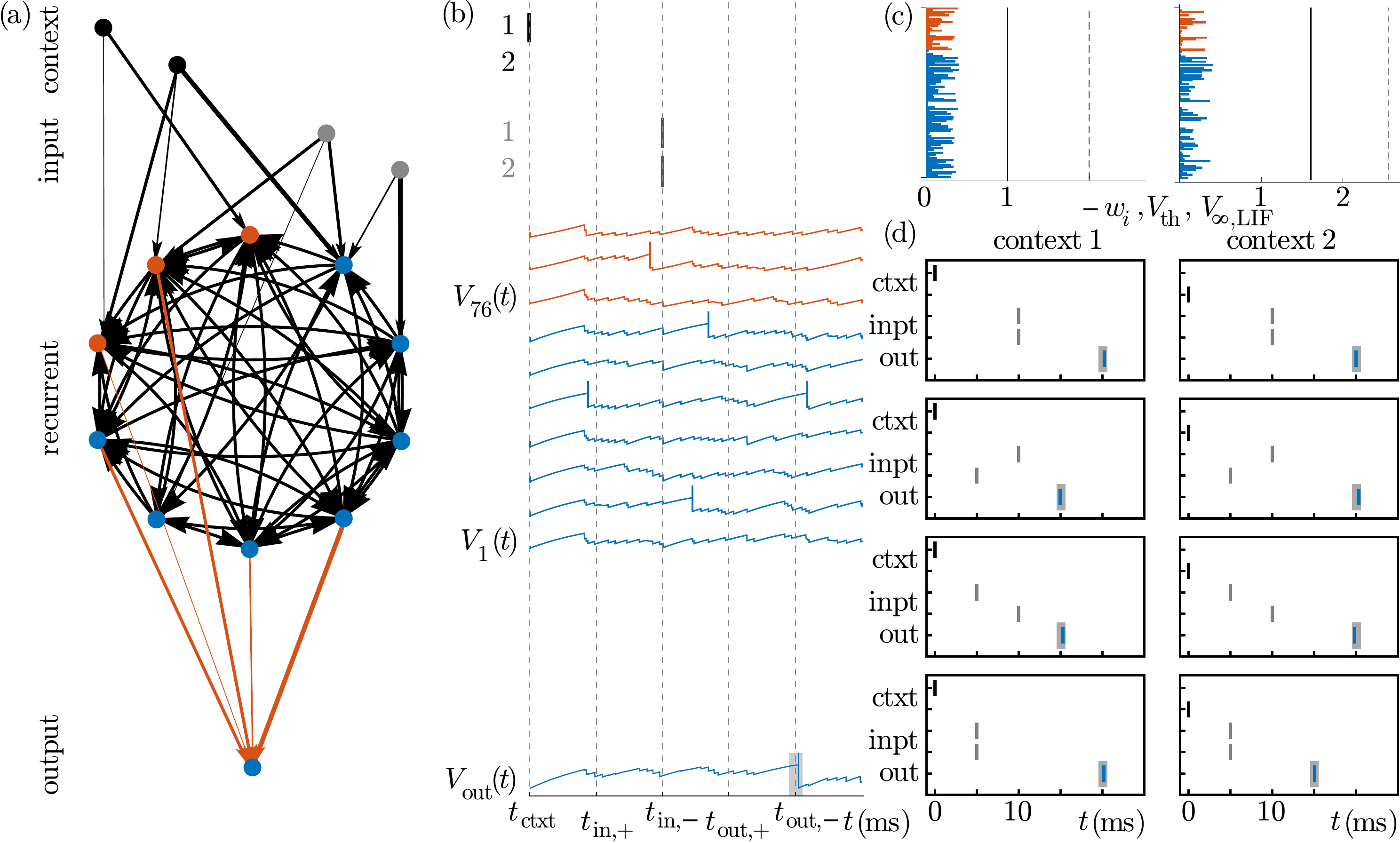}

\caption{Network for precise spike-based computations and solution of the XOR/AND
task. (a) Network architecture. The recurrent network (middle, blue
and red for LIF and XIF neurons) receives input spikes from context
(top, black) and from input neurons (gray). The output neuron (bottom)
changes its plastic weights (red) according to Finite Precision Learning
to learn the task. (b) Spiking of the context and input neurons (top),
voltage traces of recurrent neurons (middle, LIF neurons 1-7, XIF
neurons 76-78, spikes highlighted by vertical lines) and voltage traces
of the output neuron (bottom) after learning the XOR/AND task. Dashed
lines indicate the times of possible context ($t_{\text{ctxt}}$)
and input neuron ($t_{\text{in,+}}$, $t_{\text{in,-}}$) spiking
as well as the possible desired output spike times ($t_{\text{out,+}}$,
$t_{\text{out,-}}$). The output neuron sends its spike in the tolerance
window around the desired time of the specific trial (gray rectangle).
(c) Output weights $w_{i}$ from LIF (blue) and XIF neurons (red),
$\protect\vth$ (black line) and $V_{\infty,\protect\LIF}$ (gray
dashed line), before (left) and after (right) learning. The weights
have overall decreased during learning, while $\protect\vth$ (black
line) and $V_{\infty,\protect\LIF}$ have increased. The specific
weight pattern after learning is crucial for executing the task, random
weight shuffling leads to erroneous output spiking. (d) Overview of
the eight spike patterns of the task after learning. In context 1,
the system generates a temporal XOR computation, in context 2 a temporal
AND computation. The output spikes are in the desired tolerance windows
(gray rectangles) for all patterns.\label{fig:ComputingSchemexorand}}
\end{figure*}

In the following, we employ our networks for computations. In particular,
we investigate how their different phase space structures and CLVs
may be exploited in specific tasks. This requires a computational
scheme based on precise spiking, which is affected by the phase space
structure. We design a setup where one of our recurrent neural networks
acts as a kind of computational reservoir \citep{MNM02,JH04,MW12},
in the sense that it randomly nonlinearly filters its inputs. An output
neuron receives the generated spikes and learns to generate desired
outputs, see Fig.~\ref{fig:ComputingSchemexorand}a.

Inspired by experimental and computational neuroscience paradigms
\citep{maren2013contextual,mante2013context}, we assume that the
networks receive inputs from context neurons, whose spiking defines
the computation to be executed in the specific trial, and from input
neurons. Their synaptic weights as well as the recurrent ones are
static; only the output weights are learned. At the beginning of each
trial, all membrane potentials are reset to zero. The recurrent network
dynamics are therefore identical in trials with the same context and
input neuron spikes. To keep the computational scheme consistent,
we specify trains of precisely timed spikes as desired outputs.

The output neuron is an LIF neuron as used in the recurrent network.
The subthreshold dynamics of its membrane potential $V_{\text{out}}(t)$
are thus given by
\begin{align}
V_{\text{out}}(t) & =\sum_{j=1}^{N}w_{j}\sum_{k:t_{jk}<t}\e^{-\gamma_{\text{LIF}}(t-t_{jk})}\nonumber \\
 & +V_{\text{th},\text{out}}\left(-\sum_{t_{\text{sp}}<t}\e^{-\gamma_{\text{LIF}}(t-t_{\text{sp}})}\right)\nonumber \\
 & +V_{\infty,\text{out}}\left(1-\e^{-\gamma_{\text{LIF}}t}\right),\label{eq:outdyn}
\end{align}
where $w_{j}$ are the output weights, $V_{\text{th},\text{out}}$
is the threshold, $t_{\text{sp}}$ the output spikes, $V_{\infty,\text{out}}$
the asymptotic potential and $N$ the number of spiking neurons in
the recurrent network. Initially $V_{\text{th,out}}=V_{\text{th}}$,
$V_{\infty,\text{out}}=V_{\infty,\LIF}$ and the $w_{j}$ are are
drawn randomly from the uniform distribution over $[2C,0]$. We use
Finite Precision Learning \citep{MRS13} to learn the input-output
tasks. The shapes of the post-synaptic potentials in our single neuron
dynamics are different from those in ref.~\citep{MRS13} and there
is an additional constant driving term. The learning rule can be readily
adapted to this: We consider $V_{\text{out}}(t)-V_{\text{th},\text{out}}$
and cast it into the form $V_{\text{out}}(t)-V_{\text{th},\text{out}}=\sum_{k=1}^{N+2}w_{k}x_{k}(t)$.
Spikes are generated when $V_{\text{out}}(t)-V_{\text{th},\text{out}}$
reaches zero. At each time $t$, we thus have a kind of perceptron
classification task, where $w_{j}$, $V_{\text{th},\text{out}}$ and
$V_{\infty,\text{out}}$ are the ``weights'' to be learned. The
``inputs'' belonging to these weights are
\begin{align}
x_{j}(t) & =\sum_{t_{j}<t}\e^{-\gamma_{\text{LIF}}(t-t_{j})},\label{eq:xj}\\
x_{N+1}(t) & =-\sum_{t_{\text{sp}}<t}\e^{-\gamma_{\text{LIF}}(t-t_{\text{sp}})}-1,\label{eq:xNp1}\\
x_{N+2}(t) & =1-\e^{-\gamma_{\text{LIF}}t}.\label{eq:xNp2}
\end{align}
Following ref.~\citep{MRS13}, we assume a tolerance window of size
$\varepsilon$ around each desired spike (we use $\varepsilon=\SI{1}{\milli\second}$
throughout). There are now two kinds of errors: (i) undesired spikes,
i.e.~spikes out of a tolerance window or second spikes within a tolerance
window ($\Err=1$, the error time $t_{\Err}$ is the spike time) and
(ii) missing spikes within a tolerance window ($\Err=-1$, $t_{\Err}$
is the end of the tolerance window). The dynamics are stopped at the
first error and $w_{j}$, $V_{\text{th},\text{out}}$ and $V_{\infty,\text{out}}$
are corrected according to the perceptron rule,
\begin{align}
\Delta w_{j} & =-\eta\Err\sum_{k:t_{jk}<t_{\Err}}\e^{-\gamma_{\text{LIF}}(t_{\Err}-t_{jk})},\\
\Delta V_{\text{th},\text{out}} & =\eta\Err\left(\sum_{t_{\text{sp}}<t_{\Err}}\e^{-\gamma_{\text{LIF}}(t_{\Err}-t_{\text{sp}})}+1\right),\\
\Delta V_{\infty,\text{out}} & =-\eta\Err\left(1-\e^{-\gamma_{\text{LIF}}t_{\Err}}\right),
\end{align}
with learning rate $\eta$ (we use $\eta=0.01$). To focus on networks
with inhibitory neurons throughout the article, we restrict the output
weights to be inhibitory by clamping them at zero when they would
become excitatory during learning. We note that a missed spike generates
increases in $w_{j}$ and $V_{\infty,\text{out}}$ as well as a decrease
in $V_{\text{th},\text{out}}$ to foster spiking. If an undesired
spike occurs, the signs are reversed.

\subsection{Switchable temporal XOR/AND}

We exemplarily consider two tasks. In the first, the network of Fig.~\ref{fig:ComputingSchemexorand}a
learns to execute in context 1 a temporal XOR and in context 2 a temporal
AND computation, see Fig.~\ref{fig:ComputingSchemexorand}b-d. The
weights from context and input neurons to the recurrent network are
drawn randomly from the uniform distribution over $[2C,0]$. At the
beginning of a trial, at $t=\SI{0}{\milli\second}$, context neuron
1 or 2 sends a spike, specifying the context. Thereafter each input
neuron sends a spike, either at time $t_{\text{in,+}}=\SI{5}{\milli\second}$
(``$+$''-input) or at $t_{\text{in,-}}=\SI{10}{\milli\second}$
(``$-$''-input). The desired output spike is at $t_{\text{out,+}}=\SI{15}{\milli\second}$
(``$+$''-output) or at $t_{\text{out,-}}=\SI{20}{\milli\second}$
(``$-$''-output), depending on the context and the input spike
times.

The considered networks learn the task easily, whether the reservoir
consists of LIF or XIF neurons or of a mixture of both. The example
with a mixed network displayed in Fig.~\ref{fig:ComputingSchemexorand}
(same network as in Figs.~\ref{fig:rasterplot}, \ref{fig:spectra}c,
\ref{fig:pertgrowth} and \ref{fig:projection}) required 53 learning
cycles, where in each cycle the four input-desired output patterns
of both contexts were presented. The networks cannot learn the task,
if the recurrent network dynamics at the desired output times are
too similar for different contexts and input conditions. This happens
for recurrent LIF networks, if the context or input neurons have coupling
strengths that are so weak that the perturbations due to different
input timing are small. The states are then within the same flux tube
and the perturbation decays up to a time shift. In XIF and mixed networks,
the recurrent dynamics are too similar if there is insufficient time
for the perturbation to grow and spread before the first desired output.

\subsection{Detect or ignore input time differences}

\begin{figure}
\includegraphics[width=1\columnwidth]{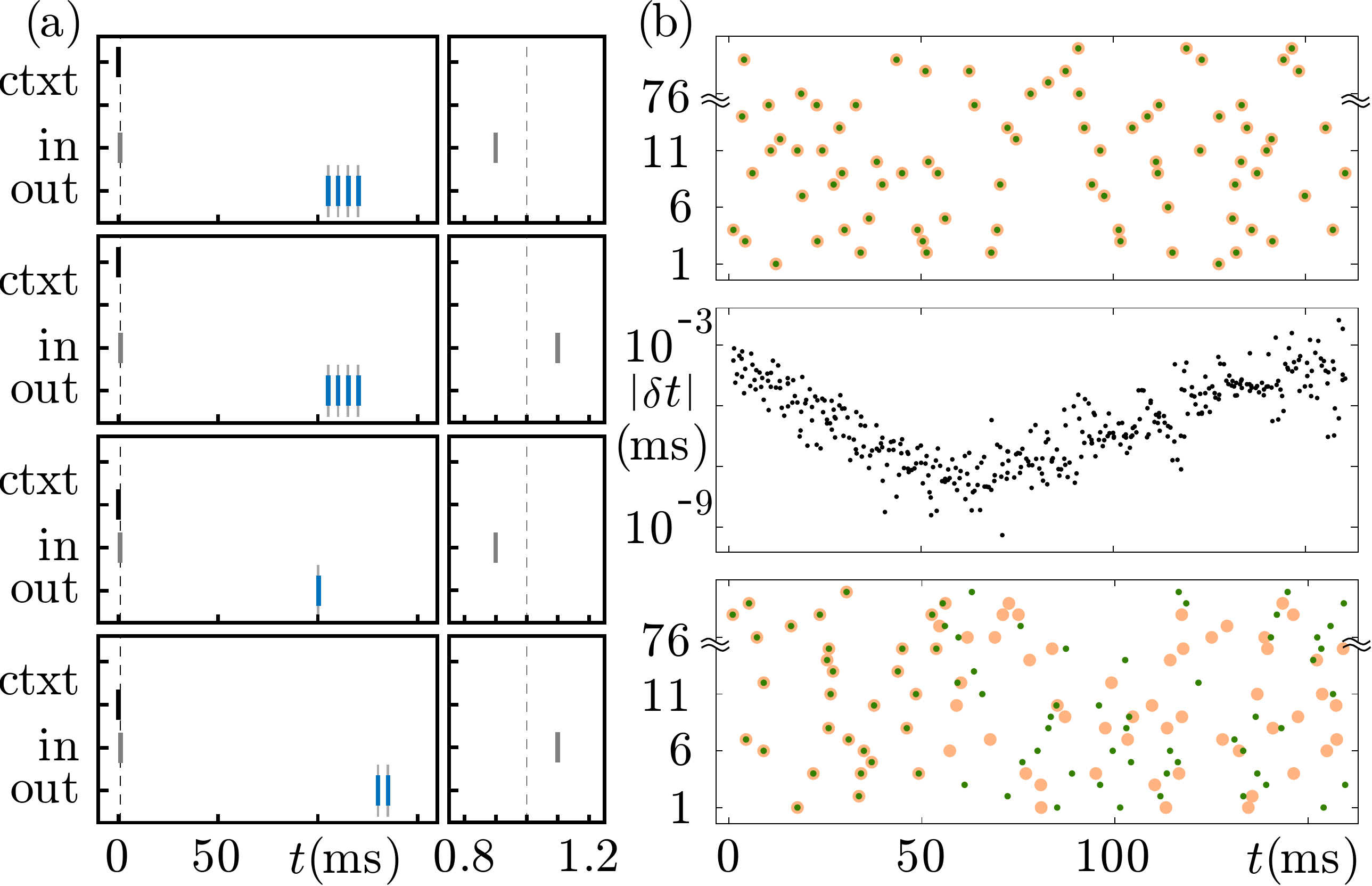}

\caption{Network trained to detect or ignore input time differences, after
learning. (a) Overview of the four spike patterns of the task. In
context 1, the network ignores the small time difference in the input
(right subpanels: closeups around input times), in context 2 it detects
and highlights it by generating different numbers and timings of output
spikes. After learning, the output spikes are in the desired tolerance
windows (gray rectangles, appearing as lines at the displayed timescale)
for all patterns. (b, top) The spiking dynamics in the recurrent reservoir
(green and yellow dots) are in context 1 similar for both input times,
due to the chosen context and input weights. This generally fosters
and here enables learning of the same output. (middle) The temporal
differences $\delta t$ between reservoir spikes display the typical
pattern of first shrinkage then growth, of perturbations along stable
CLVs, cf.~Fig.~\ref{fig:pertgrowth}d. (bottom) The different spiking
dynamics in the recurrent reservoir for different input times in context
2 allow the generation of different output.\label{fig:InputTimeDifference}}
\end{figure}

In the second task, the system has to ignore a difference in input
timing in context 1 and to detect it in context 2. The network setup
is as in Fig.~\ref{fig:ComputingSchemexorand}a, except that there
is only one input neuron. This sends a spike at $t_{\text{in},+}=t_{1}-\Delta t$
or at $t_{\text{in},-}=t_{1}+\Delta t$ ($t_{1}=\SI{1}{\milli\second},$
$\Delta t=\SI{0.1}{\milli\second}$). The output neuron shall generate
in context 1 for both input conditions the same output, a burst of
four spikes at $t=\SI{105}{\milli\second},\SI{110}{\milli\second},\SI{115}{\milli\second},\SI{120}{\milli\second}$.
In context 2 it shall detect the difference and highlight it by sending
one spike at $t=\SI{100}{\milli\second}$ (input at $t_{\text{in},+}$)
or two spikes at $t=\SI{130}{\milli\second},\SI{135}{\milli\second}$
(input at $t_{\text{in},-}$). For this task, for simplicity we assume
that the impacts of input neurons do not depend on the membrane potential,
i.e.~for them $h_{i}(V_{i}^{-})=1$. Further, we allow context and
input weights to be excitatory and inhibitory.

We find that networks with the previously chosen random parameters
of external weights drawn from $[2C,0]$ usually cannot solve the
task (criterion: no convergence within 50000 cycles). The reason is
different for pure LIF reservoirs and for reservoirs containing XIF
neurons: In a pure LIF reservoir, the small difference in input times
leads to state perturbations that are usually in the same flux tube.
These decay to a temporal shift until the time of the desired outputs.
The readout neuron thus cannot learn to generate two different output
patterns as required in context 2. In presence of XIF neurons, the
dynamics are locally unstable. The small input difference is amplified
in both contexts and the reservoir spiking is different for all four
patterns at the times of the desired output. The network therefore
has to learn four input-output relations with eleven output spikes
and silence periods in between, without being able to take advantage
of the fact that two of the four output patterns are identical. This
typically exceeds its learning capacity. We also observe for our parameters
that the dynamics of the pure LIF reservoir can leave its flux tube
due to the perturbation. If this happens only for context 2, the system
can often learn the task.

To solve the problem, we design the network such that, reliably, in
context 1 but not in context 2 the input differences leave the reservoir
spiking at the time of the desired outputs largely unaffected. This
can be achieved by choosing the context and input couplings such that
the input difference generates a state perturbation along a stable
CLV of the reservoir dynamics in context 1. In contrast, for context
2 the state perturbation should have a component in the direction
of an unstable CLV such that it is quickly amplified. The setup requires
mixed networks with both types of CLVs. We note that an alternative
approach might exploit the dichotomy of large and small perturbations,
which do and do not leave the flux tubes of pure LIF networks.

To derive appropriate weights, we compute the state perturbations
in the reservoir assuming that in the ``unperturbed'' system the
input arrives at $t_{1}$. We there have
\begin{equation}
V_{j}(t_{1}^{+})=V_{j}(t_{1}^{-})+C_{j}^{\text{\text{in}}},
\end{equation}
where $C_{j}^{\text{\text{in}}}$ is the coupling strength from the
input neuron to neuron $j$. In the ``perturbed'' system, the input
arrives shifted by $\delta t$ (here: $\delta t=\pm\Delta t$), such
that we have in linear approximation
\begin{equation}
\tilde{V}_{j}(t_{1}+\delta t^{+})=V_{j}(t_{1}^{-})+\dot{V}_{j}(t_{1}^{-})\delta t+C_{j}^{\text{in}}.
\end{equation}
To compute a perturbation in the $V_{j}(t_{1}^{+})$ that corresponds
to the perturbation due to the temporal shift of input, we propagate
the perturbed potential in linear approximation from $t_{1}+\delta t^{+}$
to $t_{1}^{+}$,
\begin{align}
\delta V_{j}(t^{+}) & \approx\tilde{V}_{j}(t_{1}+\delta t^{+})-\dot{\tilde{V}}_{j}(t_{1}+\delta t^{+})\delta t-V_{j}(t_{1}^{+})\nonumber \\
 & \approx\dot{V}_{j}(t_{1}^{-})\delta t-\left[(-\gamma_{j})\left(V_{j}(t_{1}^{-})+C_{j}^{\text{in}}\right)+I_{j}^{\text{ext}}\right]\delta t\nonumber \\
 & =\gamma_{j}C_{j}^{\text{in}}\delta t.
\end{align}
A temporal input difference that should be ignored should be proportional
to a stable CLV $\vec{v}_{i}$ at the state $\vec{V}(t_{1}^{+})$,
i.e.
\begin{equation}
\gamma_{j}C_{j}^{\text{in}}\propto v_{i,j}(\vec{V}(t_{1}^{+})).\label{eq:KoppPropLyapVekt}
\end{equation}
We choose the same recurrent network as in Figs.~\ref{fig:rasterplot},
\ref{fig:spectra}c, \ref{fig:pertgrowth} and \ref{fig:projection}
and the same CLV as in Fig.~\ref{fig:pertgrowth}d at $t=\SI{0}{\milli\second}$,
i.e.~$i=90$ and
\begin{equation}
\vec{V}(t_{1}^{+})=\vec{V}_{0},\label{eq:Vt1plus}
\end{equation}
where $\vec{V}_{0}$ is the state at which the vector was recorded.
Context input 1 determines the state at $t_{1}^{-}$ by fixing the
initial conditions of the dynamics. We choose as context input weights
\begin{align}
C_{j}^{\text{ctxt},1} & =V_{j}(0^{+})\nonumber \\
 & =e^{\gamma_{j}t_{1}}\left(V_{0,j}-C_{j}^{\text{in}}\right)+V_{\infty}\left(1-e^{\gamma_{j}t_{1}}\right),\label{eq:Init conds context 1}
\end{align}
which lead to Eq.~\eqref{eq:Vt1plus} after free propagation until
$t_{1}$ and receiving of the input $C_{j}^{\text{in}}$. To ensure
that the perturbation in context 2 has a component in the direction
of an unstable CLV, it suffices to choose a random context weight
vector, such that $\vec{V}(t_{1}^{+})\neq\vec{V}_{0}$ and $\vec{v}_{i}(\vec{V}_{0})$
is typically not a stable CLV or a linear combination of stable CLVs
at the state $\vec{V}(t_{1}^{+})$. We randomly permute the entries
of $\vec{C}^{\text{ctxt},1}$ to obtain $\vec{C}{}^{\text{ctxt},2}$.

We find that the network constructed in this way can reliably learn
the task. The example displayed in Fig.~\ref{fig:InputTimeDifference}
uses a proportionality factor of 0.01 in Eq.~\eqref{eq:KoppPropLyapVekt};
the output weights converged after 146 cycles.

\section{Discussion}

In the present article we investigate the spiking and membrane potential
statistics, the stability properties and the phase space structure
of mixed networks containing conventional LIF neurons and XIF neurons
with convex rise function. The recurrent connections are inhibitory
and the synaptic currents have infinitesimal temporal extent. We employ
two analytical mean-field approaches, one for the statistics and one
for the dynamical stability properties; numerical simulations yield
additional features of the dynamics and a better understanding of
the analytical approximations. Finally, we apply the networks for
computation with spikes, exploiting our insights into the dynamics.

We investigate networks in the balanced state. To establish it in
our networks, we introduce a voltage-dependence in the XIF neuron
inputs: below a certain potential, further input has no impact. This
simple model of a conductance-based synapse prevents XIF neurons from
switching off and provides a good-natured nonlinearity, which leaves
the dynamics analytically tractable.

The balanced state is typically investigated using spiking network
models with an excitatory and an inhibitory neuron population or with
a single population of hybrid excitatory-inhibitory or inhibitory
neurons \citep{Vreeswijk1996,Brunel00,KTADR08,MW10,deneve2016efficient}.
While detailed models of small circuits with specific abilities such
as central pattern generators commonly consider multiple neuron types
\citep{Prinz06}, studies on the impact of mixed populations of multiple
neuron types on the collective dynamics of larger networks are rare.
Ref.~\citep{savin06heterogeneous} simulated networks with excitatory
and inhibitory populations containing resonator and integrator type
neurons. These mixed networks both persistently generated activity
and quickly changed their overall rate in response to inputs, thereby
combining abilities of their pure counterparts. Refs.~\citep{Wang04ineurontypes}
and \citep{litwinkumar16ineurontypes} considered models for working
memory and visual processing with different types of interneurons
that were grouped into distinct populations with different connectivities.

We characterize the balanced dynamics of inhibitory mixed LIF and
XIF networks first from a statistical perspective, adopting a shot
noise approach, which accounts for the finite input rate and finite
size of individual inputs \citep{T88b,B06a,richardson,olmi2017exact}.
We extend this approach to XIF neurons and derive their steady-state
firing rate and voltage probability density. In contrast to the case
of LIF neurons, the final continuity equation needs to be integrated
numerically, due to the nonlinearity in the XIF input. We apply the
results to obtain neuron parameters that lead to homogeneous firing
rates for our further considered networks. We insert these rates into
the mean-field expressions of the Lyapunov exponents (LEs) and thus
analytically determine the dynamical stability properties of the network.

While networks of LIF neurons have stable dynamics \citep{ZLPT06,JMT08,ZBH09,JMT09,MW12},
we find that already one XIF neuron gives rise to a positive largest
LE indicating chaos, in contrast to the robustness against introducing
excitatory connections \citep{ZBH09,JMT09}. We give an analytical
argument for this and expand it to a mean-field estimate of the entire
Lyapunov spectrum. Simply put, the destabilizing effect of excitatory
inputs will be compensated by receiving inhibitory ones, if the latter
dominate and the period of spiking is overall increased compared to
the free neurons. If one introduces an XIF neuron there is nothing
which could counteract the increase of its perturbation through inhibitory
input other than an unlikely network backreaction triggered by its
perturbed output spikes. We note that in the phase representation
of LIF neurons used in ref.~\citep{JMT09}, in contrast to our voltage
representation an excitatory input explicitly increases a perturbation,
while an inhibitory input decreases it, unless the excitatory input
is suprathreshold \citep{MT10A,gu2018dynamics}.

While computing the largest LE is a standard procedure, few studies
have so far obtained a large part or the entire spectrum of balanced
spiking dynamics. They considered a single homogeneous or an excitatory
and an inhibitory neuron population \citep{MW10,MW12,luccioli12collective,lajoie2014structured,ullner16selfsustained,engelken2017chaotic}.
We analytically and numerically obtain the full spectrum for mixed
networks of inhibitory LIF and XIF neurons. Interestingly, we find
that it separates into two parts, in contrast to the ones reported
previously including those of networks with separate excitatory and
inhibitory populations. Furthermore, we compute the covariant Lyapunov
vectors (CLVs) of the dynamics \citep{pikovsky+politi,Kuptsov2012}.
They provide us with further insight into the phase space structure
and the approximations underlying the mean-field analysis of LEs.
The stable (unstable) CLVs are approximately aligned to the subspace
of perturbations to LIF (XIF) neurons.

Our mean-field analysis predicts that the number of negative (positive)
LEs is equal to the number of LIF (XIF) neurons. Since the underlying
arguments do not depend on the neurons' specifics, we expect this
to hold for any types of neurons with purely concave and convex rise
functions. The mean-field analysis further indicates that the size
of the LEs is approximately given by the strength of the leak and
the quotient of free and actual spike frequency. The LEs are thus
largely independent of the collective dynamics but rather reflect
properties of individual neurons. This implies in particular that
the typical perturbation growth rate does not change with network
size. It further implies that in the balanced state, where the ratio
of actual and free spike rate is low, the LEs are mainly determined
by the single neuron leak strengths, see ref.~\citep{MW12} for a
similar finding in large networks of LIF neurons with high indegree.
The result is a consequence of the linear subthreshold dynamics of
the neurons, which imply that the increase or decrease of a perturbation
is independent of the state of the neuron when receiving a spike.
We note that ref.~\citep{coombes2000chaos} defined the Lyapunov
spectrum as consisting of mean-field LEs in a numerical study on LIF
neuron networks.

Our numerical computations of the Lyapunov spectrum show that the
mean-field result is a good approximation. We explain this by analyzing
the CLVs. Furthermore, we derive an exact expression for the change
of phase space volume, which agrees with the mean-field result.

The presence of discrete events and the possibly large impact of changing
their order could in principle render the transfer of insights on
infinitesimal perturbations to finite ones difficult. Refs.~\citep{JMT09,JMT08}
studied the evolution of finite size perturbations in the pure LIF
network model with stable dynamics and showed that finite size perturbations
decay exponentially fast, while the minimal perturbation leading to
a change of event order decreases only algebraically. Thus, for sufficiently
small initial finite size perturbations the probability of a change
of event order goes to zero and no difficulties occur. For unstable
dynamics, we may expect generic interchanges of event order to be
an additional source of deviations between trajectories so that small
finite size perturbations grow as fast and larger ones at least as
fast as their infinitesimal counterparts. We therefore focus mostly
on linear stability analysis in the present article. Our numerical
simulations employ finite size perturbations and confirm the results.

To illustrate the usefulness of our findings we apply the considered
networks to neural computations. We propose a computing scheme based
on precisely timed spikes where details of the phase space structure
matter. In particular, our solution of the second task exploits details
of the network's state space, the stability or instability of the
spiking dynamics against perturbations in the direction of different
CLVs. This may be especially relevant for neuromorphic computing,
where precise spike-based schemes receive increasing interest \citep{lagorce2015stick,verzi18computing,pfeiffer2018deep,huh18gradient,zenke2018superspike}.
In our setup, the inputs are fed into a random recurrent network,
whose neurons generate precisely timed spike trains, which depend
nonlinearly on the input. In this sense, the recurrent network acts
like a random filter bank and computational reservoir. The spike trains
are read out by a spiking neuron. In contrast to previous spiking
reservoir computers \citep{MNM02,thalmeier2015universal,abbott2016building,nicola2017supervised,depasquale2018full},
we use trains of precisely timed spikes as targets. To train the readout
neuron, we use Finite Precision Learning \citep{MRS13}. It was introduced
for neurons with temporally extended input currents of either sign.
In our study we adapt it to a neuron with inhibitory, infinitesimally
short input currents and constant external drive. We note that the
general phase space structure implies that the considered networks
do not lend themselves to conventional reservoir computing: there
is no global fixed point, which could be reached by the spiking dynamics
such that sufficiently long past input is forgotten. In other words,
our networks do not have the so-called echo state property \citep{jaeger2001echo}.
We therefore introduce a forgetting mechanism by resetting the network
at the beginning of a trial.

Our findings show that by choosing appropriate numbers of LIF and
XIF neurons, one can straightforwardly construct spiking networks
with a desired number of stable and unstable directions. The obtained
CLVs allow to exploit them for computation: one can choose the input
weights such that meaningless inputs and input perturbations happen
along stable directions while meaningful ones have a component in
an unstable direction; the former ones are suppressed while the latter
ones are amplified. Our mixed networks thus combine the computational
capabilities of purely stable and purely unstable networks. It is
tempting to speculate that also in the brain the combination of different
neuron types might globally change the phase space structure and lead
to combinations of computational capabilities that can be selected
with different input vectors. While we have chosen the input weights
by hand, plasticity rules for spiking networks in the brain as well
as future artificial ones may allow to find them by learning.
\begin{acknowledgments}
We thank Arindam Saha for studying a related mixed network model during
an internship and we thank Fred Wolf, Marc Timme, Paul Tiesinga, Arindam
Saha, Anna Hellfritzsch, Diemut Regel, Felipe Kalle Kossio, Joscha
Liedke, and Rainer Engelken for many fruitful discussions. This work
was supported by the German Federal Ministry of Education and Research
BMBF through the Bernstein Network for Computational Neuroscience
(Bernstein Award 2014: 01GQ1501 and 01GQ1710).
\end{acknowledgments}

\bibliographystyle{apsrev4-1_prx}

\begin{thebibliography}{73}%
\makeatletter
\providecommand \@ifxundefined [1]{%
 \@ifx{#1\undefined}
}%
\providecommand \@ifnum [1]{%
 \ifnum #1\expandafter \@firstoftwo
 \else \expandafter \@secondoftwo
 \fi
}%
\providecommand \@ifx [1]{%
 \ifx #1\expandafter \@firstoftwo
 \else \expandafter \@secondoftwo
 \fi
}%
\providecommand \natexlab [1]{#1}%
\providecommand \enquote  [1]{``#1''}%
\providecommand \bibnamefont  [1]{#1}%
\providecommand \bibfnamefont [1]{#1}%
\providecommand \citenamefont [1]{#1}%
\providecommand \href@noop [0]{\@secondoftwo}%
\providecommand \href [0]{\begingroup \@sanitize@url \@href}%
\providecommand \@href[1]{\@@startlink{#1}\@@href}%
\providecommand \@@href[1]{\endgroup#1\@@endlink}%
\providecommand \@sanitize@url [0]{\catcode `\\12\catcode `\$12\catcode
  `\&12\catcode `\#12\catcode `\^12\catcode `\_12\catcode `\%12\relax}%
\providecommand \@@startlink[1]{}%
\providecommand \@@endlink[0]{}%
\providecommand \url  [0]{\begingroup\@sanitize@url \@url }%
\providecommand \@url [1]{\endgroup\@href {#1}{\urlprefix }}%
\providecommand \urlprefix  [0]{URL }%
\providecommand \Eprint [0]{\href }%
\providecommand \doibase [0]{http://dx.doi.org/}%
\providecommand \selectlanguage [0]{\@gobble}%
\providecommand \bibinfo  [0]{\@secondoftwo}%
\providecommand \bibfield  [0]{\@secondoftwo}%
\providecommand \translation [1]{[#1]}%
\providecommand \BibitemOpen [0]{}%
\providecommand \bibitemStop [0]{}%
\providecommand \bibitemNoStop [0]{.\EOS\space}%
\providecommand \EOS [0]{\spacefactor3000\relax}%
\providecommand \BibitemShut  [1]{\csname bibitem#1\endcsname}%
\let\auto@bib@innerbib\@empty
\bibitem [{\citenamefont {Gerstein}\ and\ \citenamefont
  {Mandelbrot}(1964)}]{GM64}%
  \BibitemOpen
  \bibfield  {author} {\bibinfo {author} {\bibfnamefont {G.}~\bibnamefont
  {Gerstein}}\ and\ \bibinfo {author} {\bibfnamefont {B.}~\bibnamefont
  {Mandelbrot}},\ }\bibfield  {title} {\emph {\bibinfo {title} {Random walk
  models for the spike activity of a single neuron},\ }}\href@noop {}
  {\bibfield  {journal} {\bibinfo  {journal} {Biophys. J.}\ }\textbf {\bibinfo
  {volume} {4}},\ \bibinfo {pages} {41} (\bibinfo {year} {1964})}\BibitemShut
  {NoStop}%
\bibitem [{\citenamefont {Shadlen}\ and\ \citenamefont
  {Newsome}(1994)}]{shadlen1994noise}%
  \BibitemOpen
  \bibfield  {author} {\bibinfo {author} {\bibfnamefont {M.~N.}\ \bibnamefont
  {Shadlen}}\ and\ \bibinfo {author} {\bibfnamefont {W.~T.}\ \bibnamefont
  {Newsome}},\ }\bibfield  {title} {\emph {\bibinfo {title} {Noise, neural
  codes and cortical organization.}\ }}\href@noop {} {\bibfield  {journal}
  {\bibinfo  {journal} {Curr. Opin. Neurobiol.}\ }\textbf {\bibinfo {volume}
  {4}},\ \bibinfo {pages} {569} (\bibinfo {year} {1994})}\BibitemShut {NoStop}%
\bibitem [{\citenamefont {van Vreeswijk}\ and\ \citenamefont
  {Sompolinsky}(1996)}]{Vreeswijk1996}%
  \BibitemOpen
  \bibfield  {author} {\bibinfo {author} {\bibfnamefont {C.}~\bibnamefont {van
  Vreeswijk}}\ and\ \bibinfo {author} {\bibfnamefont {H.}~\bibnamefont
  {Sompolinsky}},\ }\bibfield  {title} {\emph {\bibinfo {title} {Chaos in
  neuronal networks with balanced excitatory and inhibitory activity},\
  }}\href@noop {} {\bibfield  {journal} {\bibinfo  {journal} {Science}\
  }\textbf {\bibinfo {volume} {274}},\ \bibinfo {pages} {1724} (\bibinfo {year}
  {1996})}\BibitemShut {NoStop}%
\bibitem [{\citenamefont {Den\`{e}ve}\ and\ \citenamefont
  {Machens}(2016)}]{deneve2016efficient}%
  \BibitemOpen
  \bibfield  {author} {\bibinfo {author} {\bibfnamefont {S.}~\bibnamefont
  {Den\`{e}ve}}\ and\ \bibinfo {author} {\bibfnamefont {C.~K.}\ \bibnamefont
  {Machens}},\ }\bibfield  {title} {\emph {\bibinfo {title} {Efficient codes
  and balanced networks.}\ }}\href {\doibase 10.1038/nn.4243} {\bibfield
  {journal} {\bibinfo  {journal} {Nat. Neurosci.}\ }\textbf {\bibinfo {volume}
  {19}},\ \bibinfo {pages} {375} (\bibinfo {year} {2016})}\BibitemShut
  {NoStop}%
\bibitem [{\citenamefont {Pikovsky}\ and\ \citenamefont
  {Politi}(2016)}]{pikovsky+politi}%
  \BibitemOpen
  \bibfield  {author} {\bibinfo {author} {\bibfnamefont {A.}~\bibnamefont
  {Pikovsky}}\ and\ \bibinfo {author} {\bibfnamefont {A.}~\bibnamefont
  {Politi}},\ }\href@noop {} {\emph {\bibinfo {title} {Lyapunov Exponents}}}\
  (\bibinfo  {publisher} {Cambridge University Press},\ \bibinfo {address}
  {Cambridge},\ \bibinfo {year} {2016})\BibitemShut {NoStop}%
\bibitem [{\citenamefont {Kuptsov}\ and\ \citenamefont
  {Parlitz}(2012)}]{Kuptsov2012}%
  \BibitemOpen
  \bibfield  {author} {\bibinfo {author} {\bibfnamefont {P.~V.}\ \bibnamefont
  {Kuptsov}}\ and\ \bibinfo {author} {\bibfnamefont {U.}~\bibnamefont
  {Parlitz}},\ }\bibfield  {title} {\emph {\bibinfo {title} {Theory and
  computation of covariant lyapunov vectors},\ }}\href {\doibase
  10.1007/s00332-012-9126-5} {\bibfield  {journal} {\bibinfo  {journal} {J .
  Nonlinear Sci.}\ }\textbf {\bibinfo {volume} {22}},\ \bibinfo {pages} {727}
  (\bibinfo {year} {2012})}\BibitemShut {NoStop}%
\bibitem [{\citenamefont {Frederickson}\ \emph {et~al.}(1983)\citenamefont
  {Frederickson}, \citenamefont {L~Kaplan}, \citenamefont {D~Yorke},\ and\
  \citenamefont {A~Yorke}}]{KY2}%
  \BibitemOpen
  \bibfield  {author} {\bibinfo {author} {\bibfnamefont {P.}~\bibnamefont
  {Frederickson}}, \bibinfo {author} {\bibfnamefont {J.}~\bibnamefont
  {L~Kaplan}}, \bibinfo {author} {\bibfnamefont {E.}~\bibnamefont {D~Yorke}}, \
  and\ \bibinfo {author} {\bibfnamefont {J.}~\bibnamefont {A~Yorke}},\
  }\bibfield  {title} {\emph {\bibinfo {title} {The lyapunov dimension of
  strange attractors},\ }}\href@noop {} {\bibfield  {journal} {\bibinfo
  {journal} {Journal of Differential Equations}\ }\textbf {\bibinfo {volume}
  {49}},\ \bibinfo {pages} {185} (\bibinfo {year} {1983})}\BibitemShut
  {NoStop}%
\bibitem [{\citenamefont {Zillmer}\ \emph {et~al.}(2006)\citenamefont
  {Zillmer}, \citenamefont {Livi}, \citenamefont {Politi},\ and\ \citenamefont
  {Torcini}}]{ZLPT06}%
  \BibitemOpen
  \bibfield  {author} {\bibinfo {author} {\bibfnamefont {R.}~\bibnamefont
  {Zillmer}}, \bibinfo {author} {\bibfnamefont {R.}~\bibnamefont {Livi}},
  \bibinfo {author} {\bibfnamefont {A.}~\bibnamefont {Politi}}, \ and\ \bibinfo
  {author} {\bibfnamefont {A.}~\bibnamefont {Torcini}},\ }\bibfield  {title}
  {\emph {\bibinfo {title} {Desynchronization in diluted neural networks},\
  }}\href@noop {} {\bibfield  {journal} {\bibinfo  {journal} {Phys. Rev. E}\
  }\textbf {\bibinfo {volume} {74}},\ \bibinfo {pages} {036203} (\bibinfo
  {year} {2006})}\BibitemShut {NoStop}%
\bibitem [{\citenamefont {Zillmer}\ \emph {et~al.}(2009)\citenamefont
  {Zillmer}, \citenamefont {Brunel},\ and\ \citenamefont {Hansel}}]{ZBH09}%
  \BibitemOpen
  \bibfield  {author} {\bibinfo {author} {\bibfnamefont {R.}~\bibnamefont
  {Zillmer}}, \bibinfo {author} {\bibfnamefont {N.}~\bibnamefont {Brunel}}, \
  and\ \bibinfo {author} {\bibfnamefont {D.}~\bibnamefont {Hansel}},\
  }\bibfield  {title} {\emph {\bibinfo {title} {Very long transients, irregular
  firing, and chaotic dynamics in networks of randomly connected inhibitory
  integrate-and-fire neurons},\ }}\href@noop {} {\bibfield  {journal} {\bibinfo
   {journal} {Phys. Rev. E}\ }\textbf {\bibinfo {volume} {79}},\ \bibinfo
  {pages} {031909} (\bibinfo {year} {2009})}\BibitemShut {NoStop}%
\bibitem [{\citenamefont {Jahnke}\ \emph {et~al.}(2008)\citenamefont {Jahnke},
  \citenamefont {Memmesheimer},\ and\ \citenamefont {Timme}}]{JMT08}%
  \BibitemOpen
  \bibfield  {author} {\bibinfo {author} {\bibfnamefont {S.}~\bibnamefont
  {Jahnke}}, \bibinfo {author} {\bibfnamefont {R.-M.}\ \bibnamefont
  {Memmesheimer}}, \ and\ \bibinfo {author} {\bibfnamefont {M.}~\bibnamefont
  {Timme}},\ }\bibfield  {title} {\emph {\bibinfo {title} {Stable irregular
  dynamics in complex neural networks},\ }}\href@noop {} {\bibfield  {journal}
  {\bibinfo  {journal} {Phys. Rev. Lett.}\ }\textbf {\bibinfo {volume} {100}},\
  \bibinfo {pages} {048102} (\bibinfo {year} {2008})}\BibitemShut {NoStop}%
\bibitem [{\citenamefont {Jahnke}\ \emph {et~al.}(2009)\citenamefont {Jahnke},
  \citenamefont {Memmesheimer},\ and\ \citenamefont {Timme}}]{JMT09}%
  \BibitemOpen
  \bibfield  {author} {\bibinfo {author} {\bibfnamefont {S.}~\bibnamefont
  {Jahnke}}, \bibinfo {author} {\bibfnamefont {R.-M.}\ \bibnamefont
  {Memmesheimer}}, \ and\ \bibinfo {author} {\bibfnamefont {M.}~\bibnamefont
  {Timme}},\ }\bibfield  {title} {\emph {\bibinfo {title} {How chaotic is the
  balanced state?}\ }}\href@noop {} {\bibfield  {journal} {\bibinfo  {journal}
  {Front. Comput. Neurosci.}\ }\textbf {\bibinfo {volume} {3}},\ \bibinfo
  {pages} {13} (\bibinfo {year} {2009})}\BibitemShut {NoStop}%
\bibitem [{\citenamefont {Politi}\ \emph {et~al.}(1993)\citenamefont {Politi},
  \citenamefont {Livi}, \citenamefont {Oppo},\ and\ \citenamefont
  {Kapral}}]{Politi_1993}%
  \BibitemOpen
  \bibfield  {author} {\bibinfo {author} {\bibfnamefont {A.}~\bibnamefont
  {Politi}}, \bibinfo {author} {\bibfnamefont {R.}~\bibnamefont {Livi}},
  \bibinfo {author} {\bibfnamefont {G.-L.}\ \bibnamefont {Oppo}}, \ and\
  \bibinfo {author} {\bibfnamefont {R.}~\bibnamefont {Kapral}},\ }\bibfield
  {title} {\emph {\bibinfo {title} {Unpredictable behaviour in stable
  systems},\ }}\href {\doibase 10.1209/0295-5075/22/8/003} {\bibfield
  {journal} {\bibinfo  {journal} {Europhys. Lett.}\ }\textbf {\bibinfo {volume}
  {22}},\ \bibinfo {pages} {571} (\bibinfo {year} {1993})}\BibitemShut
  {NoStop}%
\bibitem [{\citenamefont {Politi}\ and\ \citenamefont
  {Torcini}(2010)}]{Politi2010}%
  \BibitemOpen
  \bibfield  {author} {\bibinfo {author} {\bibfnamefont {A.}~\bibnamefont
  {Politi}}\ and\ \bibinfo {author} {\bibfnamefont {A.}~\bibnamefont
  {Torcini}},\ }\bibinfo {title} {Stable chaos},\ in\ \href {\doibase
  10.1007/978-3-642-04629-2_6} {\emph {\bibinfo {booktitle} {Nonlinear Dynamics
  and Chaos: Advances and Perspectives}}}\ (\bibinfo  {publisher} {Springer},\
  \bibinfo {address} {Berlin, Heidelberg},\ \bibinfo {year} {2010})\ pp.\
  \bibinfo {pages} {103--129}\BibitemShut {NoStop}%
\bibitem [{\citenamefont {Monteforte}\ and\ \citenamefont {Wolf}(2012)}]{MW12}%
  \BibitemOpen
  \bibfield  {author} {\bibinfo {author} {\bibfnamefont {M.}~\bibnamefont
  {Monteforte}}\ and\ \bibinfo {author} {\bibfnamefont {F.}~\bibnamefont
  {Wolf}},\ }\bibfield  {title} {\emph {\bibinfo {title} {Dynamic flux tubes
  form reservoirs of stability in neuronal circuits},\ }}\href@noop {}
  {\bibfield  {journal} {\bibinfo  {journal} {Phys. Rev. X}\ }\textbf {\bibinfo
  {volume} {2}},\ \bibinfo {pages} {041007} (\bibinfo {year}
  {2012})}\BibitemShut {NoStop}%
\bibitem [{\citenamefont {Dayan}\ and\ \citenamefont {Abbott}(2001)}]{DA01}%
  \BibitemOpen
  \bibfield  {author} {\bibinfo {author} {\bibfnamefont {P.}~\bibnamefont
  {Dayan}}\ and\ \bibinfo {author} {\bibfnamefont {L.}~\bibnamefont {Abbott}},\
  }\href@noop {} {\emph {\bibinfo {title} {Theoretical Neuroscience:
  Computational and Mathematical Modeling of Neural Systems}}}\ (\bibinfo
  {publisher} {MIT Press},\ \bibinfo {address} {Cambridge},\ \bibinfo {year}
  {2001})\BibitemShut {NoStop}%
\bibitem [{\citenamefont {Gerstner}\ \emph {et~al.}(2014)\citenamefont
  {Gerstner}, \citenamefont {Kistler}, \citenamefont {Naud},\ and\
  \citenamefont {Paninski}}]{gerstner2014neuronal}%
  \BibitemOpen
  \bibfield  {author} {\bibinfo {author} {\bibfnamefont {W.}~\bibnamefont
  {Gerstner}}, \bibinfo {author} {\bibfnamefont {W.~M.}\ \bibnamefont
  {Kistler}}, \bibinfo {author} {\bibfnamefont {R.}~\bibnamefont {Naud}}, \
  and\ \bibinfo {author} {\bibfnamefont {L.}~\bibnamefont {Paninski}},\
  }\href@noop {} {\emph {\bibinfo {title} {Neuronal Dynamics - From single
  neurons to networks and models of cognition}}}\ (\bibinfo  {publisher}
  {Cambridge University Press},\ \bibinfo {address} {Cambridge},\ \bibinfo
  {year} {2014})\BibitemShut {NoStop}%
\bibitem [{\citenamefont {Monteforte}\ and\ \citenamefont {Wolf}(2010)}]{MW10}%
  \BibitemOpen
  \bibfield  {author} {\bibinfo {author} {\bibfnamefont {M.}~\bibnamefont
  {Monteforte}}\ and\ \bibinfo {author} {\bibfnamefont {F.}~\bibnamefont
  {Wolf}},\ }\bibfield  {title} {\emph {\bibinfo {title} {Dynamical entropy
  production in spiking neuron networks in the balanced state.}\ }}\href@noop
  {} {\bibfield  {journal} {\bibinfo  {journal} {Phys. Rev. Lett.}\ }\textbf
  {\bibinfo {volume} {105}},\ \bibinfo {pages} {268104} (\bibinfo {year}
  {2010})}\BibitemShut {NoStop}%
\bibitem [{\citenamefont {Tremblay}\ \emph {et~al.}(2016)\citenamefont
  {Tremblay}, \citenamefont {Lee},\ and\ \citenamefont
  {Rudy}}]{tremblay2016gabaergic}%
  \BibitemOpen
  \bibfield  {author} {\bibinfo {author} {\bibfnamefont {R.}~\bibnamefont
  {Tremblay}}, \bibinfo {author} {\bibfnamefont {S.}~\bibnamefont {Lee}}, \
  and\ \bibinfo {author} {\bibfnamefont {B.}~\bibnamefont {Rudy}},\ }\bibfield
  {title} {\emph {\bibinfo {title} {{GABA}ergic interneurons in the neocortex:
  From cellular properties to circuits.}\ }}\href {\doibase
  10.1016/j.neuron.2016.06.033} {\bibfield  {journal} {\bibinfo  {journal}
  {Neuron}\ }\textbf {\bibinfo {volume} {91}},\ \bibinfo {pages} {260}
  (\bibinfo {year} {2016})}\BibitemShut {NoStop}%
\bibitem [{\citenamefont {Buonomano}\ and\ \citenamefont
  {Merzenich}(1995)}]{BM95}%
  \BibitemOpen
  \bibfield  {author} {\bibinfo {author} {\bibfnamefont {D.~V.}\ \bibnamefont
  {Buonomano}}\ and\ \bibinfo {author} {\bibfnamefont {M.~M.}\ \bibnamefont
  {Merzenich}},\ }\bibfield  {title} {\emph {\bibinfo {title} {Temporal
  information transformed into a spatial code by a neural network with
  realistic properties.}\ }}\href@noop {} {\bibfield  {journal} {\bibinfo
  {journal} {Science}\ }\textbf {\bibinfo {volume} {267}},\ \bibinfo {pages}
  {1028} (\bibinfo {year} {1995})}\BibitemShut {NoStop}%
\bibitem [{\citenamefont {Dominey}(1995)}]{dominey1995complex}%
  \BibitemOpen
  \bibfield  {author} {\bibinfo {author} {\bibfnamefont {P.~F.}\ \bibnamefont
  {Dominey}},\ }\bibfield  {title} {\emph {\bibinfo {title} {Complex
  sensory-motor sequence learning based on recurrent state representation and
  reinforcement learning.}\ }}\href@noop {} {\bibfield  {journal} {\bibinfo
  {journal} {Biol. Cybern.}\ }\textbf {\bibinfo {volume} {73}},\ \bibinfo
  {pages} {265} (\bibinfo {year} {1995})}\BibitemShut {NoStop}%
\bibitem [{\citenamefont {Jaeger}\ and\ \citenamefont {Haas}(2004)}]{JH04}%
  \BibitemOpen
  \bibfield  {author} {\bibinfo {author} {\bibfnamefont {H.}~\bibnamefont
  {Jaeger}}\ and\ \bibinfo {author} {\bibfnamefont {H.}~\bibnamefont {Haas}},\
  }\bibfield  {title} {\emph {\bibinfo {title} {Harnessing nonlinearity:
  Predicting chaotic systems and saving energy in wireless communication},\
  }}\href@noop {} {\bibfield  {journal} {\bibinfo  {journal} {Science}\
  }\textbf {\bibinfo {volume} {304}},\ \bibinfo {pages} {78} (\bibinfo {year}
  {2004})}\BibitemShut {NoStop}%
\bibitem [{\citenamefont {Maass}\ \emph {et~al.}(2002)\citenamefont {Maass},
  \citenamefont {Natschl{\"a}ger},\ and\ \citenamefont {Markram}}]{MNM02}%
  \BibitemOpen
  \bibfield  {author} {\bibinfo {author} {\bibfnamefont {W.}~\bibnamefont
  {Maass}}, \bibinfo {author} {\bibfnamefont {T.}~\bibnamefont
  {Natschl{\"a}ger}}, \ and\ \bibinfo {author} {\bibfnamefont {H.}~\bibnamefont
  {Markram}},\ }\bibfield  {title} {\emph {\bibinfo {title} {Real-time
  computing without stable states: A new framework for neural computation based
  on perturbations},\ }}\href@noop {} {\bibfield  {journal} {\bibinfo
  {journal} {Neural Comput.}\ }\textbf {\bibinfo {volume} {14}},\ \bibinfo
  {pages} {2531} (\bibinfo {year} {2002})}\BibitemShut {NoStop}%
\bibitem [{\citenamefont {Westover}\ \emph {et~al.}(2002)\citenamefont
  {Westover}, \citenamefont {Eliasmith},\ and\ \citenamefont
  {Anderson}}]{westover2002linearly}%
  \BibitemOpen
  \bibfield  {author} {\bibinfo {author} {\bibfnamefont {M.~B.}\ \bibnamefont
  {Westover}}, \bibinfo {author} {\bibfnamefont {C.}~\bibnamefont {Eliasmith}},
  \ and\ \bibinfo {author} {\bibfnamefont {C.~H.}\ \bibnamefont {Anderson}},\
  }\bibfield  {title} {\emph {\bibinfo {title} {Linearly decodable functions
  from neural population codes},\ }}\href@noop {} {\bibfield  {journal}
  {\bibinfo  {journal} {Neurocomputing}\ }\textbf {\bibinfo {volume} {44-46}},\
  \bibinfo {pages} {691} (\bibinfo {year} {2002})}\BibitemShut {NoStop}%
\bibitem [{\citenamefont {LeCun}\ \emph {et~al.}(2015)\citenamefont {LeCun},
  \citenamefont {Bengio},\ and\ \citenamefont {Hinton}}]{lecun2015deep}%
  \BibitemOpen
  \bibfield  {author} {\bibinfo {author} {\bibfnamefont {Y.}~\bibnamefont
  {LeCun}}, \bibinfo {author} {\bibfnamefont {Y.}~\bibnamefont {Bengio}}, \
  and\ \bibinfo {author} {\bibfnamefont {G.}~\bibnamefont {Hinton}},\
  }\bibfield  {title} {\emph {\bibinfo {title} {Deep learning.}\ }}\href
  {\doibase 10.1038/nature14539} {\bibfield  {journal} {\bibinfo  {journal}
  {Nature (London)}\ }\textbf {\bibinfo {volume} {521}},\ \bibinfo {pages}
  {436} (\bibinfo {year} {2015})}\BibitemShut {NoStop}%
\bibitem [{\citenamefont {Sussillo}\ and\ \citenamefont {Abbott}(2009)}]{SA09}%
  \BibitemOpen
  \bibfield  {author} {\bibinfo {author} {\bibfnamefont {D.}~\bibnamefont
  {Sussillo}}\ and\ \bibinfo {author} {\bibfnamefont {L.~F.}\ \bibnamefont
  {Abbott}},\ }\bibfield  {title} {\emph {\bibinfo {title} {Generating coherent
  patterns of activity from chaotic neural networks.}\ }}\href {\doibase
  10.1016/j.neuron.2009.07.018} {\bibfield  {journal} {\bibinfo  {journal}
  {Neuron}\ }\textbf {\bibinfo {volume} {63}},\ \bibinfo {pages} {544}
  (\bibinfo {year} {2009})}\BibitemShut {NoStop}%
\bibitem [{\citenamefont {Ponulak}\ and\ \citenamefont
  {Kasi{\'n}ski}(2010)}]{PK10}%
  \BibitemOpen
  \bibfield  {author} {\bibinfo {author} {\bibfnamefont {F.}~\bibnamefont
  {Ponulak}}\ and\ \bibinfo {author} {\bibfnamefont {A.}~\bibnamefont
  {Kasi{\'n}ski}},\ }\bibfield  {title} {\emph {\bibinfo {title} {Supervised
  learning in spiking neural networks with resume: sequence learning,
  classification, and spike shifting.}\ }}\href {\doibase
  10.1162/neco.2009.11-08-901} {\bibfield  {journal} {\bibinfo  {journal}
  {Neural Comput.}\ }\textbf {\bibinfo {volume} {22}},\ \bibinfo {pages} {467}
  (\bibinfo {year} {2010})}\BibitemShut {NoStop}%
\bibitem [{\citenamefont {Florian}(2012)}]{Florian2012}%
  \BibitemOpen
  \bibfield  {author} {\bibinfo {author} {\bibfnamefont {R.~V.}\ \bibnamefont
  {Florian}},\ }\bibfield  {title} {\emph {\bibinfo {title} {The chronotron:
  {A} neuron that learns to fire temporally precise spike patterns.}\ }}\href
  {\doibase 10.1371/journal.pone.0040233} {\bibfield  {journal} {\bibinfo
  {journal} {PLOS One}\ }\textbf {\bibinfo {volume} {7}},\ \bibinfo {pages}
  {e40233} (\bibinfo {year} {2012})}\BibitemShut {NoStop}%
\bibitem [{\citenamefont {Mohemmed}\ \emph {et~al.}(2012)\citenamefont
  {Mohemmed}, \citenamefont {Schliebs}, \citenamefont {Matsuda},\ and\
  \citenamefont {Kasabov}}]{MSM+12}%
  \BibitemOpen
  \bibfield  {author} {\bibinfo {author} {\bibfnamefont {A.}~\bibnamefont
  {Mohemmed}}, \bibinfo {author} {\bibfnamefont {S.}~\bibnamefont {Schliebs}},
  \bibinfo {author} {\bibfnamefont {S.}~\bibnamefont {Matsuda}}, \ and\
  \bibinfo {author} {\bibfnamefont {N.}~\bibnamefont {Kasabov}},\ }\bibfield
  {title} {\emph {\bibinfo {title} {Span: spike pattern association neuron for
  learning spatio-temporal spike patterns.}\ }}\href {\doibase
  10.1142/S0129065712500128} {\bibfield  {journal} {\bibinfo  {journal} {Int.
  J. Neural Syst.}\ }\textbf {\bibinfo {volume} {22}},\ \bibinfo {pages}
  {1250012} (\bibinfo {year} {2012})}\BibitemShut {NoStop}%
\bibitem [{\citenamefont {Xu}\ \emph {et~al.}(2013)\citenamefont {Xu},
  \citenamefont {Zeng},\ and\ \citenamefont {Zhong}}]{XZZ13}%
  \BibitemOpen
  \bibfield  {author} {\bibinfo {author} {\bibfnamefont {Y.}~\bibnamefont
  {Xu}}, \bibinfo {author} {\bibfnamefont {X.}~\bibnamefont {Zeng}}, \ and\
  \bibinfo {author} {\bibfnamefont {S.}~\bibnamefont {Zhong}},\ }\bibfield
  {title} {\emph {\bibinfo {title} {A new supervised learning algorithm for
  spiking neurons},\ }}\href@noop {} {\bibfield  {journal} {\bibinfo  {journal}
  {Neural Comput.}\ }\textbf {\bibinfo {volume} {25}},\ \bibinfo {pages} {1}
  (\bibinfo {year} {2013})}\BibitemShut {NoStop}%
\bibitem [{\citenamefont {Memmesheimer}\ \emph {et~al.}(2014)\citenamefont
  {Memmesheimer}, \citenamefont {Rubin}, \citenamefont {{\"O}lveczky},\ and\
  \citenamefont {Sompolinsky}}]{MRS13}%
  \BibitemOpen
  \bibfield  {author} {\bibinfo {author} {\bibfnamefont {R.-M.}\ \bibnamefont
  {Memmesheimer}}, \bibinfo {author} {\bibfnamefont {R.}~\bibnamefont {Rubin}},
  \bibinfo {author} {\bibfnamefont {B.}~\bibnamefont {{\"O}lveczky}}, \ and\
  \bibinfo {author} {\bibfnamefont {H.}~\bibnamefont {Sompolinsky}},\
  }\bibfield  {title} {\emph {\bibinfo {title} {Learning precisely timed
  spikes},\ }}\href@noop {} {\bibfield  {journal} {\bibinfo  {journal}
  {Neuron}\ }\textbf {\bibinfo {volume} {82}},\ \bibinfo {pages} {011053}
  (\bibinfo {year} {2014})}\BibitemShut {NoStop}%
\bibitem [{\citenamefont {Albers}\ \emph {et~al.}(2016)\citenamefont {Albers},
  \citenamefont {Westkott},\ and\ \citenamefont {Pawelzik}}]{albers16learning}%
  \BibitemOpen
  \bibfield  {author} {\bibinfo {author} {\bibfnamefont {C.}~\bibnamefont
  {Albers}}, \bibinfo {author} {\bibfnamefont {M.}~\bibnamefont {Westkott}}, \
  and\ \bibinfo {author} {\bibfnamefont {K.}~\bibnamefont {Pawelzik}},\
  }\bibfield  {title} {\emph {\bibinfo {title} {Learning of precise spike times
  with homeostatic membrane potential dependent synaptic plasticity},\ }}\href
  {\doibase 10.1371/journal.pone.0148948} {\bibfield  {journal} {\bibinfo
  {journal} {PLOS One}\ }\textbf {\bibinfo {volume} {11}},\ \bibinfo {pages}
  {1} (\bibinfo {year} {2016})}\BibitemShut {NoStop}%
\bibitem [{\citenamefont {Zenke}\ and\ \citenamefont
  {Ganguli}(2018)}]{zenke2018superspike}%
  \BibitemOpen
  \bibfield  {author} {\bibinfo {author} {\bibfnamefont {F.}~\bibnamefont
  {Zenke}}\ and\ \bibinfo {author} {\bibfnamefont {S.}~\bibnamefont
  {Ganguli}},\ }\bibfield  {title} {\emph {\bibinfo {title} {{SuperSpike}:
  Supervised learning in multilayer spiking neural networks},\ }}\href@noop {}
  {\bibfield  {journal} {\bibinfo  {journal} {Neural Comput.}\ }\textbf
  {\bibinfo {volume} {30}},\ \bibinfo {pages} {1514} (\bibinfo {year}
  {2018})}\BibitemShut {NoStop}%
\bibitem [{\citenamefont {Huh}\ and\ \citenamefont
  {Sejnowski}(2018)}]{huh18gradient}%
  \BibitemOpen
  \bibfield  {author} {\bibinfo {author} {\bibfnamefont {D.}~\bibnamefont
  {Huh}}\ and\ \bibinfo {author} {\bibfnamefont {T.~J.}\ \bibnamefont
  {Sejnowski}},\ }in\ \href@noop {} {\emph {\bibinfo {booktitle} {Advances in
  Neural Information Processing Systems 31}}},\ \bibinfo {editor} {edited by\
  \bibinfo {editor} {\bibfnamefont {S.}~\bibnamefont {Bengio}}, \bibinfo
  {editor} {\bibfnamefont {H.}~\bibnamefont {Wallach}}, \bibinfo {editor}
  {\bibfnamefont {H.}~\bibnamefont {Larochelle}}, \bibinfo {editor}
  {\bibfnamefont {K.}~\bibnamefont {Grauman}}, \bibinfo {editor} {\bibfnamefont
  {N.}~\bibnamefont {Cesa-Bianchi}}, \ and\ \bibinfo {editor} {\bibfnamefont
  {R.}~\bibnamefont {Garnett}}}\ (\bibinfo  {publisher} {Curran Associates,
  Inc.},\ \bibinfo {year} {2018})\ pp.\ \bibinfo {pages}
  {1439--1449}\BibitemShut {NoStop}%
\bibitem [{\citenamefont {Olmi}\ \emph {et~al.}(2017)\citenamefont {Olmi},
  \citenamefont {Angulo-Garcia}, \citenamefont {Imparato},\ and\ \citenamefont
  {Torcini}}]{olmi2017exact}%
  \BibitemOpen
  \bibfield  {author} {\bibinfo {author} {\bibfnamefont {S.}~\bibnamefont
  {Olmi}}, \bibinfo {author} {\bibfnamefont {D.}~\bibnamefont {Angulo-Garcia}},
  \bibinfo {author} {\bibfnamefont {A.}~\bibnamefont {Imparato}}, \ and\
  \bibinfo {author} {\bibfnamefont {A.}~\bibnamefont {Torcini}},\ }\bibfield
  {title} {\emph {\bibinfo {title} {Exact firing time statistics of neurons
  driven by discrete inhibitory noise.}\ }}\href {\doibase
  10.1038/s41598-017-01658-8} {\bibfield  {journal} {\bibinfo  {journal} {Sci.
  Rep.}\ }\textbf {\bibinfo {volume} {7}},\ \bibinfo {pages} {1577} (\bibinfo
  {year} {2017})}\BibitemShut {NoStop}%
\bibitem [{\citenamefont {Mirollo}\ and\ \citenamefont
  {Strogatz}(1990)}]{MS90}%
  \BibitemOpen
  \bibfield  {author} {\bibinfo {author} {\bibfnamefont {R.}~\bibnamefont
  {Mirollo}}\ and\ \bibinfo {author} {\bibfnamefont {S.}~\bibnamefont
  {Strogatz}},\ }\bibfield  {title} {\emph {\bibinfo {title} {Synchronization
  of pulse coupled biological oscillators},\ }}\href@noop {} {\bibfield
  {journal} {\bibinfo  {journal} {SIAM J. Appl. Math.}\ }\textbf {\bibinfo
  {volume} {50}},\ \bibinfo {pages} {1645} (\bibinfo {year}
  {1990})}\BibitemShut {NoStop}%
\bibitem [{\citenamefont {Memmesheimer}\ and\ \citenamefont
  {Timme}(2006)}]{MT06B}%
  \BibitemOpen
  \bibfield  {author} {\bibinfo {author} {\bibfnamefont {R.-M.}\ \bibnamefont
  {Memmesheimer}}\ and\ \bibinfo {author} {\bibfnamefont {M.}~\bibnamefont
  {Timme}},\ }\bibfield  {title} {\emph {\bibinfo {title} {Designing complex
  networks},\ }}\href@noop {} {\bibfield  {journal} {\bibinfo  {journal} {Phys.
  D}\ }\textbf {\bibinfo {volume} {224}},\ \bibinfo {pages} {182} (\bibinfo
  {year} {2006})}\BibitemShut {NoStop}%
\bibitem [{\citenamefont {Goel}\ and\ \citenamefont
  {Ermentrout}(2002)}]{goel2002synchrony}%
  \BibitemOpen
  \bibfield  {author} {\bibinfo {author} {\bibfnamefont {P.}~\bibnamefont
  {Goel}}\ and\ \bibinfo {author} {\bibfnamefont {B.}~\bibnamefont
  {Ermentrout}},\ }\bibfield  {title} {\emph {\bibinfo {title} {Synchrony,
  stability, and firing patterns in pulse-coupled oscillators},\ }}\href
  {\doibase 10.1016/s0167-2789(01)00374-8} {\bibfield  {journal} {\bibinfo
  {journal} {Phys. D}\ }\textbf {\bibinfo {volume} {163}},\ \bibinfo {pages}
  {191} (\bibinfo {year} {2002})}\BibitemShut {NoStop}%
\bibitem [{\citenamefont {Smeal}\ \emph {et~al.}(2010)\citenamefont {Smeal},
  \citenamefont {Ermentrout},\ and\ \citenamefont {White}}]{smeal2010phase}%
  \BibitemOpen
  \bibfield  {author} {\bibinfo {author} {\bibfnamefont {R.~M.}\ \bibnamefont
  {Smeal}}, \bibinfo {author} {\bibfnamefont {G.~B.}\ \bibnamefont
  {Ermentrout}}, \ and\ \bibinfo {author} {\bibfnamefont {J.~A.}\ \bibnamefont
  {White}},\ }\bibfield  {title} {\emph {\bibinfo {title} {Phase-response
  curves and synchronized neural networks.}\ }}\href {\doibase
  10.1098/rstb.2009.0292} {\bibfield  {journal} {\bibinfo  {journal} {Philos.
  Trans. R. Soc. Lond. B Biol. Sci.}\ }\textbf {\bibinfo {volume} {365}},\
  \bibinfo {pages} {2407} (\bibinfo {year} {2010})}\BibitemShut {NoStop}%
\bibitem [{\citenamefont {Viriyopase}\ \emph {et~al.}(2018)\citenamefont
  {Viriyopase}, \citenamefont {Memmesheimer},\ and\ \citenamefont
  {Gielen}}]{viriyopase18analyzing}%
  \BibitemOpen
  \bibfield  {author} {\bibinfo {author} {\bibfnamefont {A.}~\bibnamefont
  {Viriyopase}}, \bibinfo {author} {\bibfnamefont {R.-M.}\ \bibnamefont
  {Memmesheimer}}, \ and\ \bibinfo {author} {\bibfnamefont {S.}~\bibnamefont
  {Gielen}},\ }\bibfield  {title} {\emph {\bibinfo {title} {Analyzing the
  competition of gamma rhythms with delayed pulse-coupled oscillators in phase
  representation},\ }}\href {\doibase 10.1103/PhysRevE.98.022217} {\bibfield
  {journal} {\bibinfo  {journal} {Phys. Rev. E}\ }\textbf {\bibinfo {volume}
  {98}},\ \bibinfo {pages} {022217} (\bibinfo {year} {2018})}\BibitemShut
  {NoStop}%
\bibitem [{\citenamefont {van Vreeswijk}\ and\ \citenamefont
  {Sompolinsky}(1998)}]{Vreeswijk1998}%
  \BibitemOpen
  \bibfield  {author} {\bibinfo {author} {\bibfnamefont {C.}~\bibnamefont {van
  Vreeswijk}}\ and\ \bibinfo {author} {\bibfnamefont {H.}~\bibnamefont
  {Sompolinsky}},\ }\bibfield  {title} {\emph {\bibinfo {title} {Chaotic
  balanced state in a model of cortical circuits},\ }}\href@noop {} {\bibfield
  {journal} {\bibinfo  {journal} {Neural Comput.}\ }\textbf {\bibinfo {volume}
  {10}},\ \bibinfo {pages} {1321} (\bibinfo {year} {1998})}\BibitemShut
  {NoStop}%
\bibitem [{\citenamefont {Kadanoff}(2009)}]{kadanoff2009more}%
  \BibitemOpen
  \bibfield  {author} {\bibinfo {author} {\bibfnamefont {L.~P.}\ \bibnamefont
  {Kadanoff}},\ }\bibfield  {title} {\emph {\bibinfo {title} {More is the same;
  phase transitions and mean field theories},\ }}\href {\doibase
  10.1007/s10955-009-9814-1} {\bibfield  {journal} {\bibinfo  {journal} {J.
  Stat. Phys.}\ }\textbf {\bibinfo {volume} {137}},\ \bibinfo {pages} {777}
  (\bibinfo {year} {2009})}\BibitemShut {NoStop}%
\bibitem [{\citenamefont {Brunel}(2000)}]{Brunel00}%
  \BibitemOpen
  \bibfield  {author} {\bibinfo {author} {\bibfnamefont {N.}~\bibnamefont
  {Brunel}},\ }\bibfield  {title} {\emph {\bibinfo {title} {Dynamics of
  sparsely connected networks of excitatory and inhibitory spiking neurons},\
  }}\href@noop {} {\bibfield  {journal} {\bibinfo  {journal} {J. Comput.
  Neurosci.}\ }\textbf {\bibinfo {volume} {8}},\ \bibinfo {pages} {183}
  (\bibinfo {year} {2000})}\BibitemShut {NoStop}%
\bibitem [{\citenamefont {Breuer}\ \emph {et~al.}(2014)\citenamefont {Breuer},
  \citenamefont {Timme},\ and\ \citenamefont {Memmesheimer}}]{BTM13}%
  \BibitemOpen
  \bibfield  {author} {\bibinfo {author} {\bibfnamefont {D.}~\bibnamefont
  {Breuer}}, \bibinfo {author} {\bibfnamefont {M.}~\bibnamefont {Timme}}, \
  and\ \bibinfo {author} {\bibfnamefont {R.-M.}\ \bibnamefont {Memmesheimer}},\
  }\bibfield  {title} {\emph {\bibinfo {title} {Statistical physics of neural
  systems with non-additive dendritic coupling},\ }}\href@noop {} {\bibfield
  {journal} {\bibinfo  {journal} {Phys. Rev. X}\ }\textbf {\bibinfo {volume}
  {4}},\ \bibinfo {pages} {011053} (\bibinfo {year} {2014})}\BibitemShut
  {NoStop}%
\bibitem [{\citenamefont {Schuecker}\ \emph {et~al.}(2018)\citenamefont
  {Schuecker}, \citenamefont {Goedeke},\ and\ \citenamefont
  {Helias}}]{schuecker18optimal}%
  \BibitemOpen
  \bibfield  {author} {\bibinfo {author} {\bibfnamefont {J.}~\bibnamefont
  {Schuecker}}, \bibinfo {author} {\bibfnamefont {S.}~\bibnamefont {Goedeke}},
  \ and\ \bibinfo {author} {\bibfnamefont {M.}~\bibnamefont {Helias}},\
  }\bibfield  {title} {\emph {\bibinfo {title} {Optimal sequence memory in
  driven random networks},\ }}\href {\doibase 10.1103/PhysRevX.8.041029}
  {\bibfield  {journal} {\bibinfo  {journal} {Phys. Rev. X}\ }\textbf {\bibinfo
  {volume} {8}},\ \bibinfo {pages} {041029} (\bibinfo {year}
  {2018})}\BibitemShut {NoStop}%
\bibitem [{\citenamefont {Tuckwell}(1988)}]{T88b}%
  \BibitemOpen
  \bibfield  {author} {\bibinfo {author} {\bibfnamefont {H.}~\bibnamefont
  {Tuckwell}},\ }\href@noop {} {\emph {\bibinfo {title} {Introduction to
  theoretical neurobiology: Volume 2. {Nonlinear} and stochastic theories}}}\
  (\bibinfo  {publisher} {Cambridge Univ. Press},\ \bibinfo {address}
  {Cambridge},\ \bibinfo {year} {1988})\BibitemShut {NoStop}%
\bibitem [{\citenamefont {Burkitt}(2006)}]{B06a}%
  \BibitemOpen
  \bibfield  {author} {\bibinfo {author} {\bibfnamefont {A.}~\bibnamefont
  {Burkitt}},\ }\bibfield  {title} {\emph {\bibinfo {title} {A review of the
  integrate-and-fire neuron model: {I.} {Homogeneous} synaptic input},\
  }}\href@noop {} {\bibfield  {journal} {\bibinfo  {journal} {Biol. Cybern.}\
  }\textbf {\bibinfo {volume} {95}},\ \bibinfo {pages} {1} (\bibinfo {year}
  {2006})}\BibitemShut {NoStop}%
\bibitem [{\citenamefont {Richardson}\ and\ \citenamefont
  {Swarbrick}(2010)}]{richardson}%
  \BibitemOpen
  \bibfield  {author} {\bibinfo {author} {\bibfnamefont {M.~J.~E.}\
  \bibnamefont {Richardson}}\ and\ \bibinfo {author} {\bibfnamefont
  {R.}~\bibnamefont {Swarbrick}},\ }\bibfield  {title} {\emph {\bibinfo {title}
  {Firing-rate response of a neuron receiving excitatory and inhibitory
  synaptic shot noise},\ }}\href {\doibase 10.1103/PhysRevLett.105.178102}
  {\bibfield  {journal} {\bibinfo  {journal} {Phys. Rev. Lett.}\ }\textbf
  {\bibinfo {volume} {105}},\ \bibinfo {pages} {178102} (\bibinfo {year}
  {2010})}\BibitemShut {NoStop}%
\bibitem [{\citenamefont {Angulo-Garcia}\ \emph {et~al.}(2017)\citenamefont
  {Angulo-Garcia}, \citenamefont {Luccioli}, \citenamefont {Olmi},\ and\
  \citenamefont {Torcini}}]{angulo-garcia}%
  \BibitemOpen
  \bibfield  {author} {\bibinfo {author} {\bibfnamefont {D.}~\bibnamefont
  {Angulo-Garcia}}, \bibinfo {author} {\bibfnamefont {S.}~\bibnamefont
  {Luccioli}}, \bibinfo {author} {\bibfnamefont {S.}~\bibnamefont {Olmi}}, \
  and\ \bibinfo {author} {\bibfnamefont {A.}~\bibnamefont {Torcini}},\
  }\bibfield  {title} {\emph {\bibinfo {title} {Death and rebirth of neural
  activity in sparse inhibitory networks},\ }}\href@noop {} {\bibfield
  {journal} {\bibinfo  {journal} {New J. Phys.}\ }\textbf {\bibinfo {volume}
  {19}},\ \bibinfo {pages} {053011} (\bibinfo {year} {2017})}\BibitemShut
  {NoStop}%
\bibitem [{\citenamefont {Hale}\ and\ \citenamefont
  {Verduyn~Lunel}(1993)}]{hale1993introduction}%
  \BibitemOpen
  \bibfield  {author} {\bibinfo {author} {\bibfnamefont {J.~K.}\ \bibnamefont
  {Hale}}\ and\ \bibinfo {author} {\bibfnamefont {S.~M.}\ \bibnamefont
  {Verduyn~Lunel}},\ }\href {\doibase 10.1007/978-1-4612-4342-7} {\emph
  {\bibinfo {title} {Introduction to functional-differential equations}}},\
  \bibinfo {series} {Applied Mathematical Sciences}, Vol.~\bibinfo {volume}
  {99}\ (\bibinfo  {publisher} {Springer-Verlag, New York},\ \bibinfo {year}
  {1993})\BibitemShut {NoStop}%
\bibitem [{\citenamefont {Engelken}(2017)}]{engelken2017chaotic}%
  \BibitemOpen
  \bibfield  {author} {\bibinfo {author} {\bibfnamefont {R.}~\bibnamefont
  {Engelken}},\ }\emph {\bibinfo {title} {Chaotic Neural Circuit Dynamics}},\
  \href@noop {} {Ph.D. thesis},\ \bibinfo  {school} {University of
  G\"{o}ttingen} (\bibinfo {year} {2017})\BibitemShut {NoStop}%
\bibitem [{\citenamefont {{Ginelli}}\ \emph {et~al.}(2007)\citenamefont
  {{Ginelli}}, \citenamefont {{Poggi}}, \citenamefont {{Turchi}}, \citenamefont
  {{Chat{\'e}}}, \citenamefont {{Livi}},\ and\ \citenamefont
  {{Politi}}}]{ginelli}%
  \BibitemOpen
  \bibfield  {author} {\bibinfo {author} {\bibfnamefont {F.}~\bibnamefont
  {{Ginelli}}}, \bibinfo {author} {\bibfnamefont {P.}~\bibnamefont {{Poggi}}},
  \bibinfo {author} {\bibfnamefont {A.}~\bibnamefont {{Turchi}}}, \bibinfo
  {author} {\bibfnamefont {H.}~\bibnamefont {{Chat{\'e}}}}, \bibinfo {author}
  {\bibfnamefont {R.}~\bibnamefont {{Livi}}}, \ and\ \bibinfo {author}
  {\bibfnamefont {A.}~\bibnamefont {{Politi}}},\ }\bibfield  {title} {\emph
  {\bibinfo {title} {{Characterizing Dynamics with Covariant Lyapunov
  Vectors}},\ }}\href {\doibase 10.1103/PhysRevLett.99.130601} {\bibfield
  {journal} {\bibinfo  {journal} {Phys. Rev. Lett.}\ }\textbf {\bibinfo
  {volume} {99}},\ \bibinfo {eid} {130601} (\bibinfo {year}
  {2007})}\BibitemShut {NoStop}%
\bibitem [{\citenamefont {{Kramer}}\ and\ \citenamefont
  {{MacKinnon}}(1993)}]{kramer1993localization}%
  \BibitemOpen
  \bibfield  {author} {\bibinfo {author} {\bibfnamefont {B.}~\bibnamefont
  {{Kramer}}}\ and\ \bibinfo {author} {\bibfnamefont {A.}~\bibnamefont
  {{MacKinnon}}},\ }\bibfield  {title} {\emph {\bibinfo {title} {{Localization:
  theory and experiment}},\ }}\href {\doibase 10.1088/0034-4885/56/12/001}
  {\bibfield  {journal} {\bibinfo  {journal} {Rep. Prog. Phys.}\ }\textbf
  {\bibinfo {volume} {56}},\ \bibinfo {pages} {1469} (\bibinfo {year}
  {1993})}\BibitemShut {NoStop}%
\bibitem [{\citenamefont {Maren}\ \emph {et~al.}(2013)\citenamefont {Maren},
  \citenamefont {Phan},\ and\ \citenamefont {Liberzon}}]{maren2013contextual}%
  \BibitemOpen
  \bibfield  {author} {\bibinfo {author} {\bibfnamefont {S.}~\bibnamefont
  {Maren}}, \bibinfo {author} {\bibfnamefont {K.~L.}\ \bibnamefont {Phan}}, \
  and\ \bibinfo {author} {\bibfnamefont {I.}~\bibnamefont {Liberzon}},\
  }\bibfield  {title} {\emph {\bibinfo {title} {The contextual brain:
  implications for fear conditioning, extinction and psychopathology.}\ }}\href
  {\doibase 10.1038/nrn3492} {\bibfield  {journal} {\bibinfo  {journal} {Nat.
  Rev. Neurosci.}\ }\textbf {\bibinfo {volume} {14}},\ \bibinfo {pages} {417}
  (\bibinfo {year} {2013})}\BibitemShut {NoStop}%
\bibitem [{\citenamefont {Mante}\ \emph {et~al.}(2013)\citenamefont {Mante},
  \citenamefont {Sussillo}, \citenamefont {Shenoy},\ and\ \citenamefont
  {Newsome}}]{mante2013context}%
  \BibitemOpen
  \bibfield  {author} {\bibinfo {author} {\bibfnamefont {V.}~\bibnamefont
  {Mante}}, \bibinfo {author} {\bibfnamefont {D.}~\bibnamefont {Sussillo}},
  \bibinfo {author} {\bibfnamefont {K.~V.}\ \bibnamefont {Shenoy}}, \ and\
  \bibinfo {author} {\bibfnamefont {W.~T.}\ \bibnamefont {Newsome}},\
  }\bibfield  {title} {\emph {\bibinfo {title} {Context-dependent computation
  by recurrent dynamics in prefrontal cortex.}\ }}\href {\doibase
  10.1038/nature12742} {\bibfield  {journal} {\bibinfo  {journal} {Nature
  (London)}\ }\textbf {\bibinfo {volume} {503}},\ \bibinfo {pages} {78}
  (\bibinfo {year} {2013})}\BibitemShut {NoStop}%
\bibitem [{\citenamefont {Kriener}\ \emph {et~al.}(2008)\citenamefont
  {Kriener}, \citenamefont {Tetzlaff}, \citenamefont {Aertsen}, \citenamefont
  {Diesmann},\ and\ \citenamefont {Rotter}}]{KTADR08}%
  \BibitemOpen
  \bibfield  {author} {\bibinfo {author} {\bibfnamefont {B.}~\bibnamefont
  {Kriener}}, \bibinfo {author} {\bibfnamefont {T.}~\bibnamefont {Tetzlaff}},
  \bibinfo {author} {\bibfnamefont {A.}~\bibnamefont {Aertsen}}, \bibinfo
  {author} {\bibfnamefont {M.}~\bibnamefont {Diesmann}}, \ and\ \bibinfo
  {author} {\bibfnamefont {S.}~\bibnamefont {Rotter}},\ }\bibfield  {title}
  {\emph {\bibinfo {title} {Correlations and population dynamics in cortical
  networks},\ }}\href@noop {} {\bibfield  {journal} {\bibinfo  {journal}
  {Neural Comput.}\ }\textbf {\bibinfo {volume} {20}},\ \bibinfo {pages} {2185}
  (\bibinfo {year} {2008})}\BibitemShut {NoStop}%
\bibitem [{\citenamefont {Prinz}(2006)}]{Prinz06}%
  \BibitemOpen
  \bibfield  {author} {\bibinfo {author} {\bibfnamefont {A.}~\bibnamefont
  {Prinz}},\ }\bibfield  {title} {\emph {\bibinfo {title} {Insights from models
  of rhythmic motor systems},\ }}\href@noop {} {\bibfield  {journal} {\bibinfo
  {journal} {Curr. Opin. Neurobiol.}\ }\textbf {\bibinfo {volume} {16}},\
  \bibinfo {pages} {615} (\bibinfo {year} {2006})}\BibitemShut {NoStop}%
\bibitem [{\citenamefont {Savin}\ \emph {et~al.}(2006)\citenamefont {Savin},
  \citenamefont {Ignat},\ and\ \citenamefont {Muresan}}]{savin06heterogeneous}%
  \BibitemOpen
  \bibfield  {author} {\bibinfo {author} {\bibfnamefont {C.}~\bibnamefont
  {Savin}}, \bibinfo {author} {\bibfnamefont {I.}~\bibnamefont {Ignat}}, \ and\
  \bibinfo {author} {\bibfnamefont {R.}~\bibnamefont {Muresan}},\ }in\
  \href@noop {} {\emph {\bibinfo {booktitle} {IEEE 2nd International Conference
  on Intelligent Computer Communication and Processing (ICCP)}}}\ (\bibinfo
  {year} {2006})\BibitemShut {NoStop}%
\bibitem [{\citenamefont {Wang}\ \emph {et~al.}(2004)\citenamefont {Wang},
  \citenamefont {Tegn{\'e}r}, \citenamefont {Constantinidis},\ and\
  \citenamefont {Goldman-Rakic}}]{Wang04ineurontypes}%
  \BibitemOpen
  \bibfield  {author} {\bibinfo {author} {\bibfnamefont {X.-J.}\ \bibnamefont
  {Wang}}, \bibinfo {author} {\bibfnamefont {J.}~\bibnamefont {Tegn{\'e}r}},
  \bibinfo {author} {\bibfnamefont {C.}~\bibnamefont {Constantinidis}}, \ and\
  \bibinfo {author} {\bibfnamefont {P.~S.}\ \bibnamefont {Goldman-Rakic}},\
  }\bibfield  {title} {\emph {\bibinfo {title} {Division of labor among
  distinct subtypes of inhibitory neurons in a cortical microcircuit of working
  memory},\ }}\href {\doibase 10.1073/pnas.0305337101} {\bibfield  {journal}
  {\bibinfo  {journal} {Proc. Natl. Acad. Sci. U.S.A.}\ }\textbf {\bibinfo
  {volume} {101}},\ \bibinfo {pages} {1368} (\bibinfo {year}
  {2004})}\BibitemShut {NoStop}%
\bibitem [{\citenamefont {Litwin-Kumar}\ \emph {et~al.}(2016)\citenamefont
  {Litwin-Kumar}, \citenamefont {Rosenbaum},\ and\ \citenamefont
  {Doiron}}]{litwinkumar16ineurontypes}%
  \BibitemOpen
  \bibfield  {author} {\bibinfo {author} {\bibfnamefont {A.}~\bibnamefont
  {Litwin-Kumar}}, \bibinfo {author} {\bibfnamefont {R.}~\bibnamefont
  {Rosenbaum}}, \ and\ \bibinfo {author} {\bibfnamefont {B.}~\bibnamefont
  {Doiron}},\ }\bibfield  {title} {\emph {\bibinfo {title} {Inhibitory
  stabilization and visual coding in cortical circuits with multiple
  interneuron subtypes},\ }}\href {\doibase 10.1152/jn.00732.2015} {\bibfield
  {journal} {\bibinfo  {journal} {J Neurophysiol.}\ }\textbf {\bibinfo {volume}
  {115}},\ \bibinfo {pages} {1399} (\bibinfo {year} {2016})}\BibitemShut
  {NoStop}%
\bibitem [{\citenamefont {Memmesheimer}\ and\ \citenamefont
  {Timme}(2010)}]{MT10A}%
  \BibitemOpen
  \bibfield  {author} {\bibinfo {author} {\bibfnamefont {R.-M.}\ \bibnamefont
  {Memmesheimer}}\ and\ \bibinfo {author} {\bibfnamefont {M.}~\bibnamefont
  {Timme}},\ }\bibfield  {title} {\emph {\bibinfo {title} {Stable and unstable
  periodic orbits in complex networks of spiking neurons with delays},\
  }}\href@noop {} {\bibfield  {journal} {\bibinfo  {journal} {Discr. Cont. Dyn.
  Sys.}\ }\textbf {\bibinfo {volume} {28}},\ \bibinfo {pages} {1555} (\bibinfo
  {year} {2010})}\BibitemShut {NoStop}%
\bibitem [{\citenamefont {Gu}\ \emph {et~al.}(2018)\citenamefont {Gu},
  \citenamefont {Tian}, \citenamefont {Kova\u{c}i\u{c}}, \citenamefont {Zhou},\
  and\ \citenamefont {Cai}}]{gu2018dynamics}%
  \BibitemOpen
  \bibfield  {author} {\bibinfo {author} {\bibfnamefont {Q.~L.}\ \bibnamefont
  {Gu}}, \bibinfo {author} {\bibfnamefont {Z.~K.}\ \bibnamefont {Tian}},
  \bibinfo {author} {\bibfnamefont {G.}~\bibnamefont {Kova\u{c}i\u{c}}},
  \bibinfo {author} {\bibfnamefont {D.}~\bibnamefont {Zhou}}, \ and\ \bibinfo
  {author} {\bibfnamefont {D.}~\bibnamefont {Cai}},\ }\bibfield  {title} {\emph
  {\bibinfo {title} {The dynamics of balanced spiking neuronal networks under
  poisson drive is not chaotic},\ }}\href {\doibase 10.3389/fncom.2018.00047}
  {\bibfield  {journal} {\bibinfo  {journal} {Front. Comput. Neurosci.}\
  }\textbf {\bibinfo {volume} {12}} (\bibinfo {year} {2018}),\
  10.3389/fncom.2018.00047}\BibitemShut {NoStop}%
\bibitem [{\citenamefont {Luccioli}\ \emph {et~al.}(2012)\citenamefont
  {Luccioli}, \citenamefont {Olmi}, \citenamefont {Politi},\ and\ \citenamefont
  {Torcini}}]{luccioli12collective}%
  \BibitemOpen
  \bibfield  {author} {\bibinfo {author} {\bibfnamefont {S.}~\bibnamefont
  {Luccioli}}, \bibinfo {author} {\bibfnamefont {S.}~\bibnamefont {Olmi}},
  \bibinfo {author} {\bibfnamefont {A.}~\bibnamefont {Politi}}, \ and\ \bibinfo
  {author} {\bibfnamefont {A.}~\bibnamefont {Torcini}},\ }\bibfield  {title}
  {\emph {\bibinfo {title} {Collective dynamics in sparse networks},\ }}\href
  {\doibase 10.1103/PhysRevLett.109.138103} {\bibfield  {journal} {\bibinfo
  {journal} {Phys. Rev. Lett.}\ }\textbf {\bibinfo {volume} {109}},\ \bibinfo
  {pages} {138103} (\bibinfo {year} {2012})}\BibitemShut {NoStop}%
\bibitem [{\citenamefont {Lajoie}\ \emph {et~al.}(2014)\citenamefont {Lajoie},
  \citenamefont {Thivierge},\ and\ \citenamefont
  {Shea-Brown}}]{lajoie2014structured}%
  \BibitemOpen
  \bibfield  {author} {\bibinfo {author} {\bibfnamefont {G.}~\bibnamefont
  {Lajoie}}, \bibinfo {author} {\bibfnamefont {J.-P.}\ \bibnamefont
  {Thivierge}}, \ and\ \bibinfo {author} {\bibfnamefont {E.}~\bibnamefont
  {Shea-Brown}},\ }\bibfield  {title} {\emph {\bibinfo {title} {Structured
  chaos shapes spike-response noise entropy in balanced neural networks.}\
  }}\href {\doibase 10.3389/fncom.2014.00123} {\bibfield  {journal} {\bibinfo
  {journal} {Front. Comput. Neurosci.}\ }\textbf {\bibinfo {volume} {8}},\
  \bibinfo {pages} {123} (\bibinfo {year} {2014})}\BibitemShut {NoStop}%
\bibitem [{\citenamefont {Ullner}\ and\ \citenamefont
  {Politi}(2016)}]{ullner16selfsustained}%
  \BibitemOpen
  \bibfield  {author} {\bibinfo {author} {\bibfnamefont {E.}~\bibnamefont
  {Ullner}}\ and\ \bibinfo {author} {\bibfnamefont {A.}~\bibnamefont
  {Politi}},\ }\bibfield  {title} {\emph {\bibinfo {title} {Self-sustained
  irregular activity in an ensemble of neural oscillators},\ }}\href {\doibase
  10.1103/PhysRevX.6.011015} {\bibfield  {journal} {\bibinfo  {journal} {Phys.
  Rev. X}\ }\textbf {\bibinfo {volume} {6}},\ \bibinfo {pages} {011015}
  (\bibinfo {year} {2016})}\BibitemShut {NoStop}%
\bibitem [{\citenamefont {Coombes}(2000)}]{coombes2000chaos}%
  \BibitemOpen
  \bibfield  {author} {\bibinfo {author} {\bibfnamefont {S.}~\bibnamefont
  {Coombes}},\ }in\ \href {\doibase 10.1063/1.1302370} {\emph {\bibinfo
  {booktitle} {{AIP} Conference Proceedings}}}\ (\bibinfo  {publisher}
  {{AIP}},\ \bibinfo {year} {2000})\BibitemShut {NoStop}%
\bibitem [{\citenamefont {Lagorce}\ and\ \citenamefont
  {Benosman}(2015)}]{lagorce2015stick}%
  \BibitemOpen
  \bibfield  {author} {\bibinfo {author} {\bibfnamefont {X.}~\bibnamefont
  {Lagorce}}\ and\ \bibinfo {author} {\bibfnamefont {R.}~\bibnamefont
  {Benosman}},\ }\bibfield  {title} {\emph {\bibinfo {title} {{STICK}: Spike
  time interval computational kernel, a framework for general purpose
  computation using neurons, precise timing, delays, and synchrony},\ }}\href
  {\doibase 10.1162/neco_a_00783} {\bibfield  {journal} {\bibinfo  {journal}
  {Neural Comput.}\ }\textbf {\bibinfo {volume} {27}},\ \bibinfo {pages} {2261}
  (\bibinfo {year} {2015})}\BibitemShut {NoStop}%
\bibitem [{\citenamefont {Verzi}\ \emph {et~al.}(2018)\citenamefont {Verzi},
  \citenamefont {Rothganger}, \citenamefont {Parekh}, \citenamefont {Quach},
  \citenamefont {Miner}, \citenamefont {Vineyard}, \citenamefont {James},\ and\
  \citenamefont {Aimone}}]{verzi18computing}%
  \BibitemOpen
  \bibfield  {author} {\bibinfo {author} {\bibfnamefont {S.~J.}\ \bibnamefont
  {Verzi}}, \bibinfo {author} {\bibfnamefont {F.}~\bibnamefont {Rothganger}},
  \bibinfo {author} {\bibfnamefont {O.~D.}\ \bibnamefont {Parekh}}, \bibinfo
  {author} {\bibfnamefont {T.-T.}\ \bibnamefont {Quach}}, \bibinfo {author}
  {\bibfnamefont {N.~E.}\ \bibnamefont {Miner}}, \bibinfo {author}
  {\bibfnamefont {C.~M.}\ \bibnamefont {Vineyard}}, \bibinfo {author}
  {\bibfnamefont {C.~D.}\ \bibnamefont {James}}, \ and\ \bibinfo {author}
  {\bibfnamefont {J.~B.}\ \bibnamefont {Aimone}},\ }\bibfield  {title} {\emph
  {\bibinfo {title} {Computing with spikes: The advantage of fine-grained
  timing},\ }}\href {\doibase 10.1162/neco\_a\_01113} {\bibfield  {journal}
  {\bibinfo  {journal} {Neural Comput.}\ }\textbf {\bibinfo {volume} {30}},\
  \bibinfo {pages} {2660} (\bibinfo {year} {2018})}\BibitemShut {NoStop}%
\bibitem [{\citenamefont {Pfeiffer}\ and\ \citenamefont
  {Pfeil}(2018)}]{pfeiffer2018deep}%
  \BibitemOpen
  \bibfield  {author} {\bibinfo {author} {\bibfnamefont {M.}~\bibnamefont
  {Pfeiffer}}\ and\ \bibinfo {author} {\bibfnamefont {T.}~\bibnamefont
  {Pfeil}},\ }\bibfield  {title} {\emph {\bibinfo {title} {Deep learning with
  spiking neurons: Opportunities and challenges.}\ }}\href {\doibase
  10.3389/fnins.2018.00774} {\bibfield  {journal} {\bibinfo  {journal} {Front.
  Neurosci.}\ }\textbf {\bibinfo {volume} {12}},\ \bibinfo {pages} {774}
  (\bibinfo {year} {2018})}\BibitemShut {NoStop}%
\bibitem [{\citenamefont {Thalmeier}\ \emph {et~al.}(2016)\citenamefont
  {Thalmeier}, \citenamefont {Uhlmann}, \citenamefont {Kappen},\ and\
  \citenamefont {Memmesheimer}}]{thalmeier2015universal}%
  \BibitemOpen
  \bibfield  {author} {\bibinfo {author} {\bibfnamefont {D.}~\bibnamefont
  {Thalmeier}}, \bibinfo {author} {\bibfnamefont {M.}~\bibnamefont {Uhlmann}},
  \bibinfo {author} {\bibfnamefont {H.~J.}\ \bibnamefont {Kappen}}, \ and\
  \bibinfo {author} {\bibfnamefont {R.-M.}\ \bibnamefont {Memmesheimer}},\
  }\bibfield  {title} {\emph {\bibinfo {title} {Learning universal computations
  with spikes},\ }}\href@noop {} {\bibfield  {journal} {\bibinfo  {journal}
  {PLOS Comput. Biol.}\ }\textbf {\bibinfo {volume} {12}},\ \bibinfo {pages}
  {e1004895} (\bibinfo {year} {2016})}\BibitemShut {NoStop}%
\bibitem [{\citenamefont {Abbott}\ \emph {et~al.}(2016)\citenamefont {Abbott},
  \citenamefont {DePasquale},\ and\ \citenamefont
  {Memmesheimer}}]{abbott2016building}%
  \BibitemOpen
  \bibfield  {author} {\bibinfo {author} {\bibfnamefont {L.}~\bibnamefont
  {Abbott}}, \bibinfo {author} {\bibfnamefont {B.}~\bibnamefont {DePasquale}},
  \ and\ \bibinfo {author} {\bibfnamefont {R.-M.}\ \bibnamefont
  {Memmesheimer}},\ }\bibfield  {title} {\emph {\bibinfo {title} {Building
  functional networks of spiking model neurons},\ }}\href@noop {} {\bibfield
  {journal} {\bibinfo  {journal} {Nat. Neurosci.}\ }\textbf {\bibinfo {volume}
  {19}},\ \bibinfo {pages} {350} (\bibinfo {year} {2016})}\BibitemShut
  {NoStop}%
\bibitem [{\citenamefont {Nicola}\ and\ \citenamefont
  {Clopath}(2017)}]{nicola2017supervised}%
  \BibitemOpen
  \bibfield  {author} {\bibinfo {author} {\bibfnamefont {W.}~\bibnamefont
  {Nicola}}\ and\ \bibinfo {author} {\bibfnamefont {C.}~\bibnamefont
  {Clopath}},\ }\bibfield  {title} {\emph {\bibinfo {title} {Supervised
  learning in spiking neural networks with force training.}\ }}\href {\doibase
  10.1038/s41467-017-01827-3} {\bibfield  {journal} {\bibinfo  {journal} {Nat.
  Commun.}\ }\textbf {\bibinfo {volume} {8}},\ \bibinfo {pages} {2208}
  (\bibinfo {year} {2017})}\BibitemShut {NoStop}%
\bibitem [{\citenamefont {DePasquale}\ \emph {et~al.}(2018)\citenamefont
  {DePasquale}, \citenamefont {Cueva}, \citenamefont {Rajan}, \citenamefont
  {Escola},\ and\ \citenamefont {Abbott}}]{depasquale2018full}%
  \BibitemOpen
  \bibfield  {author} {\bibinfo {author} {\bibfnamefont {B.}~\bibnamefont
  {DePasquale}}, \bibinfo {author} {\bibfnamefont {C.~J.}\ \bibnamefont
  {Cueva}}, \bibinfo {author} {\bibfnamefont {K.}~\bibnamefont {Rajan}},
  \bibinfo {author} {\bibfnamefont {G.~S.}\ \bibnamefont {Escola}}, \ and\
  \bibinfo {author} {\bibfnamefont {L.~F.}\ \bibnamefont {Abbott}},\ }\bibfield
   {title} {\emph {\bibinfo {title} {full-force: A target-based method for
  training recurrent networks.}\ }}\href {\doibase
  10.1371/journal.pone.0191527} {\bibfield  {journal} {\bibinfo  {journal}
  {PLOS One}\ }\textbf {\bibinfo {volume} {13}},\ \bibinfo {pages} {e0191527}
  (\bibinfo {year} {2018})}\BibitemShut {NoStop}%
\bibitem [{\citenamefont {Jaeger}(2001)}]{jaeger2001echo}%
  \BibitemOpen
  \bibfield  {author} {\bibinfo {author} {\bibfnamefont {H.}~\bibnamefont
  {Jaeger}},\ }\bibfield  {title} {\emph {\bibinfo {title} {The ``echo state''
  approach to analysing and training recurrent neural networks-with an erratum
  note},\ }}\href@noop {} {\bibfield  {journal} {\bibinfo  {journal} {Bonn,
  Germany: German National Research Center for Information Technology GMD
  Technical Report}\ }\textbf {\bibinfo {volume} {148}},\ \bibinfo {pages} {34}
  (\bibinfo {year} {2001})}\BibitemShut {NoStop}%
\end{thebibliography}
\providecommand{\noopsort}[1]{}\providecommand{\singleletter}[1]{#1}%

\section*{Appendices}

\makeatletter \def\p@subsection{} \makeatother

\subsection{Voltage probability distribution of LIF neurons\label{subsec:Voltage-probability-distribution}}

\begin{figure}
\includegraphics[width=1\columnwidth]{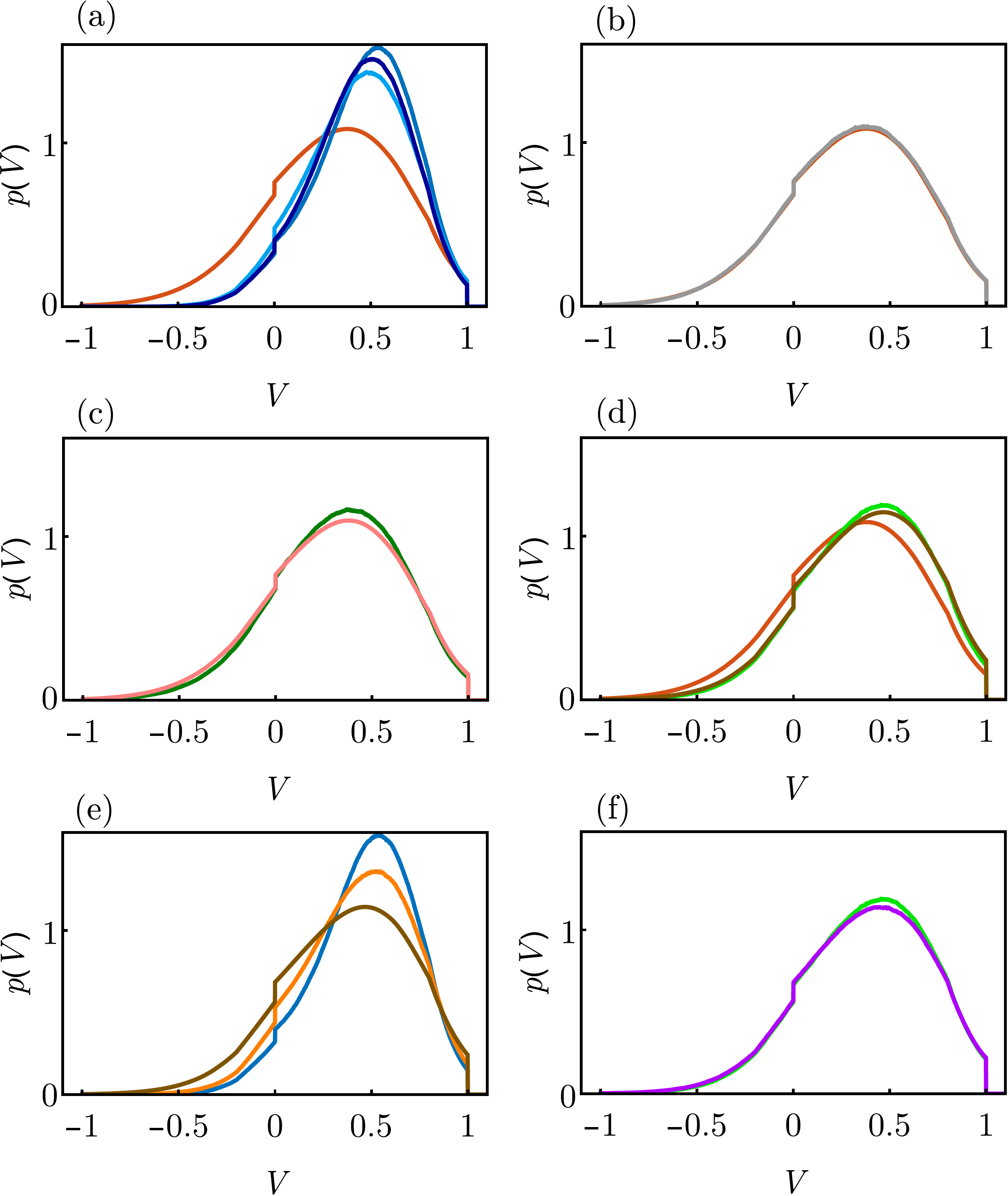}

\caption{Comparison of different numerically sampled and analytical voltage
probability densities $p(V)$ of LIF neurons. (a) Analytical density
from Eqs.~\eqref{eq:Continuity equation stationary} and \eqref{eq:GLIF self-consistent}
(red) and numerically sampled $p(V)$ of three neurons receiving different
sets of $K=50$ spike trains from the network of Fig.~\ref{fig:rasterplot}
(different shades of blue). (b) Analytical estimate (red, mostly covered)
and numerically sampled $p(V)$ of an LIF neuron receiving Poisson
input with the same rate (gray). (c) Similar plot as in (a) with a
single numerically sampled distribution of an LIF neuron receiving
$K=50$ spike trains from a larger network of $N=10000$ neurons (dark
green) and an additional analytically estimated $p(V)$ using Eqs.~\eqref{eq:Continuity equation stationary}
and \eqref{eq:rho analytic final result} with input rate $r$ set
to the rate of the superposed network spike trains (light red). (d)
$p(V)$ of an LIF neuron receiving $K=50$ time-shifted spike trains
from the network of Fig.~\ref{fig:rasterplot} (light green), analytical
estimate with the same input rate (brown) and analytical estimate
as in (a) (red). (e) $p(V)$ of LIF neurons receiving $K=50$ spike
trains from the network of Fig.~\ref{fig:rasterplot} (blue), Gamma
process input with the same rate and CV (orange) and Poisson input
with the same rate (brown). (f) $p(V)$ of LIF neurons receiving shifted
spike trains as in (d) (light green) and Gamma process input with
the same rate and CV (purple).\label{fig:LIFdistributions-1}}
\end{figure}

In the following, we further discuss the discrepancy between the voltage
probability density of an LIF neuron obtained by the shot noise approach
and the one observed if the inputs are spike trains recorded in the
network of Fig.~\ref{fig:rasterplot}, cf.~Figs.~\ref{fig:probdensity}a
and \ref{fig:LIFdistributions-1}a. The analytical density obtained
by the shot-noise approach, Eq.~\eqref{eq:Continuity equation stationary}
with rate given by Eq.~\eqref{eq:GLIF self-consistent}, matches
that of an LIF neuron receiving a Poisson spike train with the same
rate and spike impact strength, see Fig.~\ref{fig:LIFdistributions-1}b.
Hence, we can attribute the observed discrepancy for LIF neurons with
network spike train input to deviations of the spike trains' rate
and the assumed Poisson statistics.

We expect that the discrepancy is mainly caused by spatial correlations
that arise in a rather dense network of $N=100$ neurons with an indegree
of $K=50$. To substantiate this we reduce the correlations in two
ways: First, we use spike trains from a sparse network with $N=10000$
and $K=50$ to generate the neuron input. Secondly, we randomly shift
the individual spike trains of the original $N=100$ network in time
before superposing them to generate the input; this eliminates spatial
correlations while keeping the temporal correlations of the individual
spike trains intact. Fig.~\ref{fig:LIFdistributions-1}c,d shows
that both manipulations strongly reduce the discrepancy to the analytical
density. Some of the remaining discrepancy is due to the difference
between the network spike rate and the result of Eq.~\eqref{eq:GLIF self-consistent},
see Fig.~\ref{fig:LIFdistributions-1}d.

Finally we explore the impact of the reduced variability of the inter-spike
intervals. For this, we use Poisson and Gamma process input spike
trains. The latter are completely characterized by their rate and
the coefficient of variation of the inter-spike-interval distribution,
which we match to those of the superposed spike trains of the network
of Fig.~\ref{fig:rasterplot}. The quality of approximation increases
when taking into account the reduced variability, see Fig.~\ref{fig:LIFdistributions-1}e.
Also if the input is a superposition of shifted spike trains, accounting
for it still slightly improves the similarity between the resulting
$p(V)$, compare Fig.~\ref{fig:LIFdistributions-1}f and Fig.~\ref{fig:LIFdistributions-1}d
(which matches the result for Poisson input).

\subsection{Mean-field Lyapunov exponents\label{subsec:Meanfield-1}}

In the following we compute the mean-field LEs. For this, we describe
the dynamics by a sequence of discrete maps from the state at a time
(strictly speaking: infinitesimally) shortly after generation of a
spike to the state at a time shortly after generation of the next
spike. We take a stroboscopic map approach, i.e.~the times remain
unchanged if a small perturbation is applied to the dynamics. The
dynamics of small perturbations are encoded in the Jacobian matrices
at each time point. Specifically, for our discrete description we
need the single spike Jacobians \citep{MW10,MW12} $J(k)$. They generally
describe the linear evolution of infinitesimal perturbations from
time $t_{k}^{+}=t_{k}+\varepsilon$ (with $\varepsilon>0$ arbitrarily
small) shortly after the $k$th spike event in a network to time $t_{k+1}^{+}$
shortly after the next spike. For our effective single neuron dynamics,
they reduce to scalar factors
\begin{equation}
J(k)=\frac{\partial V(t_{k+1}^{+})}{\partial V(t_{k}^{+})},\label{eq:ssjac-MF-1}
\end{equation}
where the relevant events are the spike generations of the considered
neuron. To compute $J(k)$, we first recall that the free evolution
between spikes until time $t_{k+1}^{-}=t_{k+1}-\varepsilon$ yields
\begin{align}
V(t_{k+1}^{-})= & V(t_{k}^{+})\mathrm{e}^{-\gamma(t_{k+1}-t_{k})}\nonumber \\
 & +\frac{I^{\mathrm{ext}}}{\gamma}\left[1-\mathrm{e}^{-\gamma(t_{k+1}-t_{k})}\right],\label{eq:ssjac0-MF-1}
\end{align}
see Eq.~\eqref{eq:IF}. At $t_{k+1}$ the neuron is reset,
\begin{align}
V(t_{k+1}^{+})= & V(t_{k+1}^{-})-V_{\text{th}},\label{eq:event LIF XIF-MF-1}
\end{align}
and $V(t_{k+1}^{-})=\vth$ implies an implicit dependence of $t_{k+1}$
on $V(t_{k}^{+})$ via
\begin{align}
\vth= & V(t_{k}^{+})\mathrm{e}^{-\gamma(t_{k+1}-t_{k})}\nonumber \\
 & +\frac{I^{\mathrm{ext}}}{\gamma}\left[1-\mathrm{e}^{-\gamma(t_{k+1}-t_{k})}\right].\label{eq:ssjac1new-MF-1}
\end{align}
We now consider the evolution of an infinitesimal perturbation of
the membrane potential. According to Eqs.~\eqref{eq:ssjac0-MF-1},
\eqref{eq:event LIF XIF-MF-1}, the perturbation $\delta V(t_{k}^{+})$
changes until $t_{k+1}^{-}$ by a factor $\mathrm{e}^{-\gamma(t_{k+1}-t_{k})}$.
Further, it generates a perturbation $\delta t_{k+1}=\bigl(\partial t_{k+1}/\partial V(t_{k}^{+})\bigr)\delta V(t_{k}^{+})$
of $t_{k+1}$. The different evaluation time before the spike event
results in an additional membrane potential change $\dot{V}(t_{k+1}^{-})\delta t_{k+1}$,
which lets the neuron reach the threshold at $\left(t_{k+1}+\delta t_{k+1}\right)^{-}$.
Since we have a stroboscopic description, we need to compensate the
time shift to obtain the state at $t_{k+1}^{+}$. This is achieved
by subtracting $-\dot{V}(t_{k+1}^{+})\delta t_{k+1}$. A perturbation
$\delta V(t_{k}^{+})$ of the state at $t_{k}^{+}$ thus generates
at time $t_{k+1}^{+}$ a perturbation
\begin{align}
\delta V(t_{k+1}^{+})= & \mathrm{e}^{-\gamma(t_{k+1}-t_{k})}\delta V(t_{k}^{+})\nonumber \\
 & +\left[\dot{V}(t_{k+1}^{-})-\dot{V}(t_{k+1}^{+})\right]\nonumber \\
 & \times\frac{\partial t_{k+1}}{\partial V(t_{k}^{+})}\delta V(t_{k}^{+})\label{eq:Perturbation propagation MF-1}
\end{align}
and the resulting mean-field Jacobian reads
\begin{align}
J_{\mf}(k)= & \mathrm{e}^{-\gamma(t_{k+1}-t_{k})}\nonumber \\
 & +\left[\dot{V}(t_{k+1}^{-})-\dot{V}(t_{k+1}^{+})\right]\frac{\partial t_{k+1}}{\partial V(t_{k}^{+})}.\label{eq:Single Spike Jacobian Step 1 MF-1}
\end{align}
Application of the implicit function theorem, 
\begin{equation}
\frac{\partial t_{k+1}}{\partial V(t_{k}^{+})}=-\frac{1}{\dot{V}(t_{k+1}^{-})}\frac{\partial V(t_{k+1}^{-})}{\partial V(t_{k}^{+})},\label{eq:Implicit partial derivative of spike time-1}
\end{equation}
and inserting Eqs.~\eqref{eq:IF}, \eqref{eq:ssjac0-MF-1}, \eqref{eq:event LIF XIF-MF-1},
\eqref{eq:ssjac1new-MF-1}, \eqref{eq:ffreeXIF} or \eqref{eq:ffreeLIF}
results in
\begin{align}
J_{\mf}(k) & =\frac{\dot{V}(t_{k+1}^{+})}{\dot{V}(t_{k+1}^{-})}\mathrm{e}^{-\gamma(t_{k+1}-t_{k})}\nonumber \\
 & =\exp\left(\frac{\gamma}{\rho_{\text{free}}}-\gamma(t_{k+1}-t_{k})\right).\label{eq:Mean field Jacobian final-1}
\end{align}
We note that another, equivalent derivation of $J(k)$ first computes
the voltages at a fixed time $t^{\prime}$ between $t_{k+1}$ and
$t_{k+2}$ in terms of the voltages at another fixed time $t$ between
$t_{k}$ and $t_{k+1}$. Taking derivatives leads to the Jacobian
for the dynamical evolution from $t$ to $t^{\prime}$. The limits
$t\searrow t_{k}$ and $t^{\prime}\searrow t_{k+1}$ then yield $J(k)$.

The growth rate of perturbations and thus the mean-field LE are given
by the long-term average of Eq.~\eqref{eq:Mean field Jacobian final-1},
\begin{align}
\lambda_{\mf} & =\lim_{L\to\infty}\frac{1}{t_{L}}\sum_{k=0}^{L-1}\ln\left|J_{\mf}(k)\right|\nonumber \\
 & =\lim_{L\to\infty}\frac{1}{t_{L}}\sum_{k=0}^{L-1}\frac{\gamma}{\rho_{\mathrm{\text{free}}}}-\gamma\nonumber \\
 & =-\gamma\left(1-\frac{\rho}{\rho_{\text{free}}}\right).\label{eq:meanfieldexponent-1}
\end{align}

\subsection{Network single spike Jacobian\label{subsec:Single-Spike-Jacobian-1}}

The components of the single spike Jacobian $J(k)$ are given by
\begin{equation}
J_{ij}(k)=\frac{\partial V_{i}(t_{k+1}^{+})}{\partial V_{j}(t_{k}^{+})}.\label{eq:ssjac-1}
\end{equation}
To compute them, as in our mean-field approach we need to take into
account the decay of perturbations between spikes as well as the reset
of the neuron sending the $(k+1)$th spike, say neuron $l$. $V_{l}(t_{k+1}^{-})=\vth$
implies an implicit dependence of the spike time $t_{k+1}$ on $V_{l}(t_{k}^{+})$
as in Eq.~\eqref{eq:ssjac1new-MF-1}. Additionally, we now have to
include the jump-like potential change by $C_{il}h_{i}(V_{i}(t_{k+1}^{-}))$
that the spike induces in neuron $i$, such that
\begin{align}
V_{i}(t_{k+1}^{+})= & V_{i}(t_{k}^{+})\mathrm{e}^{-\gamma_{i}(t_{k+1}-t_{k})}\nonumber \\
 & +\frac{I_{i}^{\mathrm{ext}}}{\gamma_{i}}\left[1-\mathrm{e}^{-\gamma_{i}(t_{k+1}-t_{k})}\right]\nonumber \\
 & +C_{il}h_{i}(V_{i}(t_{k+1}^{-}))-\delta_{il}V_{\text{th}}.\label{eq:event and free dynamics LIF XIF-1}
\end{align}
The stroboscopic description yields a dependence 
\begin{align}
\delta V_{i}(t_{k+1}^{+})= & \mathrm{e}^{-\gamma_{i}(t_{k+1}-t_{k})}\delta V_{i}(t_{k}^{+})\nonumber \\
 & +\left[\dot{V}_{i}(t_{k+1}^{-})-\dot{V}_{i}(t_{k+1}^{+})\right]\nonumber \\
 & \times\frac{\partial t_{k+1}}{\partial V_{l}(t_{k}^{+})}\delta V_{l}(t_{k}^{+})
\end{align}
of the perturbation $\delta V_{i}(t_{k+1}^{+})$ on the perturbations
at the state at $t_{k}^{+}$. This is analogous to Eq.~\eqref{eq:Perturbation propagation MF-1},
with the difference that the neuron that sends the spike and determines
the shift in $t_{k+1}$ (neuron $l$) may be different from neuron
$i$. The Jacobian thus reads
\begin{align}
J_{ij}(k)= & \delta_{ij}\mathrm{e}^{-\gamma_{i}(t_{k+1}-t_{k})}\nonumber \\
 & +\delta_{jl}\left[\dot{V}_{i}(t_{k+1}^{-})-\dot{V}_{i}(t_{k+1}^{+})\right]\frac{\partial t_{k+1}}{\partial V_{l}(t_{k}^{+})}.\label{eq:Single Spike Jacobian Step 1-1}
\end{align}
Application of the implicit function theorem and inserting Eqs.~\eqref{eq:IF},
\eqref{eq:event and free dynamics LIF XIF-1}, \eqref{eq:ssjac1new-MF-1}
results in

\begin{align}
J_{ij}(k)= & \delta_{ij}\mathrm{e}^{-\gamma_{i}(t_{k+1}-t_{k})}\nonumber \\
 & +\delta_{jl}\frac{\gamma_{i}}{\gamma_{l}}\frac{\delta_{il}\vth-C_{il}h_{i}(V_{i}(t_{k+1}^{-}))}{\frac{I_{l}^{\mathrm{ext}}}{\gamma_{l}}-V_{l}(t_{k}^{+})},\label{eq:Single spike Jacobian-1}
\end{align}
for an LIF or an XIF neuron $i$.

\subsection{Volume contraction\label{subsec:Volume-contraction}}

The volume expansion and the sum of LEs are given by the time averaged
logarithms of the determinants of the Jacobians \citep{pikovsky+politi}.
We thus have
\begin{equation}
\sum_{i=1}^{N}\lambda_{i}=\lim_{L\to\infty}\frac{1}{t_{L}}\sum_{k=0}^{L-1}\ln\left|\det J(k)\right|\label{eq:sum of Lyap-1}
\end{equation}
in terms of single spike Jacobians \citep{MW12}. The specific form
of our $J(k)$ allows to split it into a diagonal matrix $\hat{J}(k)$
covering the perturbation change during subthreshold evolution and
a rank one correction,
\begin{equation}
J(k)=\hat{J}(k)+\vec{u}(k)\vec{v}(k)^{\mathrm{T}},
\end{equation}
where
\begin{align}
\hat{J}_{ij}(k) & =\delta_{ij}\mathrm{e}^{-\gamma_{i}\del{t_{k+1}-t_{k}}},\\
v_{j}(k) & =\delta_{jl},\\
u_{i}(k) & =\frac{\gamma_{i}}{\gamma_{l}}\left[\frac{\delta_{il}\vth-C_{il}h_{i}(V_{i}(t_{k+1}^{-}))}{\frac{I_{l}^{\mathrm{ext}}}{\gamma_{l}}-V_{l}(t_{k}^{+})}\right].
\end{align}
The matrix determinant lemma now allows to compute $\det J(k)$ via
\begin{equation}
\det J(k)=\left[1+\vec{v}(k)^{\mathrm{T}}\hat{J}(k)^{-1}\vec{u}(k)\right]\det\hat{J}(k).
\end{equation}
Eq.~\eqref{eq:ssjac1new-MF-1} and the relation $1+\vth/(I_{l}^{\mathrm{ext}}/\gamma_{l}-\vth)=\exp(\gamma_{l}/\rho_{\text{\text{free}},l})$
for the free spike frequency $\rho_{\text{\text{free}},l}$ of neuron
$l$ (see Eqs.~\eqref{eq:ffreeXIF}, \eqref{eq:ffreeLIF}) lead to
\begin{equation}
\det J(k)=\exp\left(\frac{\gamma_{l}}{\rho_{\text{free},l}}-\left(\sum_{i=1}^{N}\gamma_{i}\right)(t_{k+1}-t_{k})\right).\label{eq:det Jk final-1}
\end{equation}
Time averaging yields
\begin{align}
\sum_{i=1}^{N}\lambda_{i} & =\lim_{L\to\infty}\frac{1}{t_{L}}\sum_{k=0}^{L-1}\ln\det J_{k}\nonumber \\
 & =\lim_{L\to\infty}\frac{1}{t_{L}}\sum_{k=0}^{L-1}\frac{\gamma_{l(k)}}{\rho_{\mathrm{\text{free},}l(k)}}-\sum_{j=1}^{N}\gamma_{j}\nonumber \\
 & =-\sum_{j=1}^{N}\gamma_{j}\left(1-\frac{\rho_{j}}{\rho_{\text{free},j}}\right),\label{eq:sum of Lyap final-1}
\end{align}
where the index $l(k)$ denotes the neuron that spikes at time $t_{k}$
and $\rho_{j}$ is the spike rate of neuron $j$ in the network.

\subsection{Dependence of the Lyapunov spectrum on indegree and network size\label{sec:LyapSpect-Dependence-N-K}}

\noindent 
\begin{figure}
\includegraphics[scale=0.25]{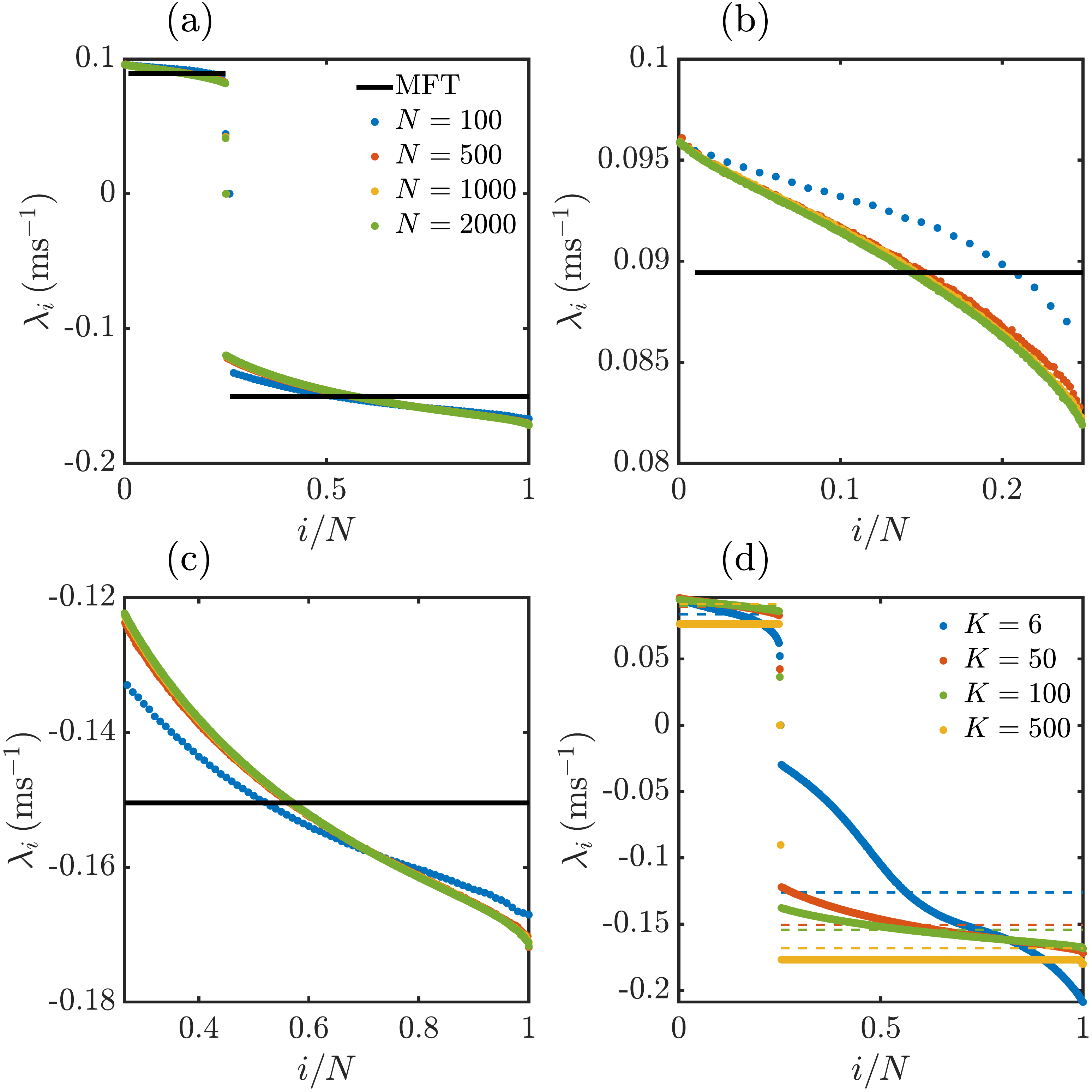}

\caption{\label{fig:kNspectra}Lyapunov spectra of networks with different
sizes and indegrees. (a) Lyapunov spectra for different network sizes
$N$ and constant indegree $K=50$ (numerical results: color coded
dots; result of the mean field theory: horizontal black lines). (b,c)
Closeups of the positive and negative parts of the spectra in (a).
(d) Lyapunov spectra for different indegrees $K$ and size $N=500$
(numerical results: color coded dots; results of the mean field theory:
color coded dashed lines). Remaining network parameters are as in
Fig.~\ref{fig:rasterplot}.}
\end{figure}
The rule that the number of negative (positive) LEs approximately
equals the number of LIF (XIF) neurons holds independent of $N$ and
$K$, see Fig.~\ref{fig:kNspectra}. Fig.~\ref{fig:kNspectra}a-c
indicates that for large $N$ and sufficiently large fixed indegree
the Lyapunov spectrum assumes a fixed shape, which differs from the
result of our mean field approach. This is because the mean field
approach neglects the nonzero off-diagonal entries of the single-spike
Jacobians, whose strength and average number $K$ do not depend on
$N$, see~App.~\ref{subsec:Single-Spike-Jacobian-1}. The shape
of the Lyapunov spectrum varies with the indegree of the network.
For larger ratios $K/N$ the positive and negative parts of the Lyapunov
spectrum become flatter, cf.~Fig.~\ref{fig:kNspectra}d. We note,
however, that also the spiking becomes more regular. We observe for
very sparse but still strongly connected networks that the Lyapunov
spectrum is no longer well approximated by our mean field theory,
cf.~Fig.~\ref{fig:kNspectra}d.

\subsection{Computing covariant Lyapunov vectors\label{sec:CLValgorithm}}

We compute the CLVs in a dynamical manner \citep{ginelli,pikovsky+politi}.
In short, if we want to compute them at $t=0$, we start sufficiently
long before with an arbitrary set of $N$ orthonormal vectors $\vec{q}_{i}(t_{0})$,
which forms a basis of the tangent space. We evolve this basis forward
until zero and further to a sufficiently long time $t_{f}$, using
the single spike Jacobians. Every few steps, we reorthonormalize the
basis. The orthogonalizations leave the first vector $\vec{q}_{1}(t)$
unchanged. It thus evolves freely (up to normalization) until it has
aligned with the first covariant Lyapunov vector at $t=0$ and thus
also at $t=t_{f}$. The second vector, $\vec{q}_{2}(t)$, is kept
orthogonal to $\vec{q}_{1}(t)$. Since it otherwise evolves freely,
$\vec{q}_{2}(0)$ will lie in the subspace of the first and the second
CLV at $t=0$, which are in general not orthogonal; the same holds
for $\vec{q}_{2}(t_{f})$ at $t=t_{f}$. Analogously $\vec{q}_{3}(0)$
will be in the subspace of $\vec{v}_{1}(0)$, $\vec{v}_{2}(0)$, and
$\vec{v}_{3}(0)$, and so on. As noted in Sec.~\ref{subsec:Lyapunov spectrum numerics},
the growth rates of the vectors already yield the LEs. In order to
find the CLVs one uses the time reversal property: we evolve the vectors
$\vec{q}_{i}(t_{f})$ along the previously taken forward trajectory
back in time until $t=0$. During this, we keep them restricted to
their respective subspaces, which are known from the forward propagation.
The vectors will then align with the least expanding directions of
their subspaces, so the backpropagated $\vec{q}_{2}(t_{f})$ will
align with $\vec{v}_{2}$, the backpropagated $\vec{q}_{3}(t_{f})$
with $\vec{v}_{3}$, and so on. We concretely implemented the simple
and efficient algorithm derived in ref.~\citep{ginelli}, which performs
the backpropagation by representing and mapping the vectors in terms
of their components in the bases $\vec{q}_{i}(t)$. After obtaining
the CLVs at $t=0$, those in the not too distant future can be obtained
using Eq.~\eqref{eq:covariant}.
\end{document}